\documentclass[graybox, nosecnum]{svmult}

\usepackage{natbib}

\usepackage{mathptmx}       % selects Times Roman as basic font
\usepackage{helvet}         % selects Helvetica as sans-serif font
\usepackage{courier}        % selects Courier as typewriter font
\usepackage{type1cm}        % activate if the above 3 fonts are
\usepackage{makeidx}         % allows index generation
\usepackage{graphicx}        % standard LaTeX graphics tool
\usepackage{multicol}        % used for the two-column index
\usepackage[bottom]{footmisc}% places footnotes at page bottom
\usepackage{hyperref}        %for hyperlinks
\usepackage{soul}            % for high-lighting of text

\bibpunct{(}{)}{;}{}{}{,}

\makeindex             % used for the subject index
\def\gtsim {>\kern-1.2em\lower1.1ex\hbox{$\sim$}~}   % Greater than sim
\def\ltsim {<\kern-1.2em\lower1.1ex\hbox{$\sim$}~}   % Less than sim

\begin{document}
%\tableofcontents{}
\title*{Chemo-Dynamical Evolution of Galaxies}
\author{Chiaki Kobayashi \thanks{Chiaki Kobayashi} and Philip Taylor}
\institute{Chiaki Kobayashi \at Centre for Astrophysics Research,
Department of Physics, Astronomy and Mathematics
University of Hertfordshire,
College Lane, Hatfield  AL10 9AB, UK \email{c.kobayashi@herts.ac.uk}
\and Philip Taylor \at Research School of Astronomy and Astrophysics, Australian National University, Canberra, ACT 2611, Australia}
\maketitle
\abstract{
Stars are fossils that retain the history of their host galaxies. Elements heavier than helium are created inside stars and are ejected when they die. From the spatial distribution of elements in galaxies, it is therefore possible to constrain the physical processes during galaxy formation and evolution. This approach, Galactic archaeology, has been popularly used for our Milky Way Galaxy with a vast amount of data from Gaia satellite and multi-object spectrographs to understand the origins of sub-structures of the Milky Way. Thanks to integral field units, this approach can also be applied to external galaxies from nearby to distant universe with the James Webb Space Telescope. In order to interpret these observational data, it is necessary to compare with theoretical predictions, namely chemodynamical simulations of galaxies, which include detailed chemical enrichment into hydrodynamical simulations from cosmological initial conditions. These simulations can predict the evolution of internal structures (e.g., metallicity radial gradients) as well as that of scaling relations (e.g., the mass-metallicity relations). After explaining the formula and assumptions, we will show some example results, and discuss future prospects.
}

\section{Introduction}

\begin{figure}[t]
\begin{center}
\includegraphics[width=8cm]{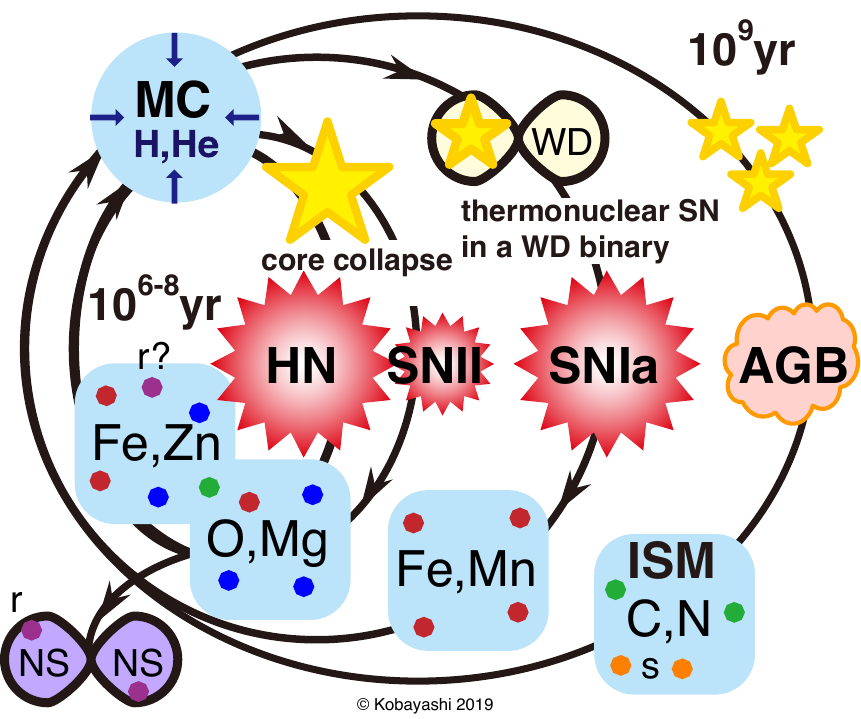}
\caption{\label{fig:intro}
Schematic view of chemical enrichment in galaxies.
}
\end{center}
\end{figure}

Explaining the origin of the elements is one of the scientific triumphs linking nuclear physics with astrophysics. As Fred Hoyle predicted, carbon and heavier elements (`metals' in astrophysics) were not produced during the Big Bang but instead created inside stars. The so-called $\alpha$ elements (O, Ne, Mg, Si, S, Ar, and Ca) are mainly produced by core-collapse supernovae, while iron-peak elements (Cr, Mn, Fe, and Ni) are more produced by thermonuclear explosions, observed as Type Ia supernovae (SNe Ia; \citealt{kob20sr}, hereafter K20). The production depends on the mass of white dwarf (WD) progenitors, and a large fraction of SNe Ia should come from near-Chandrasekhar (Ch) mass explosions (see \citealt{kob20ia} for constraining the relative contribution between near-Ch and sub-Ch mass SNe Ia).
Among core-collapse supernovae, hypernovae ($\gtsim 10^{52}$ erg) produce a significant amount of Fe as well as Co and Zn, and a significant fraction of massive stars ($\gtsim 20M_\odot$) should explode as hypernovae in order to explain the Galactic chemical evolution (GCE; \citealt{kob06}).

Heavier elements are produced by neutron-capture processes. The slow neutron-capture process (s-process) occurs in asymptotic giant branch (AGB) stars \citep[e.g.,][]{busso99,her05,cri11,kar14}, while the astronomical sites of rapid neutron-capture process (r-process) have been debated. The possible sites are neutron-star (NS) mergers \citep[NSMs,][]{wan14,jus15}, magneto-rotational supernovae \citep[MRSNe,][]{nis15,mos18,rei21}, magneto-rotational hypernovae \citep[MRHNe,][]{yon21a}, 
accretion disks/collapsars \citep{sie19},
and common envelope jet supernovae \citep{gri22}. Light neutron-capture elements (e.g., Sr) are also produced by electron-capture supernovae (ECSNe, \citealt{wan13ec}), $\nu$-driven winds \citep{arc07,wan13nu}, and rotating massive stars \citep{fri16,lim18}.

The cycles of chemical enrichment are schematically shown in Figure \ref{fig:intro}, where each cycle produces different elements and isotopes with different timescales.
In a galaxy, not only the total amount of metals, i.e. metallicity $Z\equiv\Sigma_{i>{\rm He}}\,m_i/m$, but also elemental abundance ratios, [X/Fe]$\equiv{\log N_X/N_{\rm Fe}/(N_{X,\odot}/N_{{\rm Fe},\odot}})$, evolve as a function of time. 
Therefore, we can use all of this information as fossils to study the formation and evolutionary histories of the galaxy. This approach is called Galactic archaeology, and several on-going and future surveys with multi-object spectrographs (MOS; e.g., APOGEE, HERMES-GALAH, Gaia-ESO, DESI, WEAVE, 4MOST, MOONS, Subaru Prime Focus Spectrograph (PFS), and Maunakea Spectroscopic Explorer (MSE)) are producing a vast amount of observational data of elemental abundances.
Moreover, integral field unit (IFU) spectrographs (e.g., SAURON, SINFONI, CALIFA, SAMI, MaNGA, KMOS, MUSE, HECTOR, and NIRSpec on JWST) allow us to measure metallicity and some elemental abundance ratios within galaxies. It is now possible to apply the same approach not only to our own Milky Way but also to other types of galaxies or distant galaxies. Let us call this extra-galactic archaeology.

While the evolution of the dark matter in the standard $\Lambda$ cold dark matter (CDM) cosmology is reasonably well understood, how galaxies form and evolve is still much less certain because of the complexity of the baryon physics such as star formation and feedback. The thermal energy ejected from supernovae to the interstellar medium (ISM) suppresses star formation, while the production of heavy elements in these supernovae enhances gas cooling. Since these processes affect each other, galaxy formation and evolution are complicated, and have to be solved consistently with a numerical simulation. Feedback from central super-massive black holes (SMBHs) is also found to be very important for explaining the observed properties of galaxies \citep[e.g.,][]{cro06,spr05agn,hop06,dub12,tay14}.

Since the 90’s, the development of high performance computers and computational techniques has made it possible to simulate the formation and evolutionary history of galaxies, not only of isolated systems \citep[e.g.,][]{bur87,kat92,ste94,mih96,kaw03,kob04} but also for cosmological simulations of individual galaxies
\citep[e.g.,][]{nav94} or of the galaxy population as a whole \citep[e.g.,][]{cen99,kob07}.
Thanks to the public release of hydrodynamical simulation codes such as Gadget \citep{spr01}, RAMSES \citep{tey02}, and AREPO \citep{spr10}, it became easier to run galaxy simulations, and even simulation data are made public (e.g., EAGLE, Illustris).
Needless to say, the simulation results highly depend on the input baryon physics, namely nuclear astrophysics, for predicting chemical abundances.
In this review we first discuss how the yields are constrained with observations using a simple chemical evolution model, and then how chemical enrichment can be calculated in more sophisticated, chemodynamical simulations of galaxies.

\section{Galactic chemical evolution}
\label{sec:gce}

Galactic chemical evolution (GCE) has been calculated analytically and numerically since the 70's \citep[e.g.,][hereafter K00]{tinsley80,pra93,tim95,pagel97,chi97,matteucci01,kob00} basically integrating the following equation:
\begin{equation}\label{eq:gce}
\frac{d(Z_if_{\rm g})}{dt}=E_{\rm SW}+E_{\rm SNcc}+E_{\rm SNIa}+E_{\rm NSM}-Z_i\psi+Z_{i,{\rm inflow}}R_{\rm inflow}-Z_iR_{\rm outflow}
\end{equation}
where the mass fraction of each element $i$ in gas-phase ($f_{\rm g}$ denotes the gas fraction, or the gas mass in the system considered with a unit mass) increases via element ejections from stellar winds ($E_{\rm SW}$), core-collapse supernovae ($E_{\rm SNcc}$), Type Ia supernovae ($E_{\rm SNIa}$), and neutron star mergers ($E_{\rm NSM}$). It also decreases by star formation (with a rate $\psi$), and can be modified by inflow (with a rate $R_{\rm inflow}$) and outflow (with a rate $R_{\rm outflow}$) of gas in/from the system considered.
It is assumed that the chemical composition of gas is instantaneously well mixed in the system (called an one-zone model), but the instantaneous recycling approximation is not adopted nowadays \citep{matteucci21}.
The model with $R_{\rm inflow}=R_{\rm outflow}=0$ is called a closed-box model, but is not realistic in any observed galaxies.
The initial conditions are $f_{{\rm g},0}=1$ (a closed system) or $f_{{\rm g},0}=0$ (an open system) with the chemical composition ($Z_{i,0}$) from the Big Bang nucleosynthesis ($Z_{i,{\rm BBN}}$). External enrichment is often neglected, assuming $Z_{i,{\rm inflow}}=Z_{i,{\rm BBN}}$.
In Eq.(\ref{eq:gce}) the first two terms depend only on nucleosynthesis yields, while the third and fourth terms also depend on modelling of the progenitor binary systems, which is uncertain. The last three terms are galactic terms, and should be determined from galactic dynamics, but are assumed with analytic formulae in GCE models.
The stars formed at a given time $t$ have the initial metallicity, $Z(t)$, which is equal to the metallicity of the ISM from which the stars form\footnote{Later, integrated metallicity of stellar populations in a galaxy will be described as $Z_*$.}.

\subsection{The metal ejection terms in GCE}

The ejection terms are given by integrating the contributions with various initial mass ($m$) and metallicity ($Z$) of progenitor stars. Hence, the first assumption is the initial mass function (IMF), $\phi$.
The IMF is often assumed to have time/metallicity-invariant mass spectrum normalized to unity at $m_\ell \leq m \leq m_u$ as
\begin{equation}
\phi(m) \propto m^{-x} ,
\end{equation}
\begin{equation}
\int_{m_\ell}^{m_u} m^{-x} dm= \frac{1}{1-x} (m_u^{1-x}-m_\ell^{1-x}) = 1 .
\end{equation}
\citet{sal55} found the slope of the power-law, $x=1.35$, from observation of nearby stars.
See \citet{tinsley80} for a different definition of the slope, $\alpha=x+1$.
\citet{kro93} updated this result finding three slopes at different mass ranges, which are summarised as $x=-0.7, 0.3$, and $1.3$ for $m/M_\odot \ltsim 0.08, 0.08 \ltsim m/M_\odot \ltsim0.5$, $0.05\ltsim m/M_\odot \ltsim 150$, respectively \citep{kro08}. However, the author suggested that the slope of the massive end should be $x=1.7$ in the solar neighborhood, which has been used in some GCE works \citep[e.g.,][]{rom10}.
In most of results in this paper the Kroupa IMF (with the massive-end slope $x=1.3$) is adopted for a mass range from $m_\ell=0.01 M_\odot$ to $m_u=120 M_\odot$, while \citet{cha03}'s IMF is often used in hydrodynamical simulations.
Note that there are a few claims for an IMF variation from observations \citep[e.g.,][]{vandokkum10,gunawardhana11,cappellari13}.

Then the ejection terms for single stars can be calculated as follows\footnote{The metals in Eq.(\ref{eq:e_sw}) and Eq.(\ref{eq:e_cc}) are called `unprocessed' and `processed' metals, respectively. Our supernova yield table contains `processed' metals only, the AGB yield table contains net yields, so that the mass loss with star particle's chemical composition should also be included with Eq.(\ref{eq:e_sw}).}.
\begin{equation}
E_{\rm SW}=\int_{m_t}^{m_u}\,(1-w_m-p_{z_im,{\rm II}})\,Z_i(t-\tau_m)\,
\psi(t-\tau_m)\,\phi(m)~dm ,
\label{eq:e_sw}
\end{equation}
\begin{equation}
E_{\rm SNcc}=\int_{m_t}^{m_u}\,p_{z_im,{\rm II}}\,
\psi(t-\tau_m)\,\phi(m)~dm .
\label{eq:e_cc}
\end{equation}
The lower mass limit for integrals is the turning off mass $m_t$ at $t$, which
is the mass of the star with the main-sequence lifetime $\tau_m=t$.
$\tau_m$ is a lifetime of a star with $m$, and also depends on $Z$ of the star.
$w_m$ is the remnant mass fraction, 
which is the mass fraction of a WD (for stars with initial masses of $m\ltsim8M_\odot$), a NS (for $\sim 8-20M_\odot$) or a black hole (BH) (for $\gtsim20M_\odot$). 
$p_{z_im,{\rm II}}$ denotes the nucleosynthesis yields of core-collapse supernovae, given as a function of $m$, $Z$, and explosion energy.

The additional ejection terms from binary systems are more complicated. Suppose that the event rates are given, the enrichment can be calculated as
\begin{equation}
E_{\rm Ia}=m_{\rm CO}\,p_{z_im,{\rm Ia}}\,{\cal R}_{t,{\rm Ia}} ,
\label{eq:e_ia}
\end{equation}
\begin{equation}
E_{\rm NSM}=m_{\rm NSM\,ejecta}\,p_{z_im,{\rm NSM}}\,{\cal R}_{t,{\rm NSM}} .
\label{eq:e_nsm}
\end{equation}
${\cal R}_{t,{\rm x}}$ are the rates of SNe Ia and NSMs per unit time per unit stellar mass formed in a population of stars with a coeval chemical composition and age (simple stellar population, SSP), and called the delay-time distribution (DTD).
For the DTDs, simple analytic formula were also proposed \citep[][and references therein]{matteucci21}. However, the functions are significantly different from what are obtained from binary population synthesis (BPS), ignore metallicity dependence during binary evolution, and do not include the effects of supernova kicks for NSMs. \citet{ded04} was the first work that combined BPS to GCE.

All of the matter in the WD ($m_{\rm CO}$) is ejected from SNe Ia (except for a sub-class called SNe Iax, see \citealt{kob15}); for Ch-mass SNe Ia, $m_{\rm CO}=1.38M_\odot$.
On the other hand, only a small fraction of matter is ejected from a NSM ($m_{\rm NSM\,ejecta}\sim0.01M_\odot$), depending on the mass ratios, the equation of state of the NSs, and the spin of BHs \citep{kob22}.
These are provided together with the nucleosynthesis yields ($p_{z_im,{\rm Ia}}$ and $p_{z_im,{\rm NSM}}$).

For SNe Ia from the single-degenerate systems, \cite{kob98} proposed another analytic formula based on binary calculation:
\begin{equation}
{\cal R}_{t,{\rm Ia}}=b~
\int_{\max[m_{{\rm p},\ell}(Z),\,m_t]}^{m_{{\rm p},u}(Z)}\,
\frac{1}{m}\,\phi(m)~dm~
\int_{\max[m_{{\rm d},\ell}(Z),\,m_t]}^{m_{{\rm d},u}(Z)}\,
\frac{1}{m}\,\psi(t-\tau_m)\,\phi_{\rm d}(m)~dm .
\label{eq:r_ia}
\end{equation}
The first integral is for the primary star, and the mass range is for the stars that can produce $\sim 1M_\odot$ of C+O WDs in binaries, which is set to be $\sim 3-8M_\odot$.
The second integral is for the secondary star, and the mass range depends on the optically thick winds from the WD, which is about $\sim1M_\odot$ and $\sim3M_\odot$ for the red-giant+WD systems and main-sequence+WD systems, respectively, depending on $Z$.

The metallicity dependence on the SN Ia DTD is very important to reproduce the observed [$\alpha$/Fe]--[Fe/H] relation in the solar neighbourhood \citep[see][for the detailed parameters and the resultant DTDs]{kob09}.
These parameters are for Ch-mass explosions. Eq.(\ref{eq:r_ia}) can also be used for SNe Iax and sub-Ch mass explosion triggered by slow H accretion \citep[see][for the detailed parameters]{kob15}, but not for the double-degenerate systems (which are likely to cause sub-Ch mass explosions).
DTDs of these sub-classes of SNe Ia have been predicted by various BPS models, but none of these models can reproduce the observations because the total rate is too low and/or the typical timescale is too short \citep[see][for more details]{kob22}.

For NSMs, K20 used a metallicity-dependent DTD from a BPS \citep{men14,men16}.
\citet{kob22} used various BPS models and also provided new analytic formulae that can reproduce the observed [Eu/(Fe,O)]--[(Fe,O)/H] relation only with NSMs, without the r-process associated with core-collapse supernovae.
Currently, there is no BPS model that can explain the observation only with NSMs because the rate is too low and/or the timescale is too long \citep[see][for more details]{kob22}.
Other binary systems such as novae can also contribute GCE for some elements or isotopes \citep[e.g.,][]{rom19} but the DTDs are very uncertain \cite[see][for the nova DTDs from BPS]{kem22} and the yield tables are not available.

\subsection{Nucleosynthesis yields}

Nucleosynthesis yields ($p_{z_im,{\rm X}}$) are integrated over stellar lifetimes of single stars, or delay time distributions for binaries, depending on $Z$ (Eqs. \ref{eq:e_sw}-\ref{eq:e_nsm}). It is extremely important to take account of this metallicity dependence, though it is often ignored in hydrodynamical simulations. Because of this, it is also not possible to solve chemical evolution as a post-process. {\it Chemical enrichment including all elements must be followed on-the-fly.}

For core-collapse supernovae, our yields were originally calculated in \citet{kob06}, 3 models of which are replaced in \citet{kob11agb} (the identical table was also used in \citealt{nom13}), and a new set with failed supernovae is used in K20. This is based on the lack of observed progenitors at supernova locations in the HST data \citep{sma09}, and on the lack of successful explosion simulations for massive stars \citep{jan12,bur21}.
This resulted in a 20\% reduction of both the net and oxygen yields (see Table 3 of K20).
Supernova yields assumed $E_{51}\equiv E/10^{51} {\rm ergs}=1$ of explosion energy as observed for SN1987A, while hypernova yields included the observed mass dependence of the explosion energy; $E_{51}=$ 10, 10, 20, 30  for 20, 25, 30, $40M_\odot$ stars.
These supernova and hypernova yield tables are provided separately, and it is recommended to assume $\epsilon_{\rm HN}=$50\% hypernova fraction for stars with $\ge 20M_\odot$ \citep{kob06}. This fraction is expected to decrease at high metallicities due to smaller angular momentum loss, and was assumed as $\epsilon_{\rm HN}(Z)=0.5, 0.5, 0.4, 0.01$, and $0.01$ for $Z=0, 0.001, 0.004, 0.02$ in \citet{kob11mw}.

\citet{woo95} provided a yield table as a function of mass and metallicity, which was tested with a GCE model by \citet{tim95}, and has been widely used in chemical evolution studies. It would be useful to note that stellar mass loss is not included in their pre-supernova evolution calculations. Another problem is that their core-collapse supernova models tend to produce more Fe than observed because of the relatively deep mass cut, which leads to [$\alpha$/Fe]$\sim0$ in the ejecta. Therefore, the theoretical Fe yields were divided by a factor of 2 \citep{tim95,rom10}, which is obviously unphysical. This artificial reduction could mimic a shallower mass cut, but in that case the yields of other iron peak elements formed in the same layer as Fe should also be reduced. For the first problem, \citet{por98} obtained C+O core masses from stellar evolution models with mass loss and adopted to the \citet{woo95} supernova yields, which is not physically accurate either. It is also known that Mg production in the $40M_\odot$ model with $E_{51}=1$ is unreasonably small compared with that in other models. This Mg underproduction problem also exists in \citet{por98}.

\citet{lim18} also provided a yield table but as a function of stellar rotation. See \citet{kob22iau} for the impact of stellar rotation in GCE. Note that these yields do not reproduce the observed elemental abundances in the solar neighbourhood, namely iron-peak elements. This is because this yield set does not have hypernovae, and mixing during explosion is not taken into account either.

There are also pre-supernova yields in the literature that do not include explosive nucleosynthesis. These are not useful for GCE since during explosions iron-peak elements are produced and $\alpha$ element yields are also largely modified, although they can be used for studying isotopic ratios of light elements.

The yields for AGB stars were originally calculated in \citet{kar10} and \citet{kar16}, but a new set with the s-process is used in K20 with optimising the mass of the partial mixing zone.
In K20, the narrow mass range of super-AGB stars is also filled with the yields from \citet{doh14a}; stars at the massive end are likely to become ECSNe, as is believed to have been the case for the progenitor of the Crab Nebula in 1054. At the low-mass end, off-centre ignition of C flame moves inward but does not reach the centre, which remains a hybrid C+O+Ne WD. This might become a sub-class of SNe Ia called SN Iax. The contribution of SN Iax is negligible in the solar neighbourhood, but can be important for dwarf spheroidal galaxies \citep{kob15,ces17}.
There are other yields in the literature, see \citet{rom10} and \citet{rom19} for comparison with GCE.

For SNe Ia, the nucleosynthesis yields of near-Ch and sub-Ch mass models are newly calculated in \citet{kob20ia}, which used the same 2D code as in \citet{leu18} and \citet{leu20} but with more realistic, solar-scaled initial composition. The initial composition gives significantly different (Ni, Mn)/Fe ratios, compared with the classical W7 model \citep{nom97ia} or more recent delayed detonation model \citep{sei13}. When constraining the progenitors of SNe Ia from the observed Mn abundances in the solar neighborhood, it is important to use the latest yields of SNe Ia.
See \citet{blondin22} for a complete comparison including many other yields.

The r-process yields are taken from the following references in K20: $8.8M_\odot$ model in \citet[][2D]{wan13ec} for electron capture supernovae, $1.2-2.4M_\odot$ models in \citet{wan13nu} for $\nu$-driven winds, $1.3M_\odot$+$1.3M_\odot$ model in \citet[][3D-GR]{wan14} for neutron star mergers, and $25M_\odot$ ``b11tw1.00'' model in \citet[][axisymmetric MHD]{nis15} for magnetorotational supernovae.
See \citet{kob22} for the GCE models including the neutron star merger yields from \citet{jus15} with various masses of compact objects.
These yields are uncertain depending on 1) initial conditions (e.g., $M$, $Z$, rotation, magnetic fields), 2) the quality of the base hydrodynamical simulations (3 dimensional general relativity or not), 3) neutrino physics, and 4) nuclear physics (e.g., reaction rates, fission modelling).

\begin{figure}[t]
\begin{center}
\includegraphics[width=0.7\textwidth]{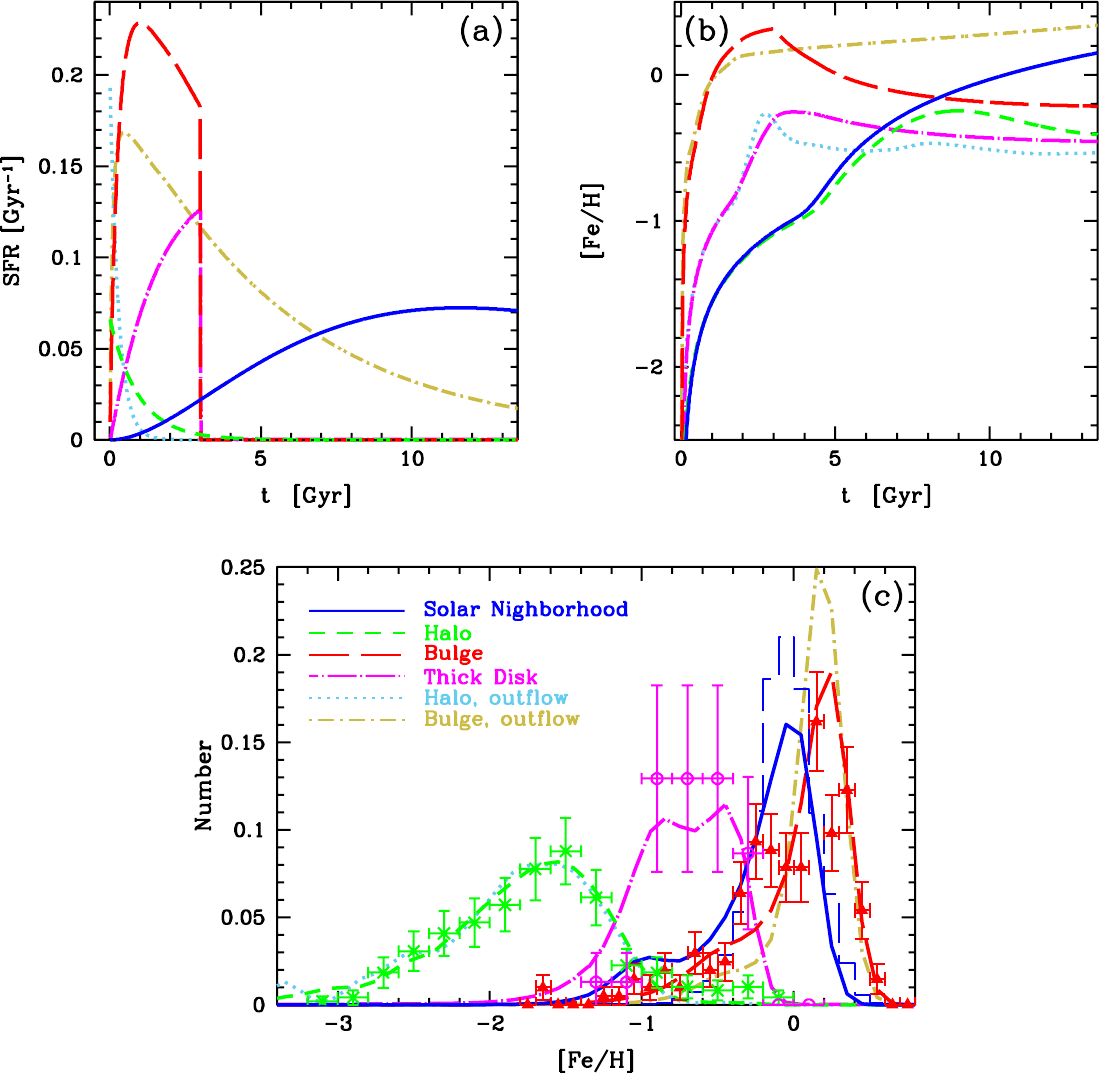}
\caption{\label{fig:mdf}
Star formation histories (panel a), age-metallicity relations (panel b), and metallicity distribution functions (panel c) for the 
solar neighborhood (blue solid lines),
halo (green short-dashed lines),
halo with stronger outflow (light-blue dotted lines),
bulge (red long-dashed lines),
bulge with outflow (olive dot-short-dashed lines)
and thick disk (magenta dot-long-dashed lines).
Figure is taken from \citet{kob20sr}, see the reference for the observational data sources and the model details.
}
\end{center}
\end{figure}

\begin{figure}[t]
\begin{center}
\includegraphics[width=0.7\textwidth]{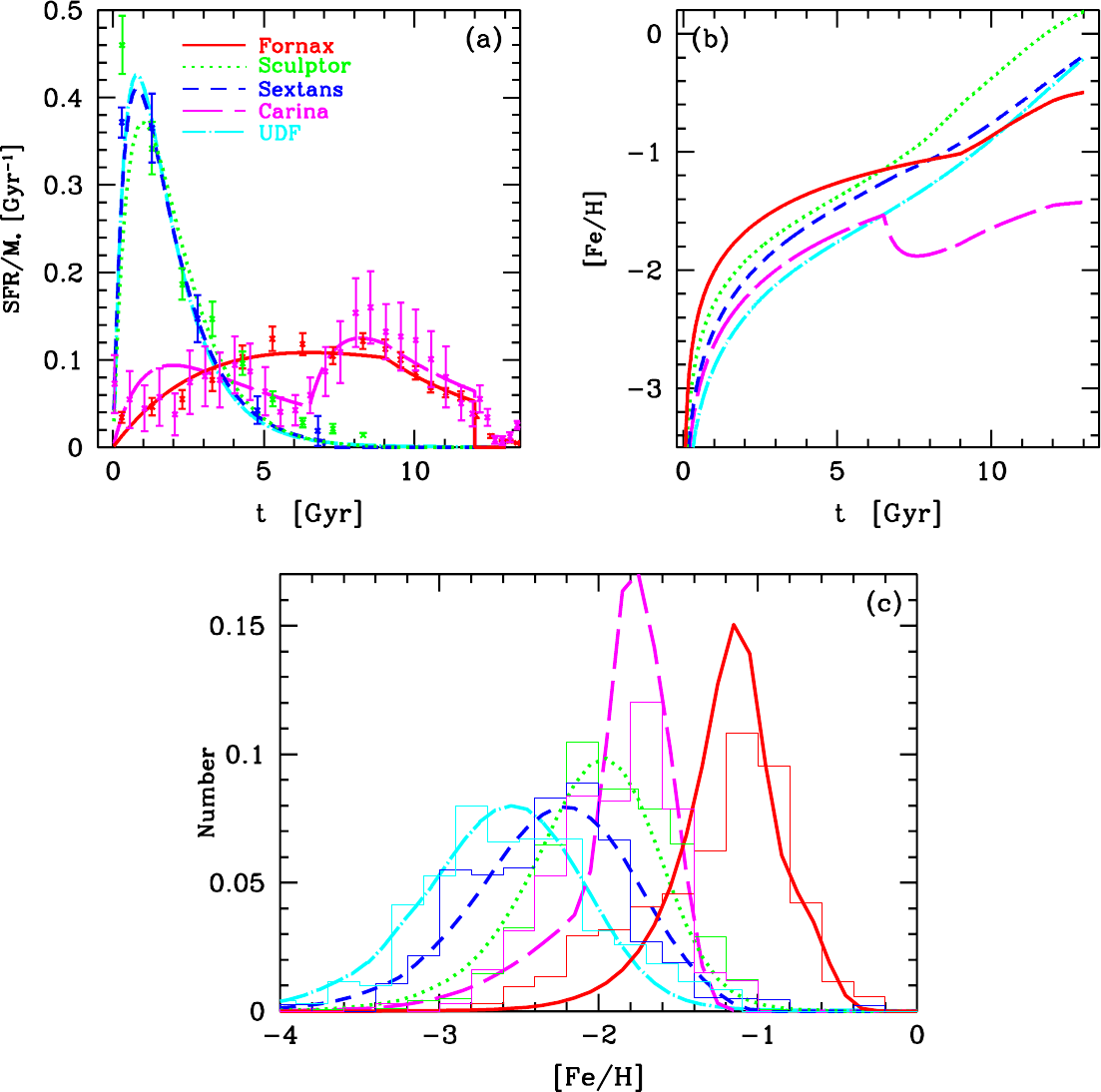}
\caption{\label{fig:mdf2}
The same as Fig. \ref{fig:mdf} but for dwarf spheroidal galaxies, Fornax (red solid lines), Sculptor (green long-dashed lines), Sextans (blue short-dashed lines), and Carina (magenta dotted lines), and ultra-faint dwarf (UFD) galaxies (cyan dot-dashed lines).
Figure is updated from \citet{kob20ia}, see the reference for the observational data sources and the model details.
The observational data for the UFD is provided by A. Ji (priv. comm.).
}
\end{center}
\end{figure}

\subsection{The galactic terms in GCE}

Star formation rate is usually modelled as $\psi\!=\!\frac{1}{\tau_{\rm s}}f_{\rm g}^{\,n}$ but often $n=1$ is adopted, and thus it is proportional to the gas fraction. ${\tau_{\rm s}}$ denotes a timescale of star formation, and $1/{\tau_{\rm s}}$ gives the efficiency of star formation.
The inflow rate is often assumed to be exponential as, $R_{\rm inflow}\!\!=\!\frac{1}{\tau_{\rm i}}\exp\frac{-t}{\tau_{\rm i}}$, but occasionally as $R_{\rm inflow}\!\!=\!\frac{t}{\tau_{\rm i}^2}\exp\frac{-t}{\tau_{\rm i}}$ in \citet{pagel97} and in the solar neighbourhood model of K20.
The outflow rate is assumed to be proportional to the star formation rate as $R_{\rm outflow}\!\!=\!\frac{1}{\tau_{\rm o}}f_{\rm g}^{\,n}\propto\psi$, which is reasonable if the outflow is driven by supernova feedback. In addition, star formation is quenched ($\psi=0$) at a given epoch $t_{\rm w}$, which corresponds to a galactic wind driven by the feedback from active galactic nuclei (AGN).

The timescales are determined to match the observed metallicity distribution function (MDF) of stars in each system modelled. 
Figures \ref{fig:mdf} and \ref{fig:mdf2} show the assumed star formation history and the resultant metallicity evolution and MDF.
The parameter sets that have very similar MDFs give almost identical tracks of elemental abundance ratios (Fig.\,A1 of \citealt{kob20ia}). This means that, for the given MDF, elemental abundance tracks do not depend so much on the star formation history.
Therefore GCE can be used to constrain nuclear astrophysics, if the MDF is known, in Galactic archaeology.
Without knowing the MDFs, star formation histories are unconstrained, and elemental abundance tracks can vary depending on the star formation histories.
Provided that nuclear astrophysics is known, elemental abundance measurements can be used to constrain the star formation histories through GCE, which is the case for extra-galactic archaeology.

\subsection{The origin of elements}

\begin{figure*}[t]
\begin{center}
  \includegraphics[width=0.95\textwidth]{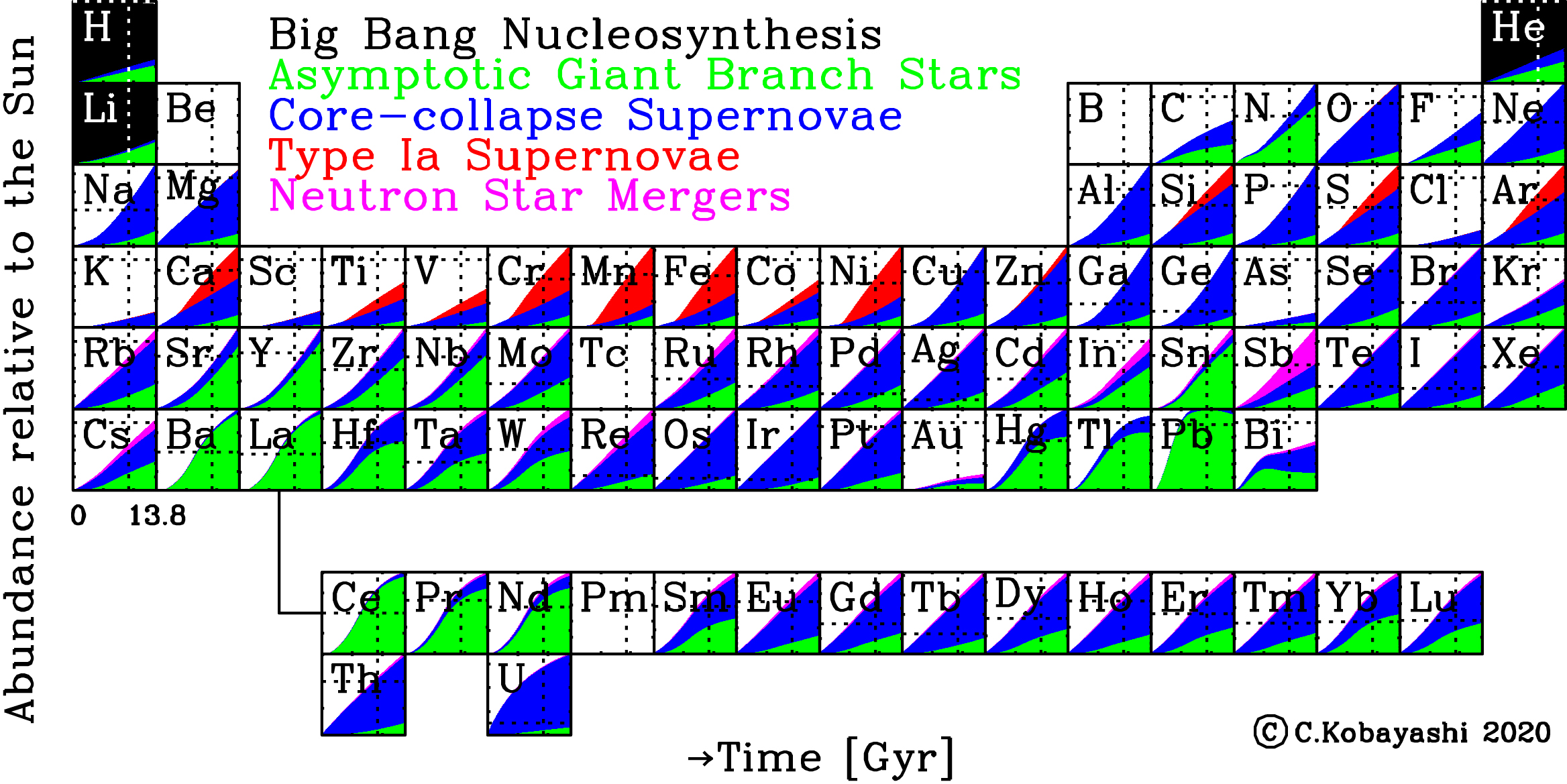}
\caption{The time evolution (in Gyr) of the origin of elements in the periodic table: Big Bang nucleosynthesis (black), AGB stars (green), core-collapse supernovae including SNe II, HNe, ECSNe, and MRSNe (blue), SNe Ia (red), and NSMs (magenta). The amounts returned via stellar mass loss are also included for AGB stars and core-collapse supernovae depending on the progenitor mass. The dotted lines indicate the observed solar values.
Figure is taken from \citet{kob20sr}.
}
\label{fig:origin}
\end{center}
\end{figure*}

Using the K20 GCE model for the solar neighbourhood, we summarize the origin of elements in the form of a periodic table. In each box of Figure \ref{fig:origin}, the contribution from each chemical enrichment source is plotted as a function of time: Big Bang nucleosynthesis (black), AGB stars (green), core-collapse supernovae including SNe II, HNe, ECSNe, and MRSNe (blue), SNe Ia (red), and NSMs (magenta).
It is important to note that the amounts returned via stellar mass loss are also included for AGB stars and core-collapse supernovae depending on the progenitor star mass (Eq.\ref{eq:e_sw}).
The x-axis of each box shows time from $t=0$ (Big Bang) to $13.8$ Gyrs, while the y-axis shows the linear abundance relative to the Sun, $X/X_\odot$.
The dotted lines indicate the observed solar values, i.e., $X/X_\odot=1$ and $4.6$ Gyr for the age of the Sun;
the solar abundances are taken from \citet{asp09}, except for $A_\odot({\rm O}) = 8.76,  A_\odot({\rm Th}) = 0.22$, and $A_\odot({\rm U}) = -0.02$ (\S 2.2 of K20 for the details).
The adopted star formation history is similar to the observed cosmic star formation rate history, and thus this figure can also be interpreted as the origin of elements in the universe, which can be summarized as follows:
\begin{itemize}
\item H and most of He are produced in Big Bang nucleosynthesis. As noted, the green and blue areas also include the amounts returned to the ISM via stellar mass loss in addition to He newly synthesized in stars.
After tiny production in Big Bang nucleosynthesis,
Be and B are supposed to be produced by cosmic rays \citep{pra93}, 
which are not included in the K20 model.
\item The Li model is very uncertain because the initial abundance and nucleosynthesis yields are uncertain. Li is supposed to be produced also by cosmic rays and novae, which are not included in the K20 model. The observed Li abundances show an increasing trend from very low metallicities to the solar metallicity, which could be explained by cosmic rays. Then the observation shows a decreasing trend from the solar metallicities to the super-solar metallicities, which might be caused by the reduction of the nova rate \citep{gri19}; this is also shown in theoretical calculation with binary population synthesis \citep{kem22}, where the nova rate becomes higher due to smaller stellar radii and higher remnant masses at low metallicities.
\item 49\% of C, 51\% of F, and 74\% of N are produced by AGB stars\footnote{In extra-galactic studies, the N production from AGB stars is referred as ``secondary'', which is confusing, and is a {\it primary} production from freshly synthesized $^{12}$C. For massive stars, the N yield depends on the metallicity of progenitor stars for {\it secondary} production, and can also be enhanced by stellar rotation for {\it primary} production \citep{kob11agb}.} (at $t=9.2$ Gyr).
For the elements from Ne to Ge, the newly synthesized amounts are very small for AGB stars, and the small green areas are mostly for mass loss.
\item $\alpha$ elements (O, Ne, Mg, Si, S, Ar, and Ca) are mainly produced by core-collapse supernovae, but 22\% of Si, 29\% of S, 34\% of Ar, and 39\% of Ca come from SNe Ia. 
These fractions would become higher with sub-Ch-mass SNe Ia \citep{kob20ia} instead of 100\% Ch-mass SNe Ia adopted in the K20 model.
\item A large fraction of Cr, Mn, Fe, and Ni are produced by SNe Ia. In classical works, most of Fe was thought to be produced by SNe Ia, but the fraction is only 60\% in the K20 model, and the rest is mainly produced by HNe. The inclusion of HNe is very important as it changes the cooling and star formation histories of the universe significantly \citep{kob07}.
Co, Cu, Zn, Ga, and Ge are largely produced by HNe.
In the K20 model, 50\% of stars at $\ge 20M_\odot$ are assumed to explode as hypernovae, and the rest of stars at $> 30M_\odot$ become failed supernovae.
\item Among neutron-capture elements, as predicted from nucleosynthesis yields, AGB stars are the main enrichment source for the s-process elements at the second (Ba) and third (Pb) peaks. 
\item 32\% of Sr, 22\% of Y, and 44\% of Zr can be produced from ECSNe, which are included in the blue areas, even with the adopted conservative mass ranges; we take the metallicity-dependent mass ranges from the theoretical calculation of super-AGB stars \citep{doh15}.
Combined with the contributions from AGB stars, it is possible to perfectly reproduce the observed trends, and no extra light element primary process (LEPP) is needed \citep[but see][]{pra18}. The inclusion of $\nu$-driven winds in GCE simulation results in a strong overproduction of the abundances of the elements from Sr to Sn with respect to the observations.
\item For the heavier neutron-capture elements, contributions from both NS-NS/NS-BH mergers and MRSNe are necessary, and the latter is included in the blue areas.
Note again that the green areas include the mass-loss component, i.e., not newly produced but recycled.
Note that Tc and Pm are radioactive.
\end{itemize}

Since the Sun is slightly more metal-rich than the other stars in the solar neighborhood (see Fig.\,2 of K20), the fiducial model in K20 goes through [O/Fe]$=$[Fe/H]$=0$ slightly later compared with the Sun's age.
Thus, a slightly faster star formation timescale ($\tau_{\rm s}=4$ Gyr instead of 4.7 Gyr) is adopted in this figure.
The evolutionary tracks of [X/Fe] are almost identical.
In this model, the O and Fe abundances go though the cross of the dotted lines, meaning [O/Fe] $=$ [Fe/H] $=0$ at 4.6 Gyr ago.
This is also the case for some important elements including N, Ca, Cr, Mn, Ni, Zn, Eu, and Th. The remaining problems can be summarised as follows:
\begin{itemize}
\item
The contribution from rotating massive stars is not included in the K20 model, which can probably explain in the underproduction of C and F \citep{kob22iau}. A binary effect during stellar evolution may also increase C.
These elements could be enhanced by AGB stars as well, but the observed high F abundance in a distant galaxy strongly supported rapid production of F from Wolf-Rayet stars \citep[][at $z=4.4$ observed with ALMA]{franco21}.
\item 
Mg is slightly under-produced in the model, although at low metallicity the model [Mg/Fe] is slightly higher than observed (see Fig.\,\ref{fig:xfe-sagb}).
This may be due to a NLTE effect \citep{lind22}, or due to a binary effect. Massive stars in binaries tend to have a smaller CO core with a higher C/O ratios \citep{brown01}, which could result in a higher Mg/O ratio. Observed Mg/O ratios suggest that this binary effect should not be important at low metallicities ($\ltsim 0.1Z_\odot$).
\item
The underproduction of the elements around Ti is a long-standing problem since \citet{kob06}. It was shown that these elemental abundances can be enhanced by multi-dimensional effects (\citealt{sne16}; see also K15 model in K20). This is due to the lack of 3D simulations that can capture the production of these (less abundant) elements. 2D nucleosynthesis calculation showed an enhancement of these elements \citep{mae03,tom09}.
\item
The s-process elements are slightly overproduced even with the updated s-process yields.
Notably, Ag is over-produced by a factor of $6$, while Au is under-produced by a factor of $5$. U is also over-produced. These problems may require revising nuclear physics modelling \citep[see][for more discussion]{kob22iau}.
\end{itemize}

\begin{figure}[t]
\begin{center}
  \includegraphics[width=0.75\textwidth]{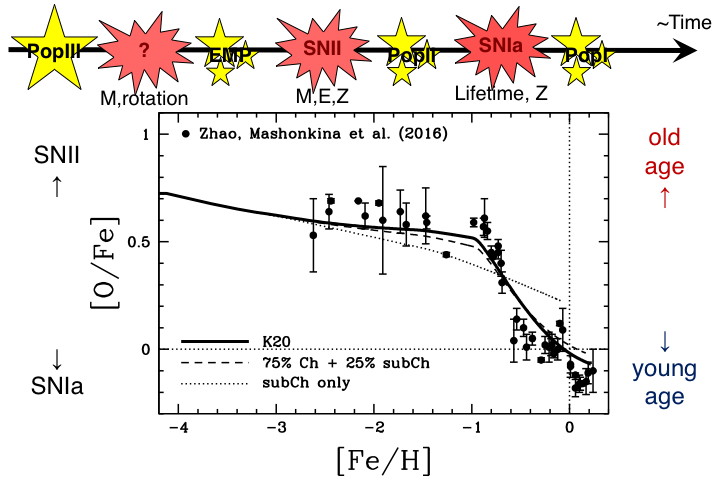}
\vspace*{-2mm}
\caption{The [O/Fe]--[Fe/H] relations in the solar neighborhood for the models with Ch-mass SNe Ia only (solid line), 75\% Ch plus 25\% sub-Ch-mass SNe Ia (dashed line), and sub-Ch-mass SNe Ia only (dotted line). The observational data (filled circles) are high-resolution non-local thermodynamic equilibrium (NLTE) abundances from \citet{zhao16}.
Figure is taken from \citet{kob20ia} with modification.
}
\label{fig:ofe}
\end{center}
\end{figure}

\subsection{The [X/Fe]--[Fe/H] relations}

Although the slopes in Figure \ref{fig:origin} show the different enrichment timescales of each element, it is easier to see the differences in the [X/Fe]--[Fe/H] diagrams\footnote{There were attempts to plot against a different element (e.g., O) to avoid the uncertainty of Fe yields \citep[e.g.,][]{cay04}, but it became rather hard to understand the plots.}.
Figure \ref{fig:ofe} shows the [$\alpha$/Fe]--[Fe/H] relation, which is probably the most important diagram in GCE. O is one of the $\alpha$ elements. 
At the beginning of the universe, the first stars (Population III stars) form and die, whose properties such as mass and rotation are uncertain and have been studied using the abundance patterns of the second generation, extremely metal-poor (EMP) stars.
Secondly, core-collapse supernovae occur, and their yields are imprinted in the abundance patterns of Population II stars in the Galactic halo. The [$\alpha$/Fe] ratio is high and stays roughly constant\footnote{There was a debate if the [O/Fe] ratio increases toward the lowest metallicity or not. UV OH line showed such an increase \citep[e.g,][]{isr98} reaching [O/Fe] $\sim 1$ at [Fe/H] $\sim -3$, which was later found to be due to 3D effects.} with a small scatter. This plateau value does not depend on the star formation history but does on the IMF.
Finally, SNe Ia occur, which produce more Fe than $\alpha$ elements, and thus the [$\alpha$/Fe] ratio decreases toward higher metallicities; this decreasing trend is seen for the Population I stars in the Galactic disk.

\begin{figure}[t]
\begin{center}
  \includegraphics[width=0.5\textwidth]{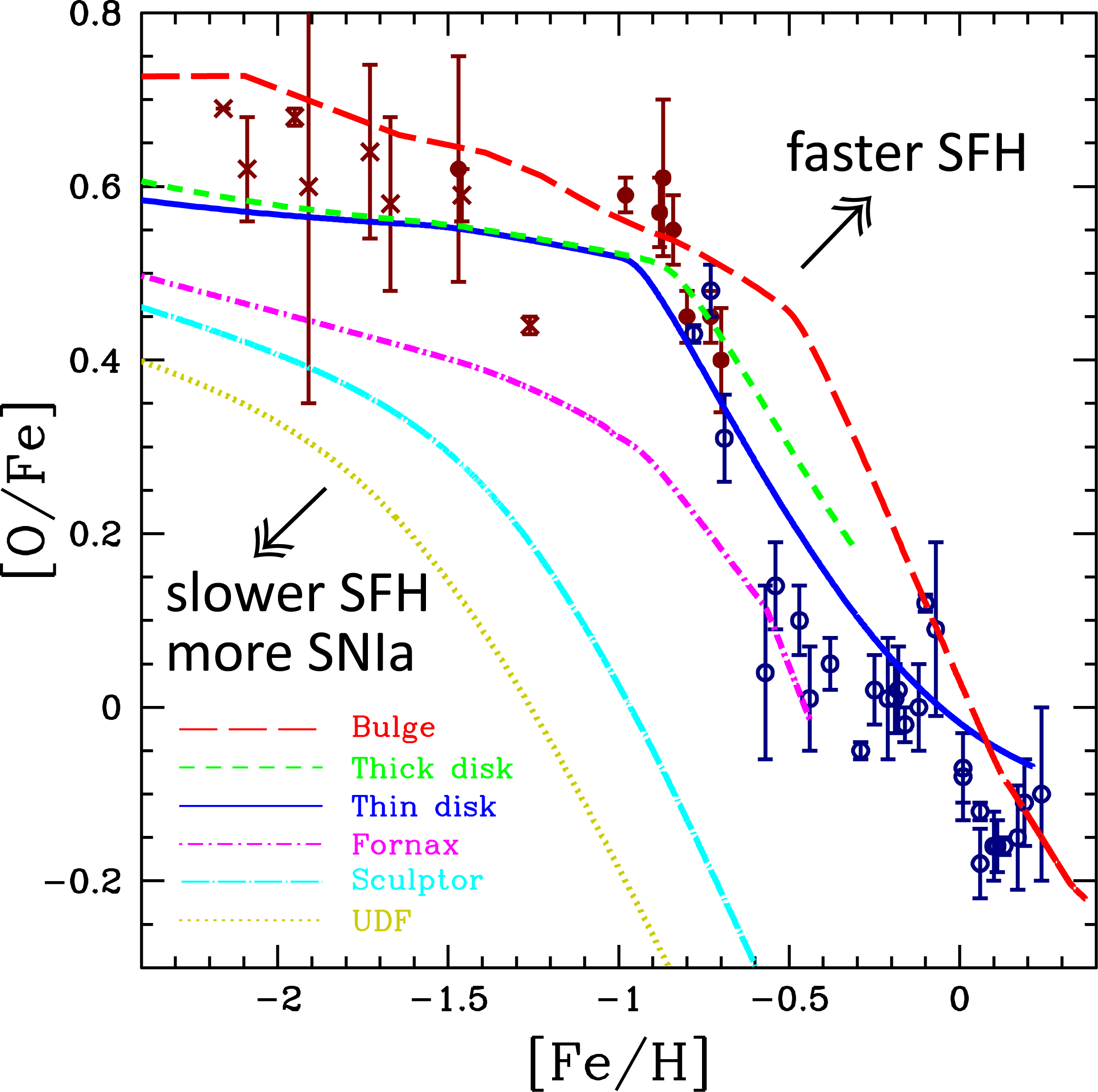}
\caption{The [O/Fe]--[Fe/H] relations in the Galactic bulge (red long-dashed line), thick disk (green short-dashed line), thin disk (blue solid line), Fornax (magenta dot-short-dashed line), Sculptor (cyan dot-long-dashed line), and ultra-faint dwarf (UFD) galaxies (olive dotted line). The star formation history of each system is constrained from its MDF in Figs.\,\ref{fig:mdf} and \ref{fig:mdf2}. See Fig.\,\ref{fig:ofe} for the observational data source for thin (navy open circles) and thick (maroon filled circles) disk stars. 
}
\label{fig:ofe-env}
\end{center}
\end{figure}

The contribution from SNe Ia depends on the progenitor binary systems, namely, the mass of the progenitor WDs. Sub-Ch mass explosions produce less Mn and Ni, and more Si, S, and Ar than near-Ch mass explosions \citep{sei13,kob20ia}.
Figure \ref{fig:ofe} shows the [O/Fe]--[Fe/H] relations with varying the fraction of sub-Ch-mass SNe Ia. Including up to 25\% sub-Ch mass contribution to the GCE (dashed line) gives a similar relation as the K20 model (solid line), while the model with 100\% sub-Ch-mass SNe Ia (dotted line) gives too low an [O/Fe] ratio compared with the observational data. For Ch-mass SNe Ia, the progenitor model is based on the single-degenerate scenario with the metallicity effect due to optically thick winds from WDs \citep{kob98}. For sub-Ch-mass SNe Ia, the observed delay-time distribution is used since the progenitors are the combination of mergers of two WDs in double degenerate systems and low accretion in single degenerate systems; \citet{kob15}'s formula are for those in single degenerate systems only.

Because of this [$\alpha$/Fe]--[Fe/H] relation, high-$\alpha$ and low-$\alpha$ are often used as a proxy of old and young ages of stars in galaxies, respectively. Note that, however, that this relation is not linear but is a plateau and decreasing trend (called a `knee'). The location of the `knee' depends on the star formation timescale, with a higher metallicity for faster star formation history \citep{mat90}, e.g., [Fe/H] $\sim -0.5, -0.8, -1, -2$ for the Galactic bulge, thick disk, thin disk, and satellite galaxies.
Figure \ref{fig:ofe-env} shows the GCE model results for various environments. The bulge (with outflow), thick disk, and solar neighborhood (thin disk) models are taken from K20 with 100\% Ch-mass SNe Ia, with the star formation histories in Fig. \ref{fig:mdf}. The models for dwarf spheroidal galaxies are taken from \citet{kob20ia} adding 100\% sub-Ch mass SNe Ia on top of the 100\% of Ch mass SNe Ia, with the star formation histories in Fig. \ref{fig:mdf2}.
Recently, a similar relation is also shown for M31 using planetary nebulae. Since Fe is not available, Ar is used as a significant fraction of Ar is produced by SNe Ia. The thin and thick disk dichotomy seems to exist, but not exactly the same as in the Milky Way, probably due to M31 experiencing an accretion of a relatively massive, gas-rich satellite galaxy \citep{arnaboldi22}.

There are GCE models that try to explain both thin and thick disk observations with two infalls \citep{chi97,spitoni19}. However, more realistic, chemodynamical simulations show that a significant number of thick disk stars are formed in satellite galaxies before they accrete onto the disk, which can be better approximated with two independent GCE models in this figure \citep[also in][]{gri17}.

\begin{figure*}[t]
\center
\includegraphics[width=0.95\textwidth]{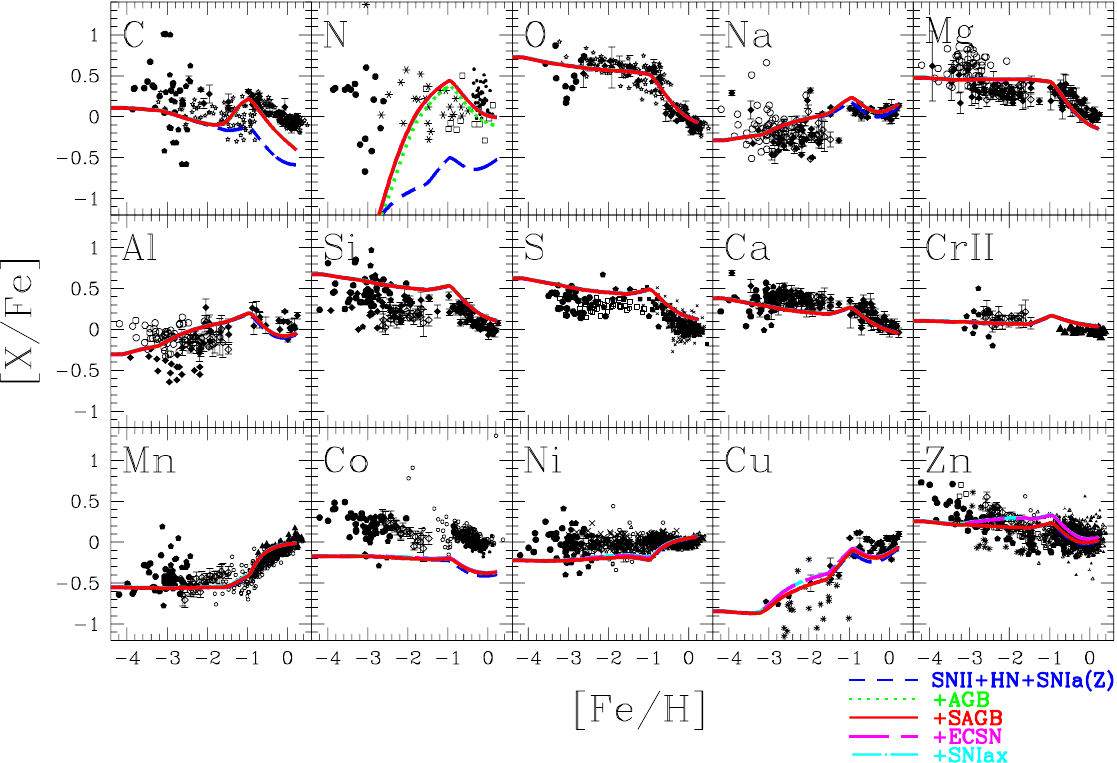}
\caption{\label{fig:xfe-sagb}
Evolution of the elemental abundances [X/Fe] from C to Zn against [Fe/H]
for the models in the solar neighborhood,
with only supernovae (without AGB and super-AGB stars, blue short-dashed line),
with AGB without super-AGB stars (green dotted lines),
with AGB and super-AGB stars (red solid line, fiducial model),
with ECSNe (magenta long-dashed lines),
and with SNe Iax (cyan dot-dashed lines).
Figure is taken from \citet{kob20sr} and see the reference for the observational data sources.
}
\end{figure*}

Figure \ref{fig:xfe-sagb} shows the evolution of elemental abundance ratios [X/Fe] against [Fe/H] from C to Zn in the solar neighborhood. 
All $\alpha$ elements (O, Mg, Si, S, and Ca) show the [$\alpha$/Fe]--[Fe/H] relations, i.e., the plateau and decreasing trend from [Fe/H] $\sim -1$ to higher metallicities.
The odd-$Z$ elements (Na, Al, and Cu) show an increasing trend toward higher metallicities due to the metallicity dependence of the core-collapse supernova yields.
See K20 for more detailed discussion on the evolutionary trends of each element, and \citet{kob22uv} for the discussion with more recent observations for Cu and Zn. In the following we focus the role of each enrichment source. 
\begin{itemize}
\item
The contribution to GCE from AGB stars (green dotted lines in Fig.\,\ref{fig:xfe-sagb}) can be seen mainly for C and N, and only slightly for Na, compared with the model that includes supernovae only (blue dashed lines).
Hence it seems not possible to explain the O--Na anti-correlation observed in globular cluster stars \citep[e.g.,][]{kraft97} with AGB stars \citep[but see][]{ventura09}.
Although AGB stars produce significant amounts of Mg isotopes, their inclusion does not affect the [Mg/Fe]--[Fe/H] relation.
\item
The contribution from super-AGB stars (red solid lines) is very small; with super-AGB stars, C abundances slightly decrease, while N abundances slightly increase.
It would be very difficult to put a constraint on super-AGB stars from the average evolutionary trends of elemental abundance ratios, but it might be possible to see some signatures of super-AGB stars in the scatters of elemental abundance ratios.
\item
With ECSNe (magenta long-dashed lines), Ni, Cu and Zn are slightly increased. These yields are in reasonable agreement with the high Ni/Fe ratio in the Crab Nebula \citep{nom87,wan09}.
\item
No difference is seen with/without SNe Iax (cyan dot-dashed lines) in the solar neighborhood because the progenitors are assumed to be hybrid WDs, which has a narrow mass range in the adopted super-AGB calculations \citet[][$\Delta M\sim0.1M_\odot$]{doh15}.
This mass range depends on convective overshooting, mass-loss, and reaction rates.
Even with the wider mass range in \citet[][$\Delta M\sim1M_\odot$]{kob15}, however, the SN Iax contribution is negligible in the solar neighborhood, but it can be important at lower metallicities such as in dwarf spheroidal galaxies with stochastic chemical enrichment \citep{ces17}.
\end{itemize}

\begin{figure*}[t]
\center
  \includegraphics[width=0.95\textwidth]{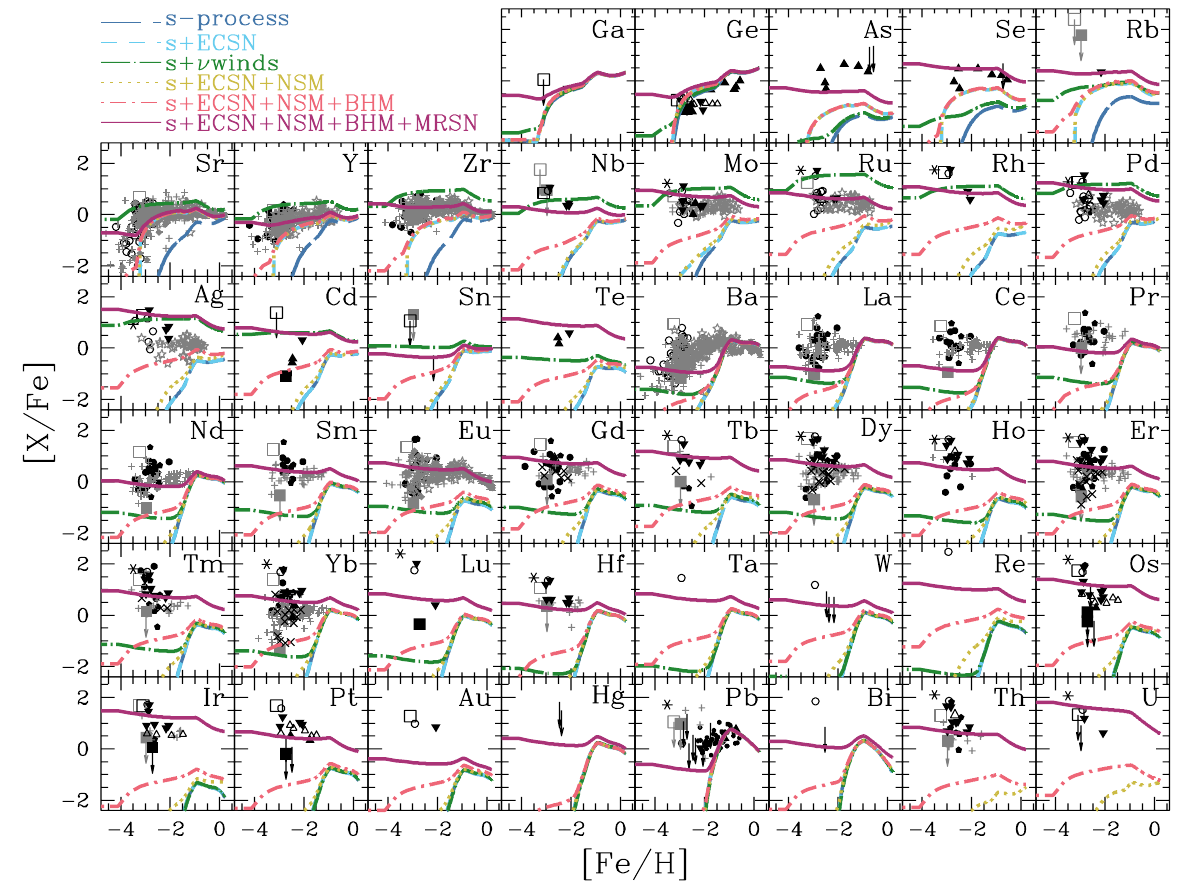}
\caption{The [X/Fe]--[Fe/H] relations for neutron capture elements, comparing to the models in the solar neighborhood, with s-process from AGB stars only (blue long-dashed lines), with s-process and ECSNe (light-blue short-dashed lines), with s-process, ECSNe, and $\nu$-driven winds (green dotted-long-dashed lines), with s-process, ECSNe, and NS-NS mergers (olive dotted lines), with s-process, ECSNe, and NS-NS/NS-BH mergers (orange dotted-short-dashed lines), with s-process, ECSNe, NS-NS/NS-BH mergers, and MRSNe (red solid lines). 
Figure is updated from \citet{kob20sr} with more observational data.
}
\label{fig:xfe}
\end{figure*}

Figure \ref{fig:xfe} shows the evolutions of neutron-capture elements as [X/Fe]--[Fe/H] relations. 
\begin{itemize}
\item
As known, AGB stars can produce the first (Sr, Y, Zr), second (Ba), and third (Pb) peak s-process elements, but no heaver elements (navy long-dashed lines).
The second-peak elements are under-produced around [Fe/H] $\sim-2$, which is eased with chemodynamical simulations, and can also be reproduced better with rotating massive stars. However, the model with rotating massive stars results in over-production of the first-peak elements \citep{kob22iau}.
\item
It is surprising that ECSNe from a narrow mass range ($\Delta M \sim 0.15-0.2M_\odot$) can produce enough of the first-peak elements; with the combination of AGB stars, it is possible to reproduce the observational data very well (cyan short-dashed lines). This means that no other light element primary process (LEPP), such as rotating massive stars, is required \citep[but see][]{ces14}. The elements from Mo to Ag seem to be overproduced, which could be tested with the UV spectrograph proposed for the VLT, CUBES.
\item
Additional production from $\nu$-driven winds leads to further over-production of these elements in the model (green dot-long-dashed lines), but this should be studied with more self-consistent calculations of supernova explosions.
\item
Neutron star mergers can produce lanthanides and actinides, but not enough (olive dotted lines); the rate is too low and the timescale is too long, according to binary population synthesis. This is not improved enough even if NS-BH mergers are included (orange dot-short-dashed lines).
It is very unlikely this problem will be solved even if the mass-dependence of the nucleosynthesis yields are included, unless the BH spins are unexpectedly high \citep{kob22}.
\item
In the GCE model with MRSNe (magenta solid lines), it is possible to reproduce a plateau at low metallicities for Eu, Pt, and Th, relative to Fe. However, even with including both MRSNe and neutron star mergers, the predicted Au abundance is more than ten times lower than observed. This underproduction is seen not only for the solar abundance but also for low metallicity stars although the observational data are very limited. UV spectroscopy with HST, or NASA's future LUVOIR, is needed to investigate this problem further.
\end{itemize}

In conclusion, an r-process associated with core-collapse supernovae, such as MRSNe, is required. The same conclusion is obtained with other GCE models and more sophisticated chemodynamical simulations \citep[e.g.,][]{hay19,van20}, as well as from the observational constraints of radioactive nuclei in the solar system \citep{wal21}. See more discussion in the later section.

It seems not to be easy to solve the missing gold problem with astrophysics; only Au yields should be increased since Pt is already in good agreement with the current model, and Ag is rather overproduced in the current model. However, there are uncertainties in nuclear physics, namely in some nuclear reaction rates and in the modelling of fission, %\citep{shi16,vas20}, 
which might be able to increase Au yields only, without increasing Pt or Ag.
It may be hard to predict the only one stable isotope of $^{197}$Au, while Pt has several stable isotopes.
The predicted Th and U abundances are after the long-term decays, to be compared with observations of metal-poor stars, and the current model does not reproduce the Th/U ratio either.

\subsection{The first galaxies}

\begin{figure}[t]
\begin{center}
\includegraphics[width=0.95\textwidth]{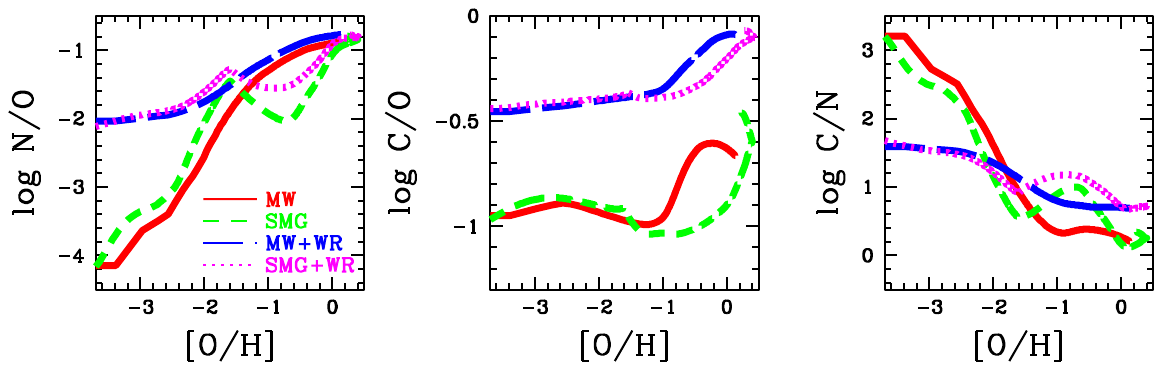}
\vspace*{-1mm}
\caption{\label{fig:wwcno}
Evolution of CNO (number) abundances in GCE models for the Milky Way (red solid and blue long-dashed lines) and for a submillimetre galaxy (green short-dashed and magenta dotted lines), which are calculated with the original K20 yields (red solid and green short-dashed lines) or the yields with Wolf-Rayet stars (blue long-dashed and magenta dotted lines; see text for the model details).
}
\end{center}
\end{figure}

The Atacama Large Millimeter/submillimeter Array (ALMA) has opened a new window to study elemental abundances and isotopic ratios of light elements at very high redshifts \citep[e.g.,][]{zhang18,franco21}, which provide us an independent constraint on stellar nucleosynthesis.
In the early universe, stellar rotation may be important at low metallicities because weaker stellar winds result in smaller angular momentum loss.

The impact of stellar rotation on stellar evolution and nucleosynthesis has been studied to explain Wolf-Rayet stars \citep[e.g.,][]{mey02,hir07,lim18}, and the observed N and $^{13}$C abundances at low metallicities have been used to constrain the effect (\citealt{chi06}; see also Fig.\,13 of \citealt{kob11agb})\footnote{However, the N/O--O/H relation can be reproduced by inhomogeneous enrichment in chemodynamical simulations (Fig. \ref{fig:cno}).}.
As discussed in the previous section, the importance of jet-like supernova explosion for iron-peak elements (Co and Zn) and r-process elements (via MRSNe) also indicates the importance of stellar rotation, although how to transport the angular momentum in stellar envelopes to stellar cores is uncertain.

Figure \ref{fig:wwcno} shows the evolution of CNO abundances in GCE models for two representative galaxies with different star formation histories: the solar neighborhood in the Milky Way, which has continues star formation, and a submillimetre galaxy (SMG), which underwent a star burst.
The SMG models use the same star formation history as in \citet{franco21}, while the red solid lines are the same as in Fig.\,8 and Fig.\,10 of K20.
As discussed, $\sim50$\% C and $\sim75$\% N comes from low- and intermediate-mass AGB stars, respectively, at later times ([O/H] $\sim-1$ for C). The rest comes from massive stars, either released before supernova explosions by stellar winds, or ejected at supernova explosions.
The N yield from massive stars increases while the C yield decreases at higher metallicities. These two effects cause the increase of N/O and C/O, and the decrease of C/N, toward high metallicities.

With rotation, both C and N yields are increased due to the interplay between the core He-burning and the H-burning shell, triggered by the rotation-induced instabilities \citep{lim18}. \citet{lim18}'s yields do not reproduce the observations of iron-peak elements because of their simple description of supernova explosions. Therefore, the C and N yields from \citet{lim18} are combined with the K20 yields in the models with Wolf-Rayet stars. High abundance of CNO enhances the production of F in He convective shell, which can explain the high F abundance observed in a submillimetre galaxy at redshift $z=4.4$ \citep{franco21}.

The evolution of isotopic ratios is also largely affected by stellar rotation. Without rotation, $^{13}$C and $^{25,26}$Mg are produced from AGB stars ($^{17}$O might be overproduced in our AGB models), while other minor isotopes are more produced from metal-rich massive stars, and thus the ratios between major and minor isotopes (e.g., $^{12}$C/$^{13}$C, $^{16}$O/$^{17,18}$O) generally decrease as a function of time/metallicity (Figs.\,17-19 of \citet{kob11agb} and Fig.\,31 of K20; see also \citealt{rom19}).
Isotopic ratios of CNO, Si, S, Cl, and Ar have been estimated for a couple of spiral galaxies around $z\sim1$ \citep[][and the references therein]{wallstrom19}, as well as for various sources in the Milky Way and nearby galaxies.

With rotation, $^{13}$C and $^{17,18}$O are enhanced. The evolution of isotopic ratios and comparison to observations can be found in Fig.\,7 of \citet{kob22iau}, which gives consistent results with \citet{rom19}'s GCE models. $^{12}$C/$^{13}$C ratio became too high, while it is possible to explain the low $^{13}$C/$^{18}$O ratio observed in submillimetre galaxies at redshift $z\sim2-3$ \citep{zhang18} without changing the IMF.

Stellar rotation also causes the weak-s process that forms s-process elements from the existing seeds at a much shorter timescale than AGB stars. This may help solving the underproduction of Ba at [Fe/H] $\ltsim -1$ (Fig.\,\ref{fig:xfe}), but causes an overproduction of light s-process elements such as Sr, Y, and Zr.
The impact of stellar rotation on nucleosynthesis is not well understood yet.

\subsection{The first chemical enrichment}
\label{sec:first}

\begin{figure}[t]
\begin{center}
\includegraphics[width=0.65\textwidth]{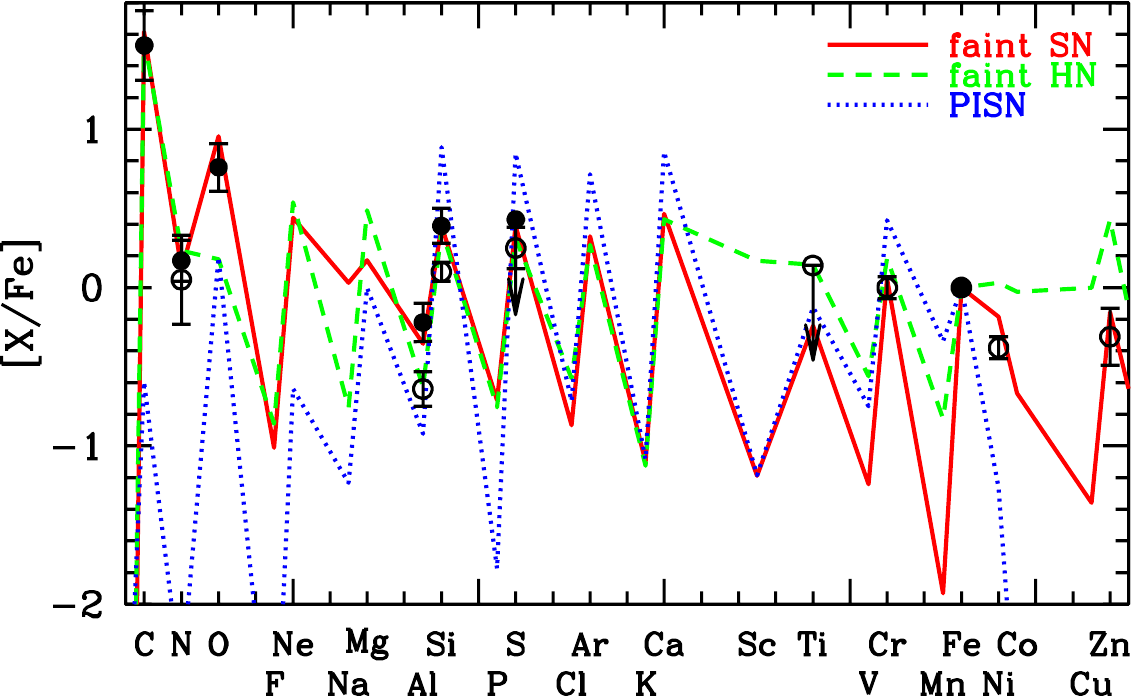}
\caption{\label{fig:cdla}
Elemental abundance patterns of the Population III supernovae.
The red solid and green short-dashed lines show the nucleosynthesis yields of faint core-collapse supernovae from $25M_\odot$ stars with mixing fallback. The blue dotted line is for pair-instability supernovae from $170M_\odot$ stars.
Stellar rotation is not included.
The dots are for the metal-poor C-rich DLA (filled circles) and peculiar DLA (open circles). 
Figure is taken from \citet{kob11dla} with modification.}
\end{center}
\end{figure}

Extremely metal-poor (EMP) stars have been an extremely useful relic in the Galactic archaeology. At the beginning of galaxy formation, stars form from a gas cloud that was enriched only by one or very small number of supernovae \citep{audouze95}, and hence the elemental abundances of EMP stars can offer observational evidence of particular types of supernovae in the past.
The expectation was that the first stars were so massive ($\sim 140-300M_\odot$) that they exploded as a pair-instability supernova \citep[PISN,][]{barkat67}, which causes high [(Si,S)/O] ratios \citep[e.g.,][]{nom13}.

However, after a half century of surveys, no star has been found with an elemental abundance pattern fitted by a PISN in the Milky Way.
Instead, it is found that quite a large fraction of massive stars become faint supernovae \citep{ume03} that give a high C/Fe ratio leaving a relatively large black hole ($\sim 5M_\odot$). 
If there were C-enhanced low-$\alpha$ stars, that would indicate black hole formation even from $10-20M_\odot$ Pop III stars \citep{kob14}.
From faint supernovae, the ejected iron mass is very small but the explosion energy can be high; faint hypernovae can explain the Zn enhancement of CEMP stars \citep{ish18}.

Similar C enhancement might also be found for damped Ly-$\alpha$ system \citep[DLA,][]{kob11dla}. Figure \ref{fig:cdla} shows the observed abundance pattern of a DLA at $z=2.34$ with [Fe/H] $\simeq -3$, comparing to the nucleosynthesis yields of a faint supernova (red solid line), a faint hypernova (green dashed line), and a PISN (blue dotted line). In recent surveys, no DLA is found with an elemental abundance pattern consistent with a PISN either.

\begin{figure}[t]
\begin{center}
  \includegraphics[width=0.65\textwidth]{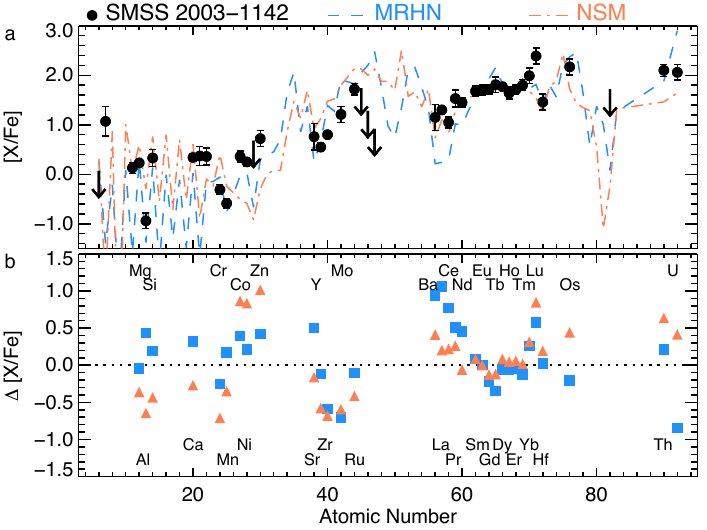}
\caption{The elemental abundance of a extremely metal-poor star SMSS J200322.54-114203.3, which has [Fe/H]$= -3.5$, comparing with mono-enrichment from a magneto-rotational hypernovae (blue dashed line and squares) and multi-enrichment including a neutron star merger (orange dot-dashed line and triangles).
The lower panel shows the differences, i.e., the observed values minus model values.
Figure is taken from \citet{yon21a}.}
\label{fig:yong}
\end{center}
\end{figure}

A small number of EMP stars show a relatively low $\alpha$ abundance, which does not necessarily mean that the ISM from which the star formed was already enriched by SNe Ia. 
The reasons that could cause low $\alpha$/Fe ratios were summarised in \citet{kob14}: (1) SNe Ia, (2) less-massive core-collapse supernovae ($\ltsim 20M_\odot$), which become more important with a low star formation rate, (3) hypernovae, although the majority of hypernovae are expected to give normal [$\alpha$/Fe] ratios ($\sim 0.4$), and (4) PISNe, which could be very important in the early universe.
Therefore, the [$\alpha$/Fe] ratio is not a perfect clock.
It is necessary to also use other elemental abundances (namely, Mn and neutron-capture elements) or isotopic ratios, with higher resolution ($>40,000$) multi-object spectroscopy on 8m class telescopes (e.g., cancelled WFMOS or planned MSE, proposed HRMOS for VLT).

EMP stars with r-process enhancement (called rII stars), namely those with enhancement of the heaviest elements (actinide-boost stars), can offer confirmation of an r-process site.
Recently, from the SkyMapper survey, \citet{yon21a} found such a star with [Fe/H] $=-3.5$, the lowest metallicity known for rII stars. The abundance pattern is shown Figure \ref{fig:yong}, which shows a clear detection of U and Th, the universal r-process pattern, normal $\alpha$ enhancement, but unexpectedly showing high Zn and N abundances ([Zn/Fe]$=0.72$, [N/Fe]$=1.07$).
This abundance pattern cannot be explained with a model with a neutron star merger (orange dot-dashed line), but instead strongly supports a magnetorotational hypernova from a $\gtsim25M_\odot$ Pop III star (blue dashed line).

There are a few magneto-hydrodynamical simulations and post-process nucleosynthesis that successfully showed enough neutron-rich ejecta to produce the 3rd peak r-process elements as in the Sun \citep{win12,mos18}. The predicted iron mass is rather small \citep{nis15,rei21}, and the astronomical object may be faint. \citet{yon21a}'s proposal is Fe-producing, magnetorotational hypernova, which is more luminous.
The r-process elements could also be formed in the accretion disk of a collapsar \citep{sie19}, but the inter ejecta that contains Fe and Zn must be ejected at the explosion as for a hypernova. As it is luminous it might be possible to detect absorption features of Au and Pt in the spectrum of a transient in future.

\begin{figure}[t]
\begin{center}
\includegraphics[width=9cm]{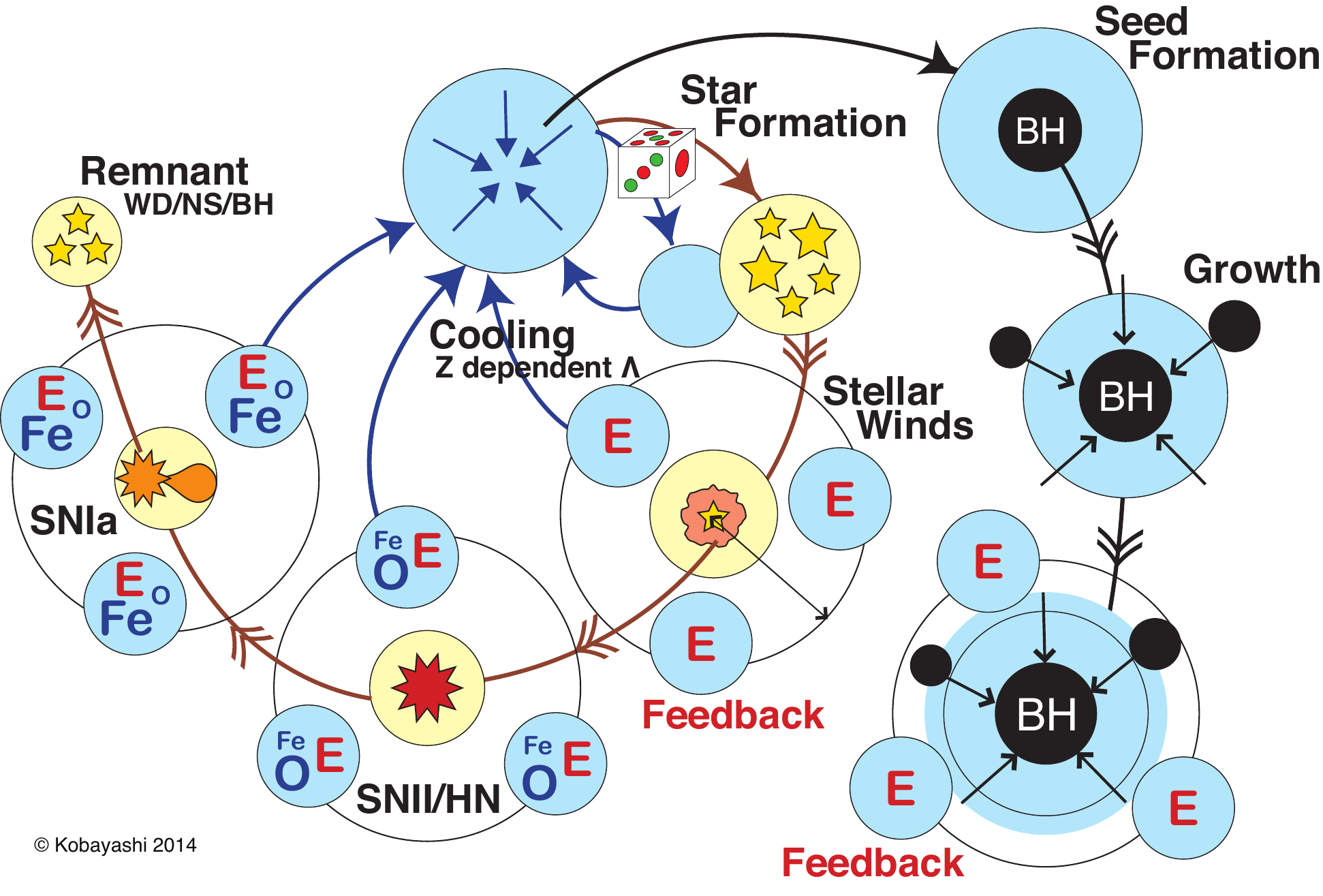}
\caption{\label{fig:hydro}
Schematic view of chemodynamical evolution of galaxies.
}
\end{center}
\end{figure}

\section{Chemodynamical evolution of galaxies}
\label{sec:hydro}

Thanks to the development of super computers and numerical techniques, it is now possible to simulate the formation and evolution of galaxies from cosmological initial conditions, i.e., initial perturbation based on $\Lambda$ cold dark matter (CDM) cosmology. Star formation, gas inflow, and outflow in Eq.(\ref{eq:gce}) are not assumed but are, in principle, given by dynamics. The baryon cycle is schematically shown in Figure \ref{fig:hydro}. Due to the limited resolution, star forming regions, supernova ejecta, and active galactic nuclei (AGN) cannot be resolved in galaxy simulations, and thus it is necessary to model star formation and the subsequent effects -- feedback -- introducing a few parameters. Fortunately, there are many observational constraints, from which it is usually possible to choose a best set of parameters. To ensure this, it is necessary to run the simulation until the present-day, $z=0$, and reproduce a number of observed relations at various redshifts.

Although hydrodynamics can be calculated with publicly available codes such as Gadget, RAMSES, and AREPO, modelling of baryon physics is the key and is different depending on the simulation teams/runs.
These are various simulations of a cosmological box with periodic boundary conditions, and containing galaxies with a wide mass range (e.g., $10^{9-12} M_\odot$ in stellar mass) at $z=0$.
In order to study massive galaxies it is necessary to increase the size of the simulation box (e.g., $\gtsim 100$ Mpc), while in order to study internal structures it is necessary to improve the spatial resolution (e.g., $\ltsim 0.5$ kpc).
The box size and resolution are chosen depending on available computational resources.
On the other hand, zoom-in techniques allow us to increase the resolution focusing a particular galaxy,
but this also requires tuning the parameters with the same resolution, comparing to a number of observations in the Milky Way.
Needless to say, elemental abundances are the most informative. There are a few zoom-in simulations for Milky Way-type galaxies \citep[e.g.,][]{brook12,few14,grand17,buck20,font20}, but in most of the cases the input nuclear astrophysics are too simple to make use of elemental abundances.

\subsection{Modelling of baryon physics}

\begin{figure}[t]
\begin{center}
  \includegraphics[width=0.4\textwidth]{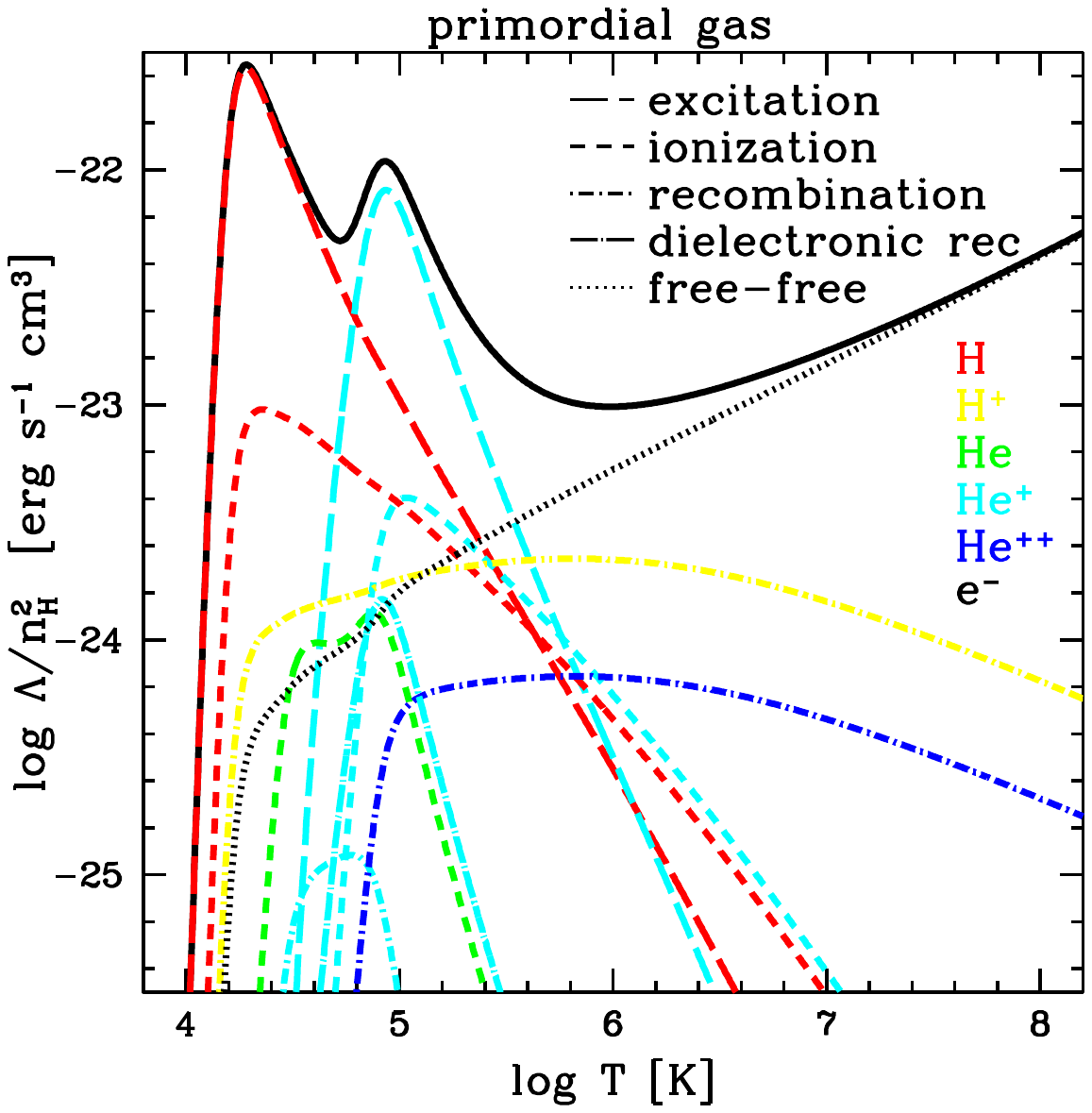}
  \includegraphics[width=0.4\textwidth]{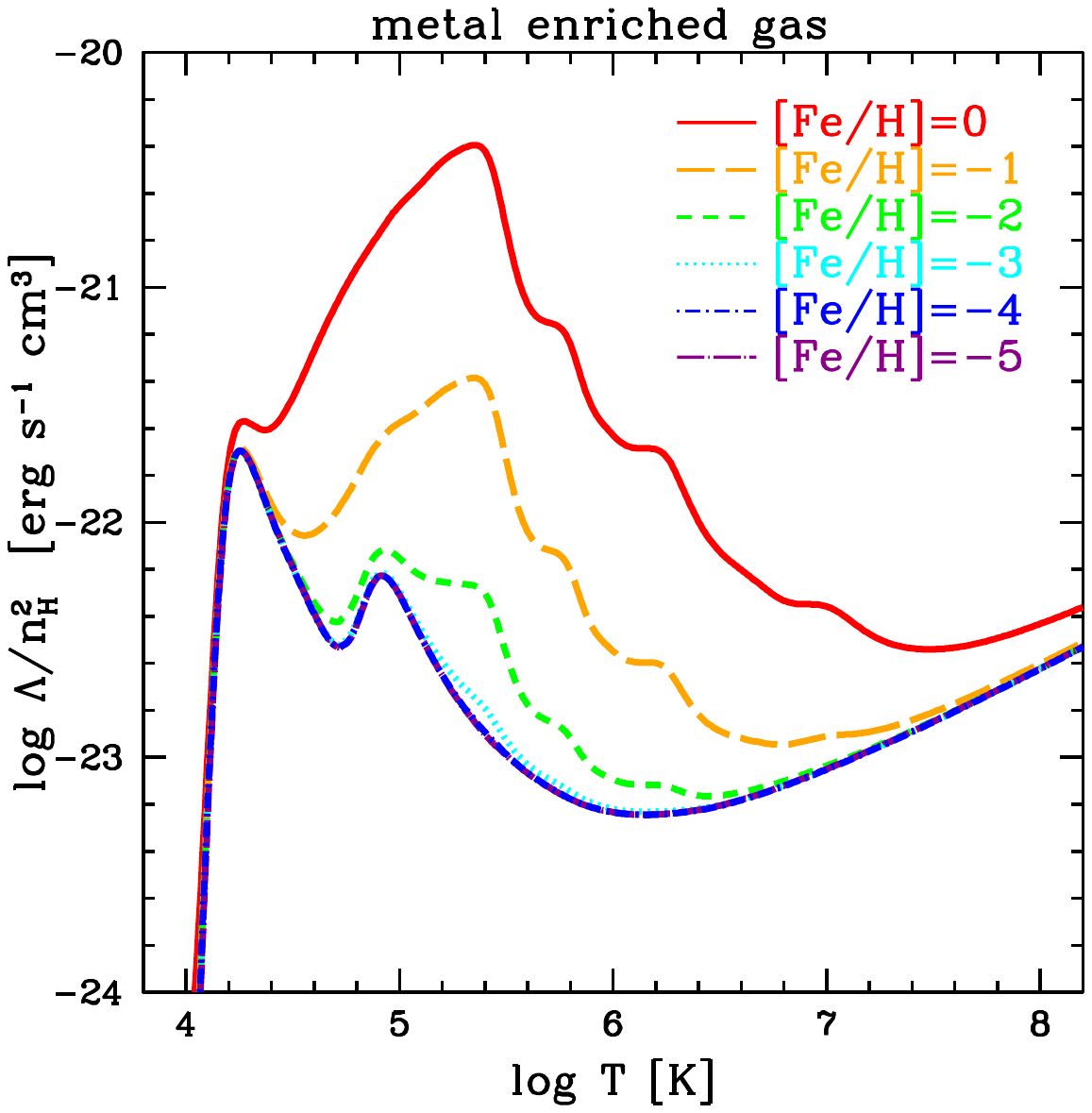}
\vspace*{-1mm}
\caption{(left) Cooling functions without photoionization.
Included processes are collisional excitation (long-dashed lines) of H$^0$ (red) and He$^+$ (cyan), collisional ionization (short-dashed lines) of H$^0$ (red), He$^0$ (green), and He$^+$ (cyan), standard recombination (short-dash dotted lines) of H$^+$ (yellow), He$^+$ (cyan), and He$^{++}$ (blue), dielectric recombination (long-dashed dotted line) of He$^+$ (cyan), and free-free emission (dotted line).
(right) Metallicity dependence of the cooling function. 
The solid, long-dashed, short dashed, dotted, short-dash dotted, and long-dash dotted lines correspond to [Fe/H] $=0,-1,-2,-3,-4$, and $-5$, respectively.
Figures are taken from \citet{kob02}.}
\label{fig:cooling}
\end{center}
\end{figure}

For the baryon physics, the first process to calculate is 
{\bf radiative cooling}, and photo-heating usually by a uniform
and evolving UV background radiation \citep{haa96}.
It would be ideal to simulate radiative feedback self-consistently but this is computationally very expensive and the available simulations are only for high-redshifts.
The cooling function depends on the chemical composition of gas, and a metallicity-dependent cooling function computed with the MAPPINGS III software \citep{sut93,kew19} is used in our code, 
assuming the [$\alpha$/Fe]--[Fe/H] relation of the solar neighborhood (Figure \ref{fig:cooling}).
Element-dependent cooling functions are also available \citep{wiersma09}.
Usually, the simulation code finds candidate gas particles (or mesh) that can form stars in a given timestep, and forms star particles with some certain mass.

\begin{figure}[t]
\begin{center}
\includegraphics[width=0.75\textwidth]{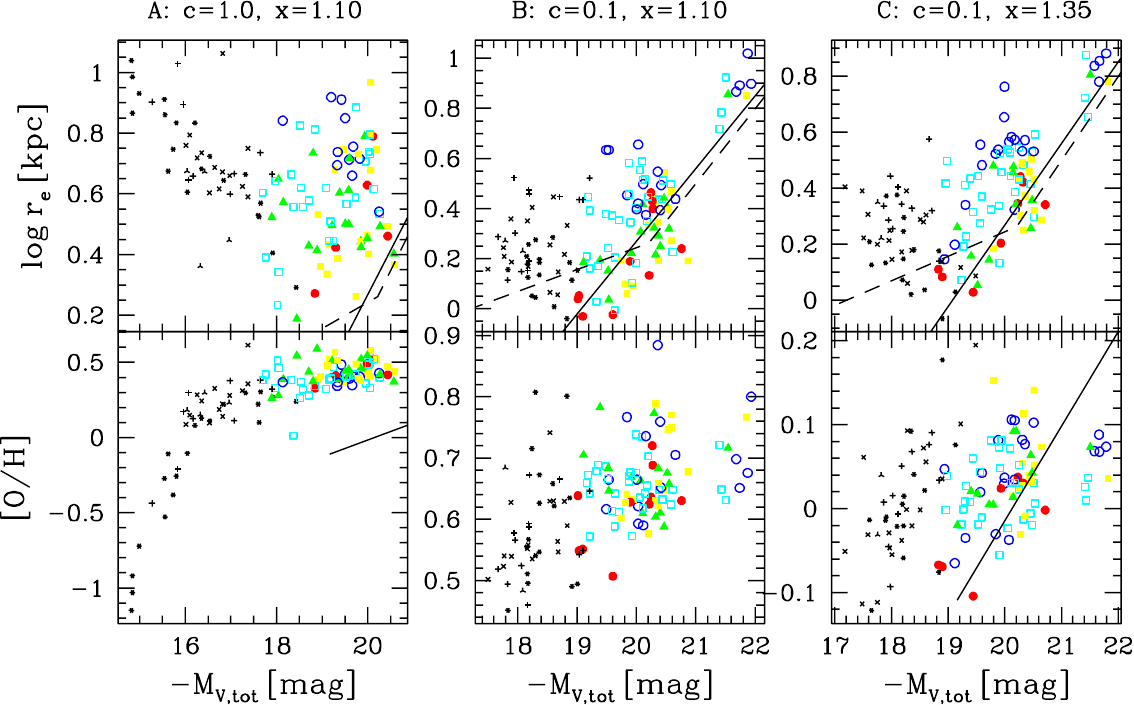}
\caption{\label{fig:csf}
Dependence of the star formation timescale parameter $c_*$ and the IMF slope $x$ on the galaxy scaling relations: size--mass (upper panels) and mass--metallicity (lower panels) relations. A single slope like Salpeter IMF assumed. The points show a set of chemodynamical simulations of early-type galaxies that undergo various merging histories. The solid and dashed lines indicate the observed relations.
Figure is taken from \citet{kob05} with modification.
}
\end{center}
\end{figure}

Our {\bf star formation} criteria are:
\begin{equation}
\mbox{converging flow, } (\nabla \cdot \mbox{\boldmath$v$})_i < 0,
\end{equation}
\begin{equation}
\mbox{rapid cooling, } t_{\rm cool} < t_{\rm dyn},
\end{equation}
\begin{equation}
\mbox{Jeans unstable gas, } t_{\rm dyn} < t_{\rm sound}.
\end{equation}
This still needs a parameter to define the star formation timescale relative to the dynamical timescale, as
\begin{equation}
t_{\rm sf} \equiv \frac{1}{c_*}t_{\rm dyn},
\end{equation}
where $c_*$ is a star formation timescale parameter. 
The parameter $c_*$ basically determines when to form stars following cosmological accretion, and has a great impact on the size--mass relation of galaxies (Figure \ref{fig:csf}). %(Fig.\,4 of \citealt{kob05}).
Alternatively, a fixed density criterion (e.g., $n_{\rm H}^*=0.1$ cm$^{-3}$) can be adopted \citep{kat92} although we found that it is better not to include it.
More sophisticated analytic formulae are also proposed based on smaller-scale simulations, including the effects of turbulence and magnetic fields \citep{fed11}.

A fractional part of the mass of the gas
  particle turns into a new star particle \citep[see][for the details]{kob07}.
Note that an individual
  star particle has a typical mass of $\sim 10^{5-7} M_\odot$, i.e.~it does not
  represent a single star but an association of many stars.  The masses of
  the stars associated with each star particle are distributed according to an
  IMF.  We adopt Kroupa IMF (with $x=1.3$), which is assumed to be
  independent of time and metallicity, and 
 limit the IMF to the mass range
  $0.01M_\odot \le m \le 120M_\odot$.
This assumption is probably valid for the resolution down to $\sim 10^4 M_\odot$.

\citet{kob04} introduced a scheme to follow the evolution of the star particle as a simple stellar population (SSP), which is defined as a single generation of coeval and chemically homogeneous stars of various masses, i.e. it consists of a number of stars with various masses but the same age and chemical composition.
The production of each element $i$ from the star particle (with an initial mass of $m_*^0$) is calculated using a very similar equation as Eq.(\ref{eq:gce}):
\begin{equation}\label{eq:e-z}
E_Z(t)= m_*^0 \left[ E_{\rm SW}+E_{\rm SNcc}+E_{\rm SNIa}+E_{\rm NSM} \right] .
\end{equation}
With this SSP method, the identical equations for GCE, Eqs.(\ref{eq:e_sw}-\ref{eq:e_nsm}) can used for $E_{\rm SW}$, $E_{\rm SNcc}$, $E_{\rm Ia}$, and $E_{\rm NSM}$.

Similarly, the energy production from the star particle is:
\begin{equation}\label{eq:e-e}
E_e(t) = m_*^0 \left[ e_{e,{\rm SW}}{\cal{R}}_{\rm SW}(t)+e_{e,{\rm SNcc}}{\cal{R}}_{\rm SNcc}(t)+e_{e,{\rm SNIa}}{\cal{R}}_{\rm SNIa}(t) \right] ,
\end{equation}
where the event rates of SWs (${\cal{R}}_{\rm SW}$) and core-collapse supernovae (${\cal{R}}_{\rm SNcc}$) can be calculated as
\begin{equation}
{\cal R}_{\rm SW}=\int_{m_t}^{m_u}\,
\frac{1}{m}\,\psi(t-\tau_m)\,\phi(m)~dm ,
\end{equation}
\begin{equation}
{\cal R}_{\rm SNcc}=\int_{\max[m_{{\rm SNcc},\ell},\,m_t]}^{\min[m_{{\rm SNcc},u},m_u]}\,
\frac{1}{m}\,\psi(t-\tau_m)\,\phi(m)~dm .
\end{equation}
The lower and upper limits are the same as IMF upper limit for stellar winds ($0.01$ and $120M_\odot$), while it is set to be $\sim8M_\odot$ (depending on $Z$) and $50M_\odot$ for core-collapse supernovae.
The energy production from SNe Ia is significant, and the rate is given by Eq.(\ref{eq:r_ia}), while the energy production from NSMs is negligible.
The energy per event (in erg) for SWs: $e_{e,{\rm SW}}=0.2 \times 10^{51} ({Z}/{Z_\odot})^{0.8}$ for $>8M_\odot$ stars, Type II supernovae: $e_{e,{\rm SNII}}=1 \times 10^{51}$, hypernovae: $e_{e,{\rm HN}}=10,10,20,30 \times 10^{51}$ for $20,25,30,40 M_\odot$ stars, respectively, and for SNe Ia: $e_{e,{\rm SNIa}}=1.3 \times 10^{51}$ erg.

\begin{figure}[t]
\begin{center}
\includegraphics[width=0.8\textwidth]{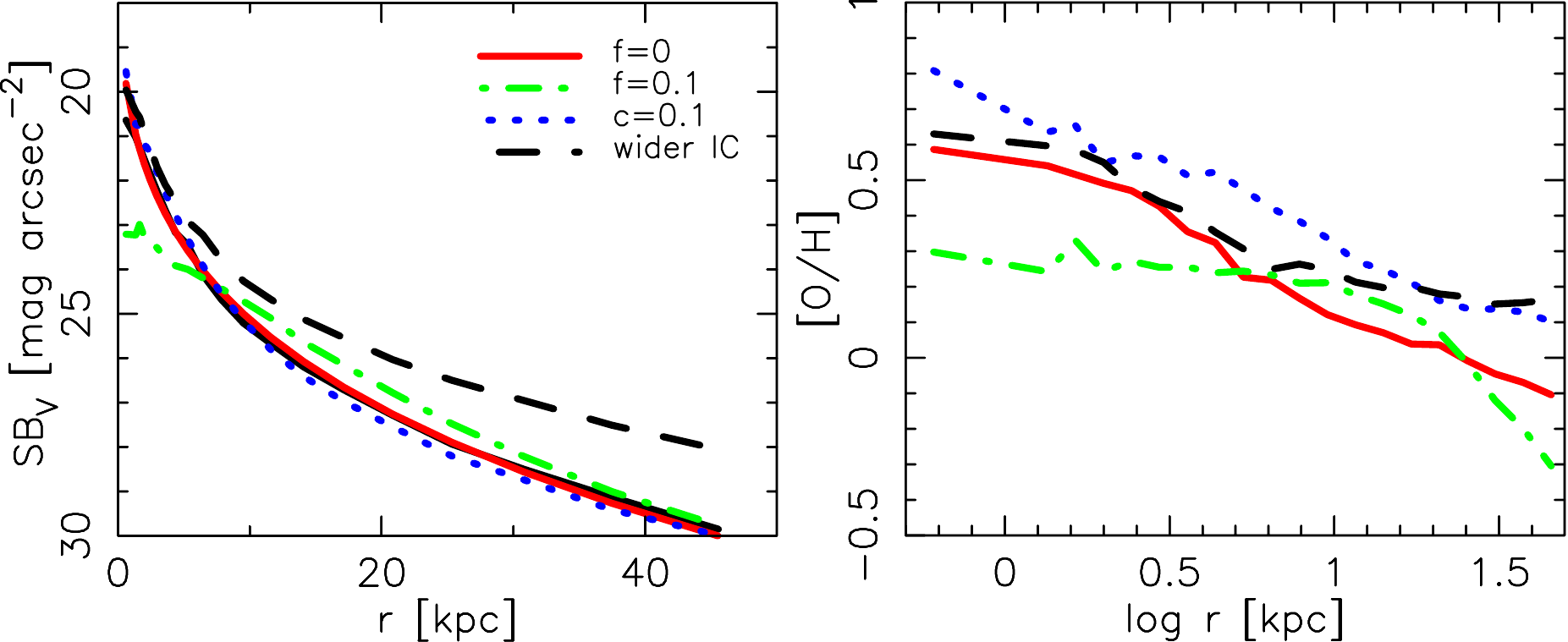}
\caption{\label{fig:kin}
Dependence of the star formation timescale parameter $c_*$ and the kinetic feedback fraction $f_{\rm kin}$ on the internal structure of galaxies: surface brightness profile (left panel) and the stellar metallicity radial gradient (right panel) for a chemodynamical simulation of an early-type galaxy with the initial perturbation. A single slope with $x=1.1$ adopted. 
Figure is taken from \citet{kob04} with modification.
}
\end{center}
\end{figure}

We distribute this {\bf feedback} metal mass and energy from a star particle $j$
isotropically to a fixed number, $N_{\rm FB}$, of neighbour gas particles $\ell$, weighted by the smoothing kernel as:
\begin{equation}\label{eq:fb-z}
\frac{d(Z_{i,k}\,m_{{\rm g},k})}{dt}= \sum_j \left[W_{kj} E_{Z,j}(t) / \sum_\ell^{N_{\rm FB}} W_{\ell j} \right]
\end{equation}
and
\begin{equation}\label{eq:fb-e}
\frac{du_k}{dt}= (1-f_{\rm kin}) \sum_j \left[W_{kj} E_{e,j}(t) / \sum_\ell^{N_{\rm FB}} W_{\ell j} \right]
\end{equation}
where $u_k$ denotes the internal energy of a gas particle $k$, and $f_{\rm kin}$ denotes the kinetic energy fraction \citep{nav93}, i.e., the fraction of energy that is distributed as a random velocity kick to the gas particle.
Note that in the first sum the number of neighbour star particles is not fixed to $N_{\rm FB}$.
To calculate these equations, the feedback neighbour search needs to be done twice in order to ensure proper
mass and energy conservation; first to compute the sum of weights for the
normalization, and a second time for the actual distribution.

\begin{figure}[t]
\center
\includegraphics[width=0.95\textwidth]{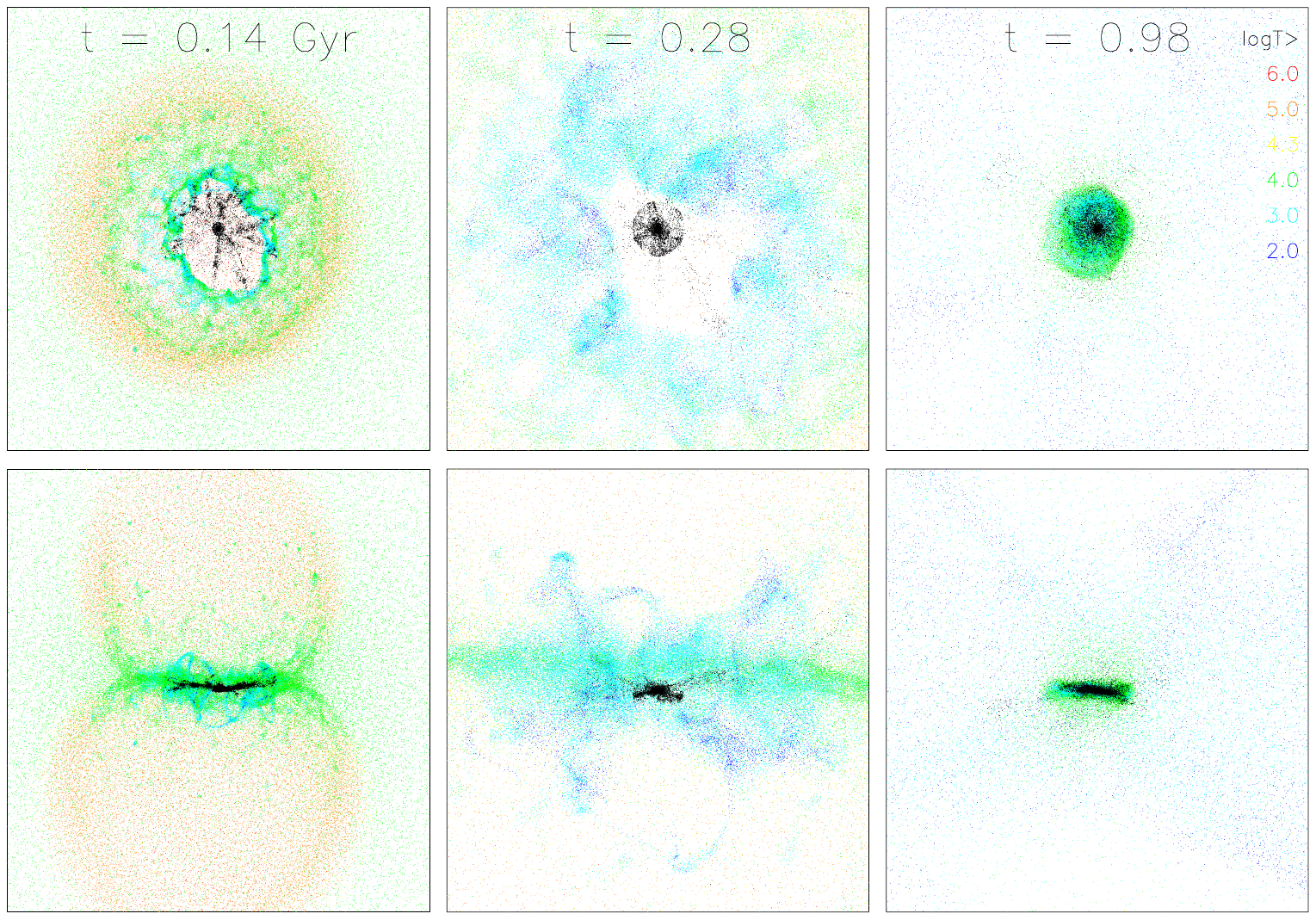}
\caption{Time evolution of the formation of an isolated disk galaxy in a halo of mass of $10^{10} h^{-1}M_\odot$. The black points show star particles, while the gas particles are colour-coded according to their temperature. Each panel is $20$ kpc on a side. The upper row shows face-on projections, the lower row gives edge-on views.
Figure is taken from \citet{kob07}.
}
\label{fig:diskmap}
\end{figure}

The parameter $N_{\rm FB}$ determines the average energy distributed to gas particles; a large value of $N_{\rm FB}$ leads to inefficient feedback as the ejected energy radiatively cools away shortly, while a small value of $N_{\rm FB}$ results in a only small fraction of matter affected by feedback.
Alternatively, with good resolution in zoom-in simulations, feedback neighbors could be chosen within a fixed radius ($r_{\rm FB}$), which affects a number of observations including the MDF (Fig.\,12 of \citealt{kob11mw}).
The parameter $f_{\rm kin}$ has a great impact on surface brightness profiles and radial metallicity gradients in galaxies, and $f_{\rm kin}=0$, i.e., purely thermal feedback, gives the best match to the observations (Figure \ref{fig:kin}). %(Fig.\,14 of \citealt{kob04}).
Note that various feedback methods are proposed such as
the stochastic feedback \citep[][used for EAGLE]{dal08} and the mechanical feedback \citep[][used for FIRE]{hop18}.
Feedback modelling can have an impact on chemical evolution of galaxies (D. Ibrahim et al., in prep.).

Figure \ref{fig:diskmap} shows a test simulation of an isolated disk galaxy with supernova feedback.
The initial condition is a rotating gas cloud with total mass of $10^{10}h^{-1}M_\odot$, and non-isotropic infall of gas form a dense disc, where stars form. As a result of the energy input, the low-density hot-gas region expands in a bipolar flow (left panels) ejecting metals outside the galaxy. In the disk plane, the hot gas region expands and forms a dense shell where stars keep on forming. The energy from these stars can quickly propagate to the surrounding low-density region. After the galactic wind forms (middle panel), the gas density becomes so low at the centre that star formation is terminated for a while. Because of radiative cooling, a part of the ejected, metal-enhanced gas, however, returns, again settling in the disk where it fuels new star formation, but this secondary star formation is not as strong as the initial starburst. Although some small bubbles are forming in the galaxy in this stage, not much gas and metals are ejected by them from the disk (right panels).

\begin{figure}[t]
\begin{center}
\includegraphics[width=0.45\textwidth]{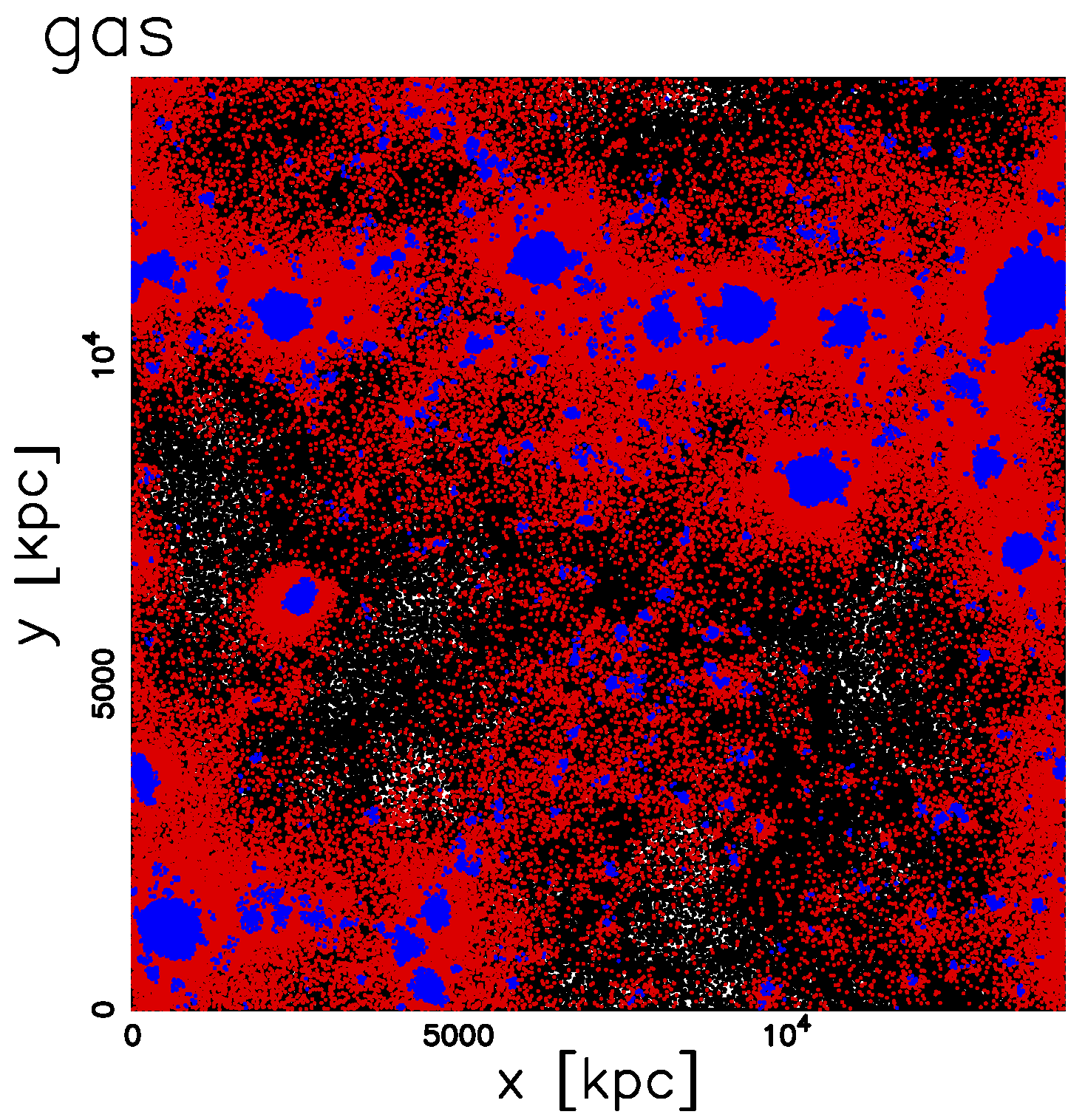}
\includegraphics[width=0.45\textwidth]{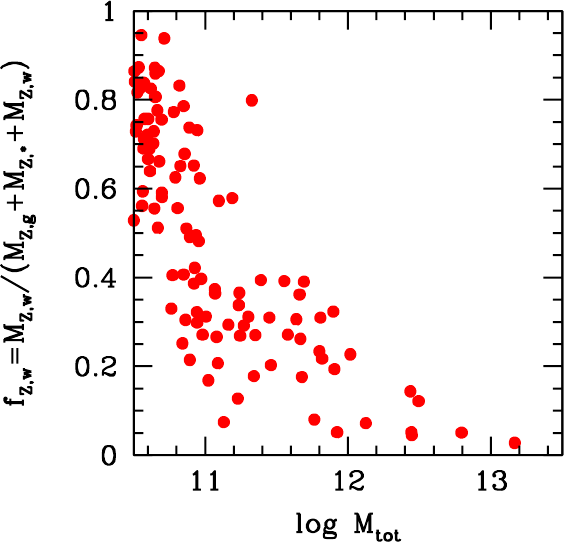}
\caption{\label{fig:gw}
(Left) Spatial distribution of ISM (blue) and wind (red) particles at $z = 0$.
(Right) Ejected metal fraction $f_{Z,{\rm w}} = M_{Z,{\rm w}}/(M_{Z,{\rm g}} + M_{Z,*} + M_{Z,{\rm w}})$ against total mass. 
Figures are taken from \citet{kob07} with modification.
}
\end{center}
\end{figure}

However, galactic winds become much weaker in more massive galaxies. Figure \ref{fig:gw} shows the fraction of metal loss as a function of galaxy mass in a cosmological simulation including hypernova feedback. Most of metals are inside the galaxy at $M_{\rm tot}\gtsim 10^{12}h^{-1}M_\odot$.
Eq.(\ref{eq:fb-z}) gives positive feedback that enhances radiative cooling, while Eq.(\ref{eq:fb-e}) gives negative feedback that suppresses star formation.
The mass ejection in Eq.(\ref{eq:fb-z}) never becomes zero, while the energy production in Eq.(\ref{eq:fb-e}) becomes small after $\sim35$ Myrs.
Therefore, it is not easy to control star formation histories depending on the mass/size of galaxies within this framework.
In particular, it is not possible to quench star formation in massive galaxies, since low-mass stars keep returning their envelope mass into the ISM for a long timescale, which will cool and keep forming stars.

\begin{figure}[t]
\begin{center}
\includegraphics[width=0.5\textwidth]{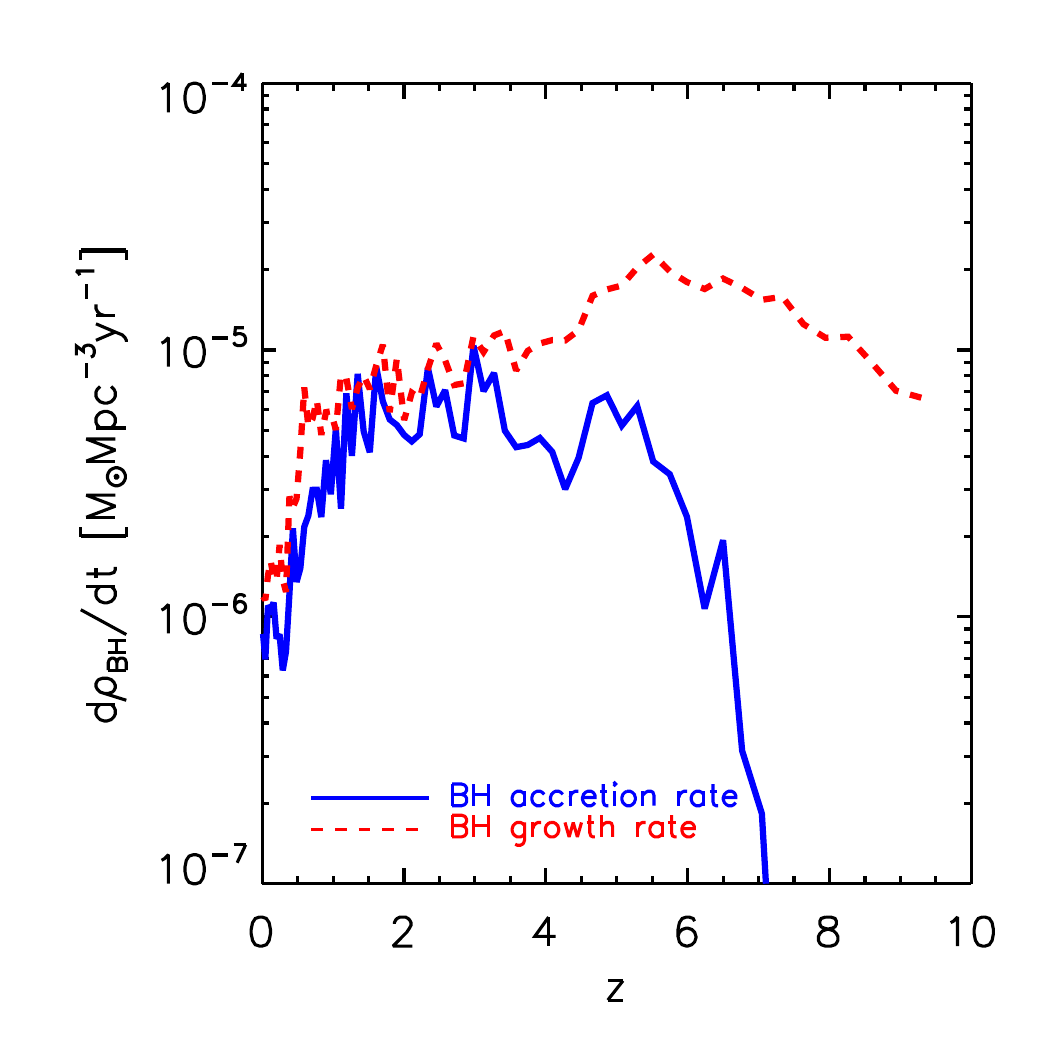}
\caption{\label{fig:agn}
Cosmic BH accretion rate density (solid line) and BH growth rate density (including BH formation, dashed line) with redshift. The difference of these rates shows the BH merger rate.
Figure is taken from \citet{tay14} with modification.
}
\end{center}
\end{figure}

Therefore, in order to reproduce observed properties of massive galaxies, additional feedback was required, and the discovery of the co-evolution of super-massive black holes and host galaxies \citep{magorrian98} provided a solution -- {\bf AGN feedback}.
Modelling of AGN feedback consists of (1) seed formation, (2) growth by mergers and gas accretion, and (3) thermal and/or kinetic feedback (see \citealt{tay14} for the details).
In summary, we introduced a seeding model where seed black holes originate from the formation of the first stars, which is different from the `standard' model by \citet{spr05agn} and from most large-scale hydrodynamical simulations. %\cite{dub12,vog14,sch15,pillepich18a,dave19}.
In our AGN model, 
seed black holes are generated if the metallicity of the gas cloud is zero ($Z=0$) and the density is higher than a threshold density ($\rho_{\rm g} > \rho_{\rm c}$). 
The growth of black holes is roughly the same as in other cosmological simulations, and is calculated with Bondi-Hoyle accretion, swallowing of ambient gas particles, and merging with other black holes.
Since we start from relatively small seeds, the black hole growth is driven by mergers at $z\gtsim 3$ (Figure \ref{fig:agn}), which could be detected by gravitational waves. %(Fig.\,2 of \citealt{tay14}).
On a very small scale, it is not easy to merge two black holes, and an additional time-delay is applied in more recent simulations (P. Taylor et al. in prep.).

Proportional to the accretion rate, thermal energy is distributed to the surrounding gas, which is also the same as in many other simulations.
In more recent simulations, non-isotropic distribution of feedback area is used to mimic the small-scale jet \citep[e.g.,][]{dipanjan18}.
There are a few parameters in our chemodynamical simulation code, but \citet{tay14} constrained the model parameters from observations, and determined the best parameter set: $\alpha=1$, $\epsilon_{\,\rm f}=0.25$, $M_{\rm seed}= 10^3h^{-1}M_\odot$, and $\rho_{\rm c}=0.1 h^2 m_{\rm H}\,\textrm{cm}^{-3}$.
Our black holes seeds are indeed the debris of the first stars although we explored a parameter space of $10^{1-5}h^{-1}M_\odot$.
This is not the only one channel for seeding, and the direct collapse of primordial gas and/or the collapse of dense stellar clusters ($\sim 10^5M_\odot$, e.g., \citealt{madau01,woods19}) are rarer but should be included in larger volume simulations.

Nonetheless, our model can successfully drive large-scale galactic winds from massive galaxies \citep{tay15letter} and can reproduce many observations of galaxies with stellar masses of $\sim 10^{9-12}M_\odot$ \citep{tay15,tay16,tay17}.
Metals are ejected from galaxies to circum-galactic and inter-galactic medium mainly by supernova-driven winds in low-mass galaxies (as in Fig.\,\ref{fig:gw}), while by AGN-driven winds by massive galaxies \citep{tay20}. The fraction of metals ejected by AGN is only a few percent of the total produced metals, and thus it does not affect mass-metallicity relations \citep{tay15letter}.
The movie of our fiducial run is available at \url{https://www.youtube.com/watch?v=jk5bLrVI8Tw}

\subsection{Various big simulations}

\begin{figure}[t]
\begin{center}
  \includegraphics[width=0.6\textwidth]{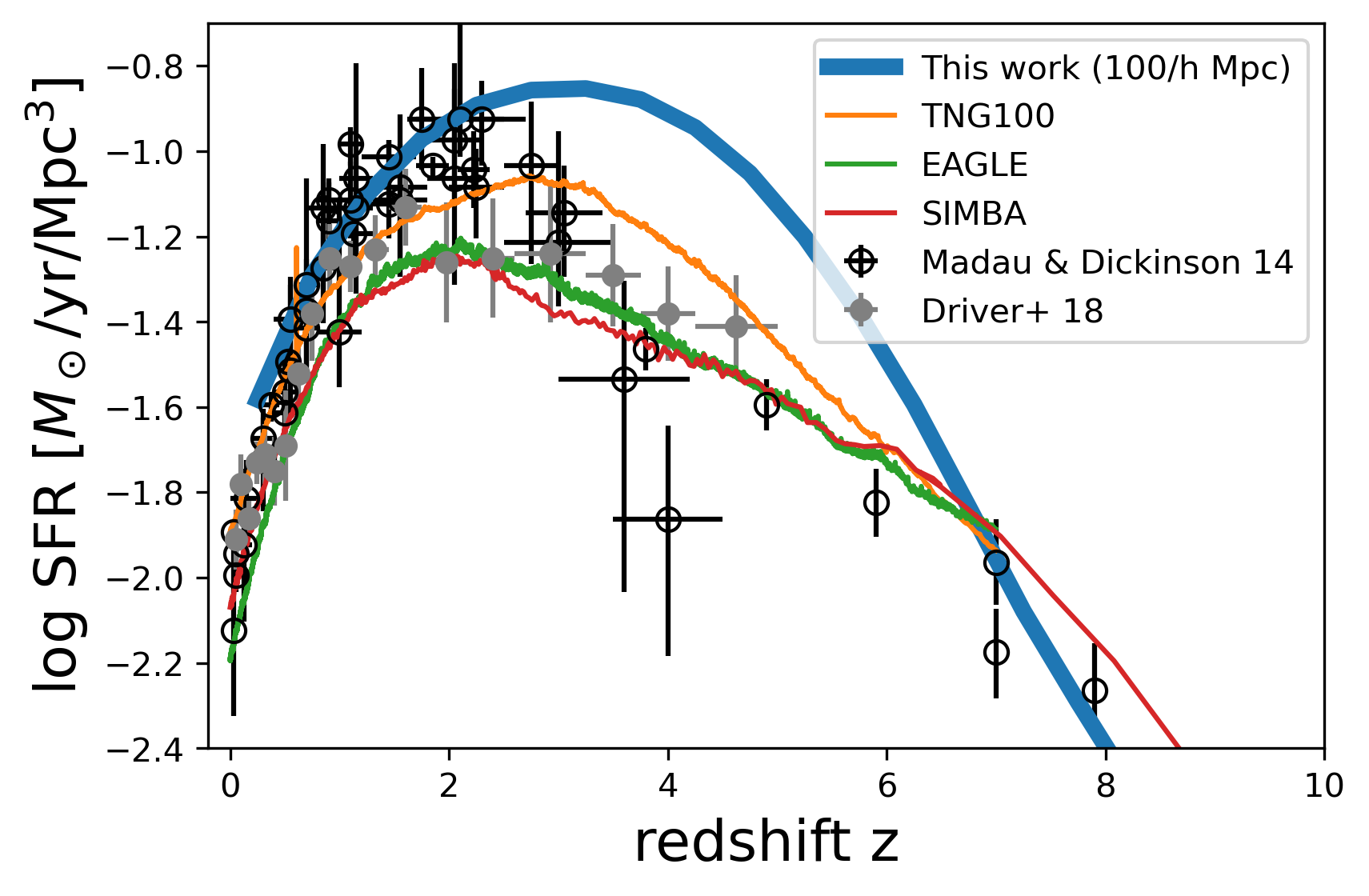}
\vspace*{-1mm}
\caption{Cosmic star formation rate history of our $100h^{-1}$ Mpc simulation, comparing to other cosmological simulations. See text for the details. The observational data (points) are taken from \citet{mad14} and \citet{driver18}. 
Note that other observations \citep[e.g.,][]{chary07,rowanrobinson16} show higher rates at high redshifts.
The data of the other simulations are provided by R. Yates and C. Lovell (priv. comm.). %taken from \citet{yates21} and \citet{lovell21}.
}
\label{fig:csfr}
\end{center}
\end{figure}

There are various large volume simulations available but the input physics is very different, which can summarised as follows.
The differences in modelling of sub-galactic physics largely affects the predicted galaxy properties, and the parameters must be calibrated using galaxy scaling relations (e.g., Figure \ref{fig:csf}) and internal structures (e.g., Figure \ref{fig:kin}) at $z=0$, before predicting observables at higher-redshifts using the simulation outputs.
Figure \ref{fig:csfr} shows a comparison of some of these simulation to our simulation, which is run by a Gadget-3 based code (\citealt{tay14}; Kobayashi \& Taylor, in prep.).
See also \citet{wang19} for the discussion on the black hole seed mass.

{\it EAGLE} \citep{sch15} -- Star formation is calculated with a metallicity and temperature dependent density threshold ($n_{\rm H}^*(Z)=0.1\left(\frac{Z}{0.002}\right)^{-0.64}$ cm$^{-3}$ and $T<10^{4.4}$K). Star particle mass is determined from a pressure law, obtained from the observed Kennicutt-Schmidt, with two free parameters ($A$ and $n$). A Chabrier IMF is adopted. Stochastic feedback from \citet{dal12} is assumed. SMBHs are grown from the seeds of $10^5M_\odot$, and the AGN feedback is thermal, independent of metallicity. The SN Ia rate decays exponential with a timescale of $\tau=2$ Gyr, with a fixed normalization $\nu$. The simulation box size is up to 100 Mpc$^3$, which has $1.81\times10^6M_\odot$ and 0.7 kpc resolution for gas.

{\it IllustrisTNG} \citep{pillepich18a} -- It is run with a magneto-hydrodynamic code \citep{pakmor11}. Star formation and pressurization of the multi-phase are treated following \citet{spr03}, which includes a density threshold $n_{\rm H}\simeq0.1$ cm$^{-3}$. A Chabrier IMF is adopted. Isotropic, kinetic (wind) feedback from \citet{spr03} is assumed (instead of bipolar winds in Illustris). SMBHs are grown from the seeds of $8\times10^5M_\odot$, and kinetic wind is assumed for AGN feedback in the case of low accretion rates as in \citet{weinberger17}. SN Ia rate is calculated with a ``simple DTD'' ($-1$ slope?) with a constant normalization $N_0$. The simulation box size is up to 302.6 Mpc$^3$, but a 110.7 Mpc$^3$ run has $1.4\times10^6M_\odot$ and 0.19 kpc resolution for gas.

{\it SIMBA} \citep{dave19} --  This uses H$_2$-based star formation rate from \citet{krumholz11}, which is based on the metallicity and local column density, assuming the efficiency of $\epsilon_*=0.02$ for the H$_2$ density. Stellar feedback is assumed to be kinetic, with decoupled wind particles, galaxy mass dependent ($\eta(M_*)$), and metal-loaded ($dZ=f_{\rm SNII}\,y_{\rm SNII}/\max[\eta,1]$). SMBHs are grown from the seeds of $1\times10^4h^{-1}M_\odot$. AGN feedback is also assumed to be kinetic for both fast and slow accretion, and purely bipolar. In addition, on-the-fly dust production and destruction is calculated. SNe Ia have two distinct delay-times, 0 and 0.7 Gyr, with two constant rates. The simulation box size is up to $100h^{-1}$ Mpc$^3$, which has $1.82\times10^7M_\odot$ and $0.5h^{-1}$ kpc resolution for gas.

{\it HORIZON-AGN} \citep{dub16} -- This uses a density threshold of $n_{\rm H}\simeq0.1$ cm$^{-3}$, thermal energy injection to model stellar feedback, and two modes of AGN feedback with thermal or bipolar outflow depending on the Eddington ratio. The simulation box size is $100h^{-1}$ Mpc$^3$, which has $1\times10^7M_\odot$, and the grid is adaptively refined down to 1 proper kpc.

{\it Magneticum} \citep{dolag17} -- Multiphase model for star formation from \citet{spr03}, and isotropic winds with 350 km s$^{-1}$ are adopted. Two modes of AGN feedback with different efficiencies of energy injection depending on the Eddington ratio are assumed. The simulation box size is $48h^{-1}$ Mpc$^3$, which has $3.7\times10^7h^{-1}M_\odot$ and $0.7h^{-1}$ kpc resolution for gas.

Similarly, there are a series of cosmological zoom-in simulations of Milky Way-type galaxies.
{\it Auriga} \citep{grand17} uses similar code as {\it Illustris} but with $n=0.13$ cm$^{-3}$ and $t_{\rm sf}=2.2$ Gyr, with resolution from $6\times10^3$ to $4\times10^5M_\odot$ for gas.
{\it FIRE-2} \citep{hop18fire} is run with the GIZMO code, which is also used for {\it SIMBA}, but with improved modelling of small-scale physics. AGN feedback is not included.
{\it ARTEMIS} \citep{font20} uses the same code as {\it EAGLE} with resolution of $2\times10^4h^{-1}M_\odot$ for gas.

\subsection{Galactic archaeology}
\label{sec:mw}

\begin{figure}[t]
\centering
\includegraphics[width=0.85\textwidth]{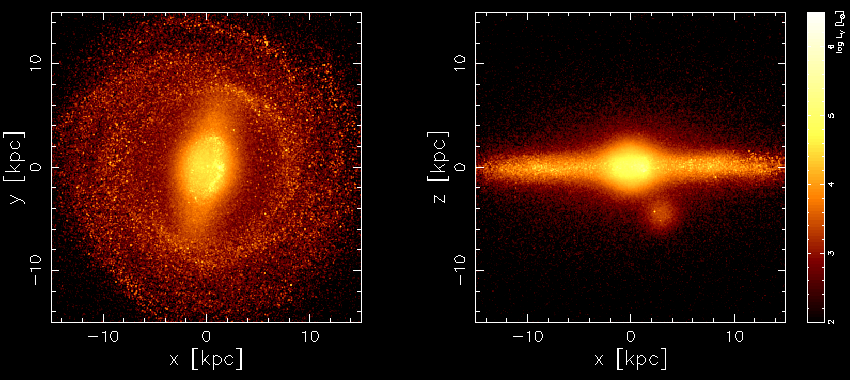}
\caption{V-band luminosity map for the edge-on (left panels) and face-on (right panels) views at $z = 0$ in 30 kpc on a side,
of our chemodynamical zoom-in simulation of a Milky Way-type galaxy.}
\label{fig:mwmap}
\end{figure}

In the Local Group, i.e., in the Milky Way and dwarf spheroidal (dSph) galaxies, the spatial distribution of detailed elemental abundances can be obtained from high-resolution spectra of individual low-mass stars\footnote{Ages of stars can also be well estimated with asteroseismology recently.}. This has been done for more than half a century, and the average trends of the observations for many elements have been used to constrain the stellar physics \citep{tim95,kob06,rom10}. The recent Galactic archaeology surveys with medium resolution MOS dramatically increased the sample and made possible to discuss the distributions of stars along the average trends, which requires more realistic, chemodynamical simulations of galaxies. Meantime, observations with different lines for each element \citep[e.g.,][]{isr98,sne16} revealed NLTE and 3D effects in the stellar atmosphere. Detailed star-by-star analysis of wide range of high-resolution spectra (including UV) is still required to obtain absolute values of elemental abundances \cite[][for the need of UV spectra]{kob22iau}.

Figure \ref{fig:mwmap} shows the edge-on and face-on maps of our simulated Milky Way-type galaxy, where V-band luminosities of star particles are calculated using stellar population synthesis models from \citet{kod97} and all detailed chemical enrichment is included.
Although there were simulations of isolated galaxies, \citet{kob11mw} was the first chemodynamical simulations that showed the evolution elements from O to Zn in a Milky Way-type galaxy from cosmological initial condition.
Our new simulation, which is run by a Gadget-3 based code (Kobayashi, in prep.), includes all stable elements, and a fully cosmological initial condition is applied \citep{sca12}.
The spatial and mass resolutions are 0.5 kpc and $3\times10^5M_\odot$, respectively.
The movie of our fiducial run is available at \url{https://star.herts.ac.uk/~chiaki/works/Aq-C-5-kro2.mpg}

\begin{figure}[t]
\begin{center}
  \includegraphics[width=0.48\textwidth]{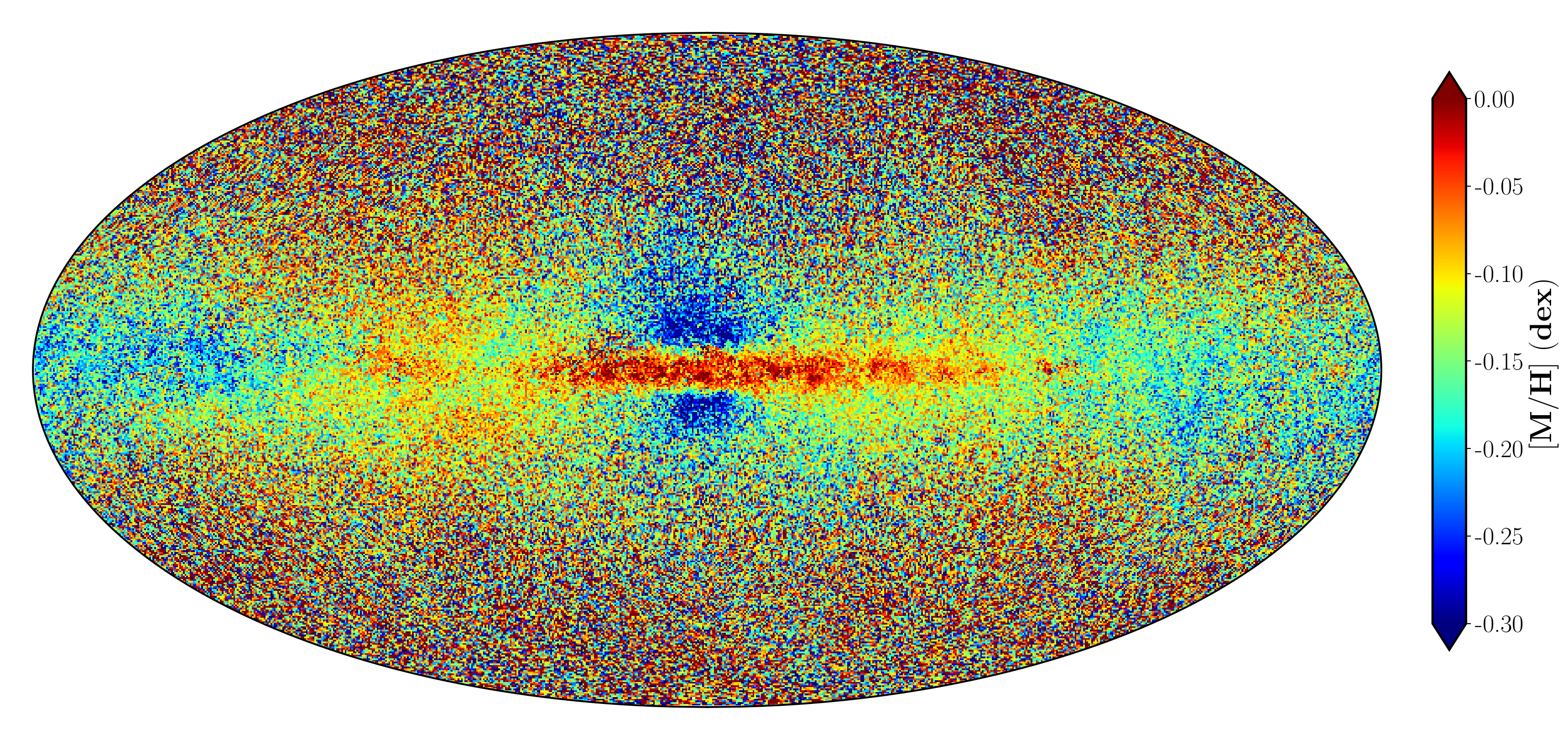}
  \includegraphics[width=0.48\textwidth]{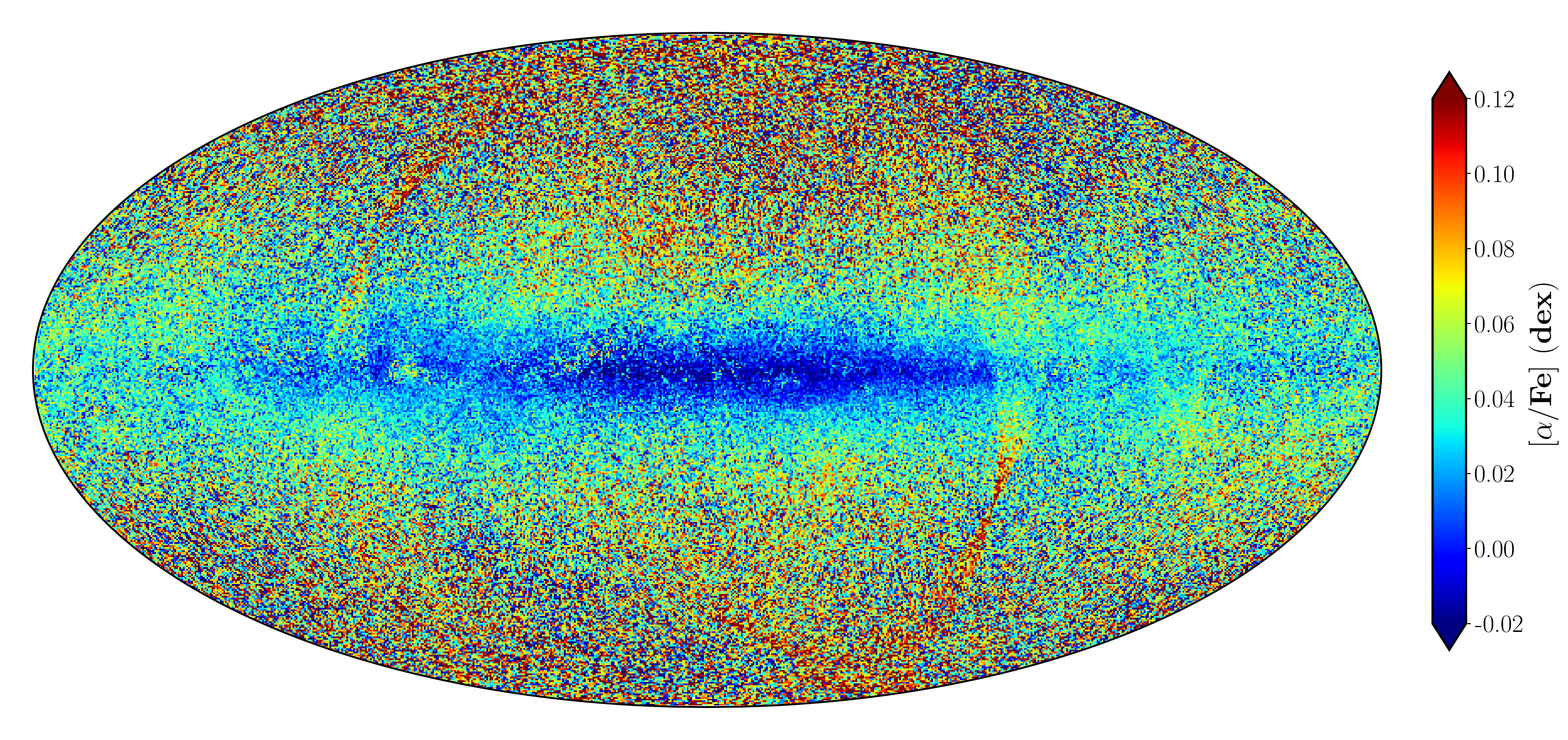}
\caption{(Left panel) Metallicity map of Gaia DR3 released on 13th June 2022, in the range of [M/H]$=-0.3$ (blue) to 0.0 (red).
(Right panel) The same as the left panel but for [$\alpha$/Fe]$=-0.02$ (blue) to 0.12 (red). Figures are taken from \citet{gaiadr3}.
}
\label{fig:gaia}
\end{center}
\end{figure}

Our simulated galaxy shows very similar maps of metallicity and [$\alpha$/Fe] ratios as observed. Figure \ref{fig:gaia} shows the observations in Gaia DR3, which contains a half billion of stars. The data show too blue outer bulge in the metallicity map, and in the [$\alpha$/Fe] map the patterns close to the Ecliptic Poles are artifacts.
There are both vertical and radial metallicity gradients; metallicity is high on the plane, and becomes higher toward galactic centre. On the other hand the [$\alpha$/Fe] ratios show a strong vertical gradient but no radial gradient except for the galactic bulge; the majority of the bulge stars have high [$\alpha$/Fe] ratios \citep{kob11mw}, which may depend on the feedback from galactic centre.
There is no chemodynamical model that includes feedback from the central SMBH in the Milky Way.

The formation of the basic structure is the same as described in \citet{kob11mw}.
The galactic bulge formed by an initial star burst, by the assembly of gas-rich sub-galaxies beyond $z=2$.
The disk formed with a longer timescale and has grown {\it inside-out}; the disk was small in the past and becomes larger at later times.
In the $\Lambda$CDM cosmology, satellite galaxies keep accreting, which interact and merge with the main body, but there is no major merger after $z=2$. Otherwise, it is not possible to keep the disk structure as in the Milky Way. 
These galaxy mergers also make the disk thicker.
Approximately one third of thick disk stars already formed in merging galaxies, which are disrupted by tidal force and accreted onto the disk plane.
In the thin disk, star formation is self-regulated, and chemical enrichment takes place following cosmological gas accretion, radial flows, and stellar migrations; these physical processes are analysed in detail in \citet{vin20}.

Inhomogeneous enrichment in chemodynamical simulations leads to a paradigm shift on the chemical evolution of galaxies.
As in a real galaxy, i) the star formation history is not a simple function of radius, ii) the ISM is not homogeneous at any time, and iii) stars migrate, which are fundamentally different from one-zone or multi-zones GCE models.
As a consequence, (1) there is no tight age--metallicity relation, namely for stars formed in merging galaxies. It is possible to form extremely metal-poor stars at a later time, from accretion of nearly primordial gas, or in isolated chemically-primitive regions.
(2) Enrichment sources with long time-delays such as AGB stars, SNe Ia, and NSMs can appear at low metallicities. This effect can naturally explain the observed N/O--O/H relation \citep{vin18no}, but is not sufficient to explain observed [Eu/(O,Fe)] ratios only with NSMs \citep{hay19}.
(3) There is a significant scatter in elemental abundance ratios at a given time/metallicity, as shown in Figs.\,16-18 of \citet{kob11mw}.

\begin{figure}[t]
\centering
\includegraphics[width=0.75\textwidth]{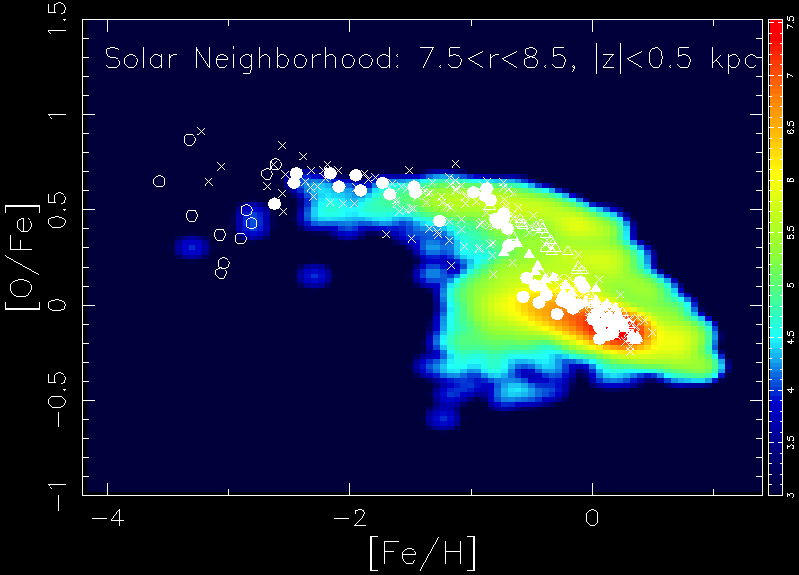}
\caption{The [O/Fe]--[Fe/H] relation in the solar neighborhood of our simulated Milky Way-type galaxy at $z=0$. Figure is the same as Fig. 10 of \citet{kob11mw} but with a newer simulation. The colour indicates the number of stars in logarithmic scale. The observational data sources are:
\citet[open and filled triangles]{ben04} for thick and thin disk stars, respectively,
\citet[open circles]{spi05},
\citet[filled circles]{zhao16}, 
\citet[crosses]{ama19b}.
}
\label{fig:mwofe}
\end{figure}

Figure \ref{fig:mwofe} shows the frequency distribution of [O/Fe] ratios in the solar neighborhood of our simulated galaxy.
The same [O/Fe]--[Fe/H] relation as in Fig. \ref{fig:ofe} is seen: the plateau at [O/Fe] $\sim 0.5$ caused by core-collapse supernovae, and the decrease of [O/Fe] from [Fe/H] $\sim -1$ to $\sim 0$ due to the delayed enrichment from SNe Ia.
The difference is that this chemodynamical model predicts not a line but a distribution (contours). A similar figure was also shown in \citet{kob11mw}, which predicted a bimodal distribution of [O/Fe] ratio at a given [Fe/H], concluding that ``this may be because the mixing of heavy elements among gas particles is not included in our model''. However, with the APOGEE survey \citep{hay15} clearly showed the bimodality of [$\alpha$/Fe], which was in fact already seen in earlier works with a much smaller sample but careful analysis \citep{fuhrmann98,ben03}. The advantage of the APOGEE survey was its large dynamic range and the change of the bimodality depending on the location within the Galaxy is also clearly shown. \citet{kob16iau} showed a similar change in the simulated galaxy, and \citet{vin20} showed a comparison to APOGEE DR16. Our simulated galaxy predicts this bimodality not only for [$\alpha$/Fe] ratios but also for most of elements from He to U; some elements even show trimodality (\citealt{kob22iau}; Kobayashi in prep.).

At [Fe/H] $\ltsim -1$ Figure \ref{fig:mwofe} also shows a significant number of stars with low [$\alpha$/Fe] ratios, which is caused by local enrichment from SNe Ia. Note that the SN Ia rate becomes almost zero from the stars below [Fe/H] $=-1.1$ in the adopted SN Ia model. Nonetheless, stars formed in less-dense regions can have low [$\alpha$/Fe] caused by SNe Ia, or by low-mass Type II supernovae ($13-15M_\odot$); this effect is important also for dSph galaxies.
The number of such low-$\alpha$ stars increases with sub-Ch mass SNe Ia (Kobayashi, in prep.).

The origin of the scatter/bimodality is discussed in \citet{kob14iau}. In chemodynamical simulations, it is possible to trace back the formation place of star particles. As a result, stars formed in merging galaxies found to have old age, low metallicity, high [$\alpha$/Fe] ratios, and relatively low [(Na,Al,Cu)/Fe] ratios, while stars formed in-situ show a tighter age-metallicity relation and low [$\alpha$/Fe] ratios. There is no age-metallicity relation for the stars that migrated, as expected.
In cosmological zoom-in simulations, stars tend to migrate outward by a few kpc, which flatten the metallicity gradient by $\sim 0.05$ dex/kpc. Radial flows steepen the gradient as chemically enriched gas moves inwards, but the average velocity is found to be only $\sim 0.7$ km/s \citep{vin20}.

\begin{figure}[t]
\begin{center}
  \includegraphics[width=0.95\textwidth]{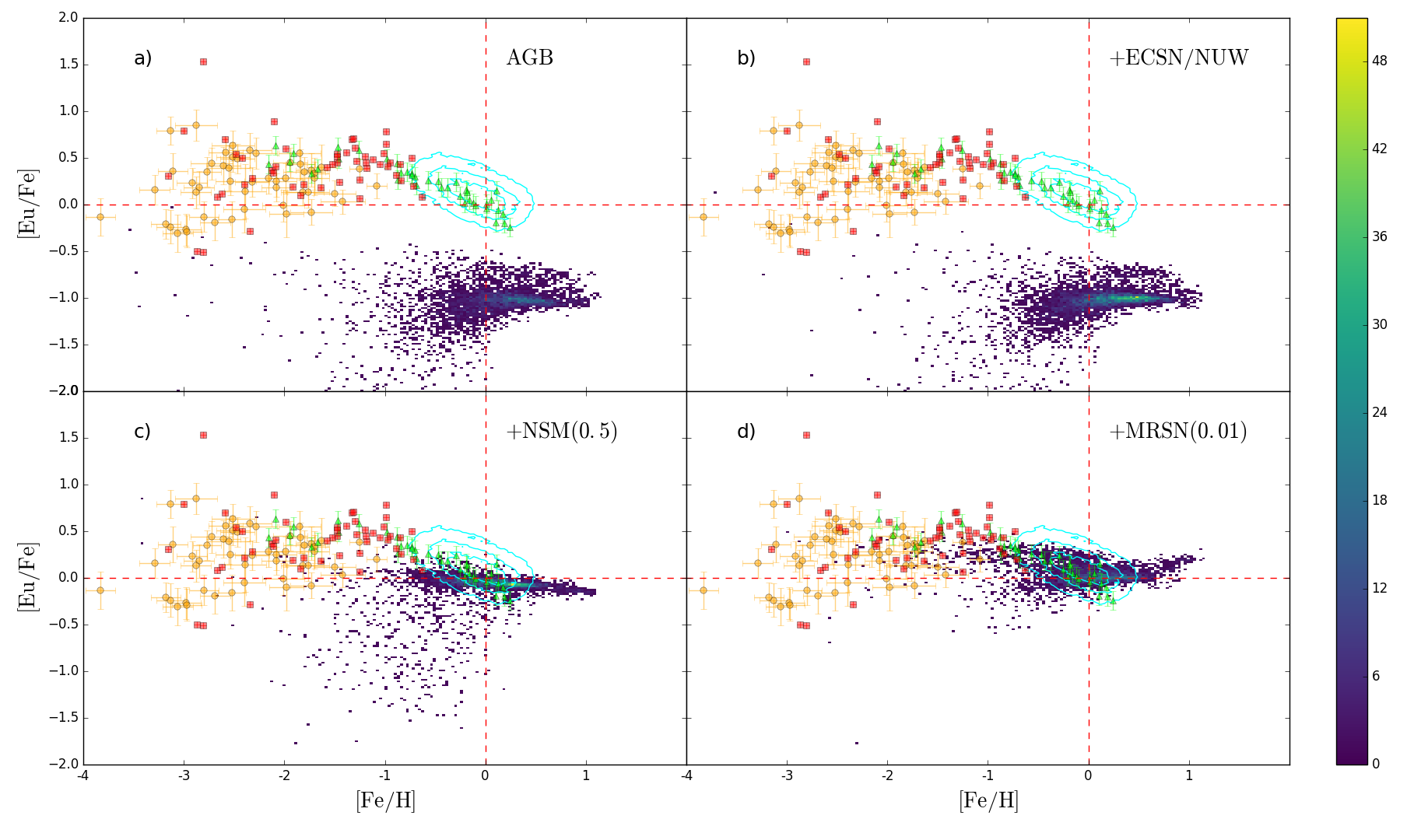}
\vspace*{-1mm}
\caption{Distribution of [Eu/Fe] against [Fe/H] for the star particles in the solar neighbourhood of our simulated Milky Way-type galaxy at $z=0$.
The panels in order show: control, ECSNe+$\nu$-driven winds, NSMs, and MRSNe. The observational data sources are: \citet[][red squares]{han16}, \citet[][orange circles]{roe14}, \citet[][green triangles]{zhao16}, and \citet[][GALAH DR2, cyan contours]{buder18}. The contours show 10, 50, and 100 stars per bin. The colour bar shows the linear number per bin of simulation data.
Figure is taken from \citet{hay19}.
}
\label{fig:mweu}
\end{center}
\end{figure}

Figure \ref{fig:mweu} shows the [Eu/Fe] ratios in the solar neighborhood of our simulated galaxy but with switching the r-process sites. With only AGB stars (panel a), Eu is not sufficiently produced. Different from one-zone GCE models, AGB contribution can be seen with a large scatter at low metallicity. 
With ECSNe or $\nu$-driven winds (panel b), Eu production is not increased enough. With NSMs (panel c), it is possible to reproduce the solar Eu/Fe ratios at the solar metallicity, but the scatter is too large at [Fe/H] $\ltsim 0$. Unlike one-zone GCE models, there are a small number of stars that have high [Eu/Fe] ratios at low metallicities due to the inhomogeneous enrichment.
The scatter can be much reduced with MRSNe (panel d) as they occur at a very short timescale. Since only a small fraction of core-collapse supernovae produce both Eu and Fe, the small scatter still remains, consistent with observations.

Very high [Eu/Fe] ratios were reported for ultrafaint dSphs \citep{ji16}, which might be caused by local enrichment with a NSM with no Fe production (and no Zn enhancement unlike \citet{yon21a}'s star). The effect of supernova kick in binary neutron star systems may help \citep{van22}. Although supernova kick is included in the delay-time distributions from BPSs, it is difficult to include it in the sub-galactic scale. In dSph galaxies, star formation takes place slowly, chemical enrichment proceeds inefficiently, and thus the inhomogeneous enrichment effect becomes even more important. Higher resolution of simulations with a new formalism/code will be required.

\subsection{Extra-galactic archaeology}
\label{sec:cosmo}

\begin{figure}[t]
\begin{center}
  \includegraphics[width=0.48\textwidth]{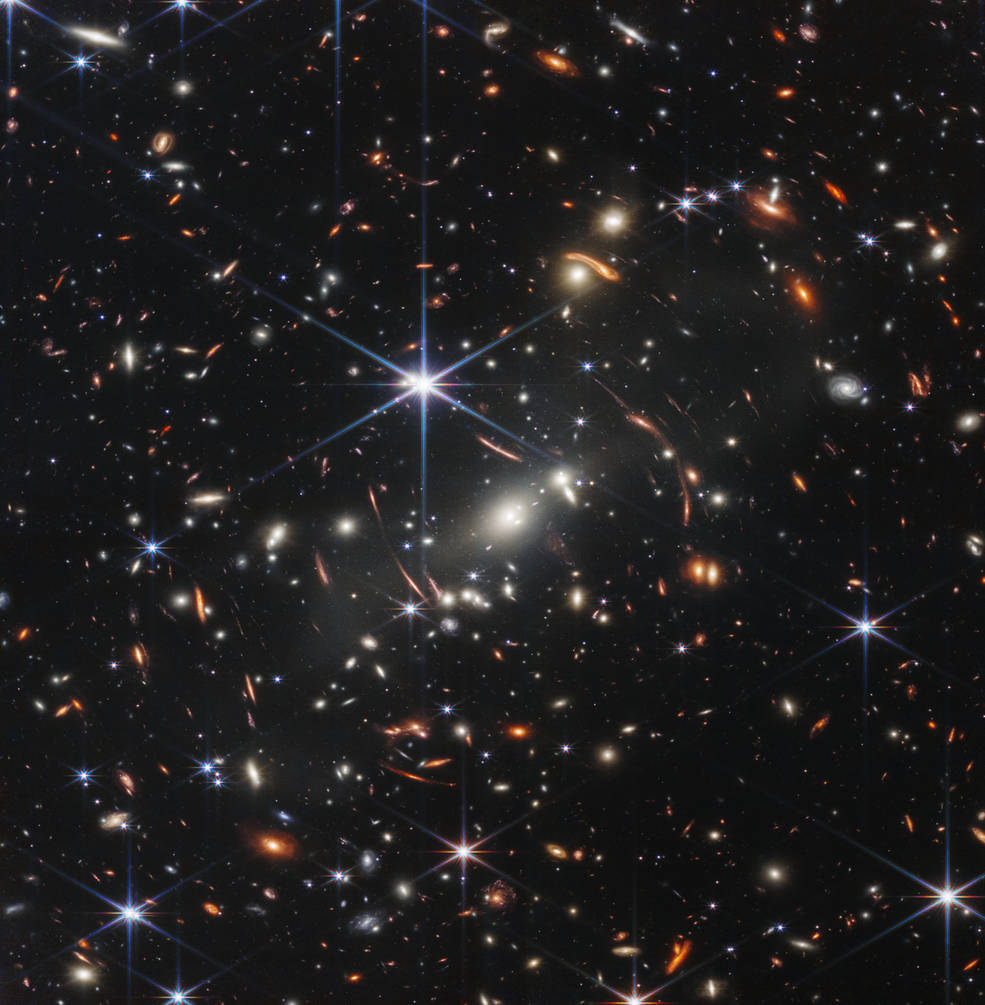}
  \includegraphics[width=0.5\textwidth]{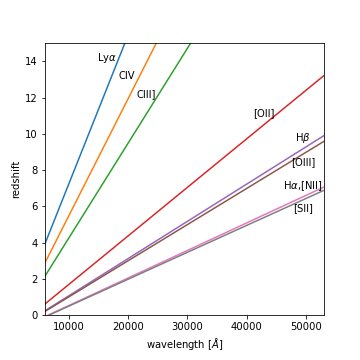}
\caption{(Left panel) The first NIRCam/JWST image of a cluster of galaxy SMAC0723 at $z=0.39$ but showing much higher redshift galaxies too, released on 11th July 2022. Figure is taken from \url{https://webbtelescope.org}.
(Right panel) Emission lines that can be obtained with NIRSpec/JWST as a function of wavelength and redshift.
}
\label{fig:jwst}
\end{center}
\end{figure}

Elemental abundances and isotopic ratios provide additional constrains on the timescales of formation and evolution of galaxies. In galaxies beyond the Local Group, it is not usually possible to resolve stars or HII regions, and metallicities of stellar populations\footnote{Multiple lines should be used to break the age-metallicity degeneracy.} or ISM are estimated from absorption or emission lines in integrated spectra. Thanks to MOS surveys since SDSS the sample is greatly increased, and thanks to IFU surveys since SAURON the spatial (projected) distributions of metallicities are also obtained.
Note that the methods for obtaining the absolute values of physical quantities are still debated \citep{worthey92,conroy13,mai19,kew19}.
For stellar populations, $\alpha$/Fe ratios have been used to estimate the formation timescale of early-type galaxies \citep{tho05,kriek16}, while Fe is not accessible in star-forming galaxies, and instead CNO abundances can be used \citep{vin18a}. The James Webb Space Telescope (JWST) will push these observations toward higher redshifts, and the wavelength coverage of which will allow us to measure CNO abundance simultaneously (Figure \ref{fig:jwst}). More elements are available for X-ray hot gas or quasar absorption line systems, although neutron capture elements are out of reach.
Isotopic ratios of light elements and some light element abundances (e.g., F) are also estimated with Atacama Large Millimeter Array (ALMA).

\begin{figure}[t]
\center
\includegraphics[width=0.95\textwidth]{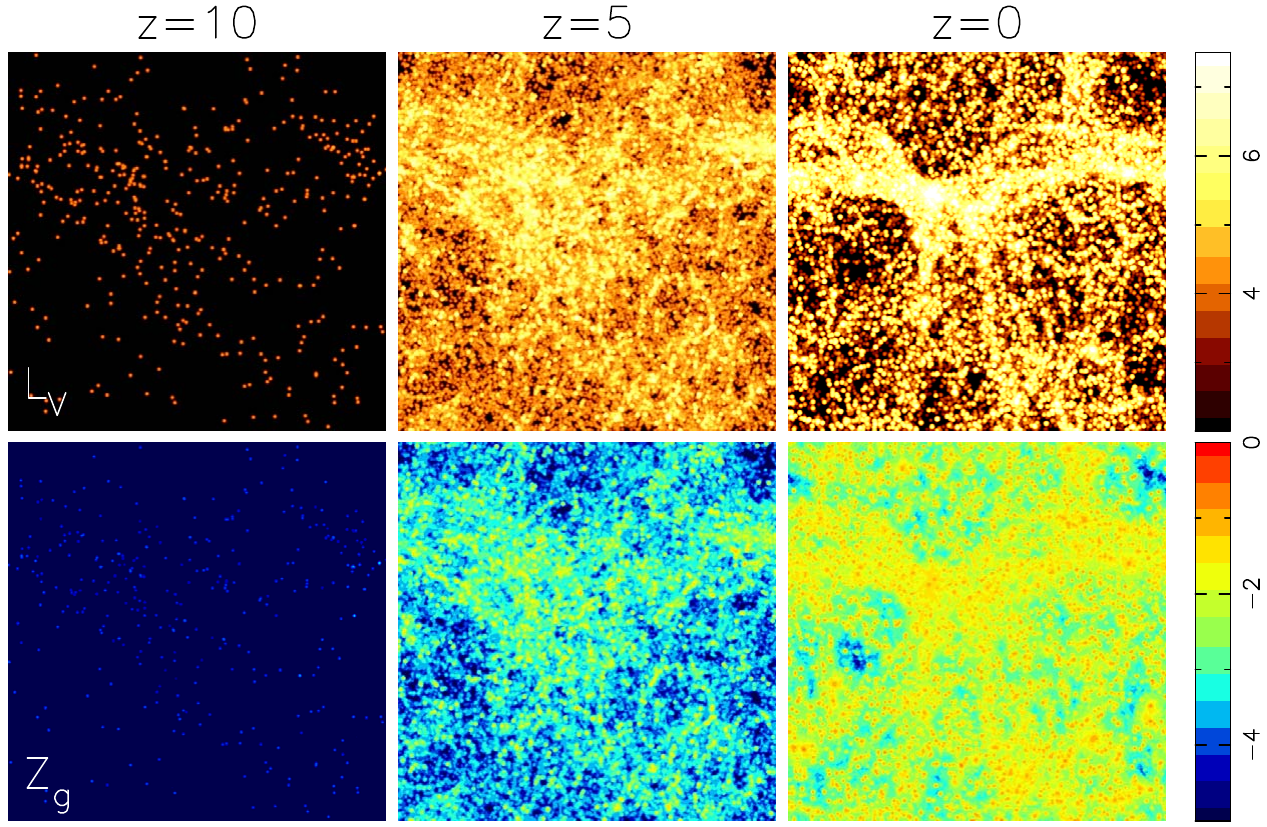}
\caption{The time evolution of our cosmological simulation in a periodic box $50h^{-1}$ Mpc on a side. We show the projected stellar $V$-luminosity (upper panels) and gas metallicity $\log Z_{\rm g}/Z_\odot$ (lower panels).
}
\label{fig:cosmicmap}
\end{figure}

Figure \ref{fig:cosmicmap} shows the evolution of stellar luminosities and metallicities in a cosmological simulation.
More massive galaxies tend to have higher metallicities, which is called {\it mass--metallicity relations} (MZR).
As already discussed, the origin of these relations are mass-dependent galactic winds by supernova feedback \citep{kob07,tay20}. More intense star formation also happens in the simulated massive galaxies. An IMF variation is also possible \citep{kob10imf}. Metal-loss by AGN feedback is not significant \citep{tay15letter}.

Various ways are possible to calculate the abundance ratio of the stellar population (or ISM) of a galaxy:
\begin{eqnarray}
\nonumber
&& \log X/Y  =
\log\left(\sum_N M_X/\sum_N M_Y\right), 
\log\left(\sum_N M \frac{M_X}{M_Y}/\sum_N M\right),
\left(\sum_N M\,\log\frac{M_X}{M_Y}\right)/\sum_N M,\\
&& \log\left(\sum_N \frac{L M_X}{M}/\sum_N \frac{L M_Y}{M}\right),
\log\left(\sum_N L \frac{M_X}{M_Y}/\sum_N L\right), 
\mathrm{or} \left(\sum_N L\,\log\frac{M_X}{M_Y}\right)/\sum_N L,
\end{eqnarray}
where $N$ is the number of stars/star particles (or HII regions/gas particles).
The first three are mass-weighted, but in order to compare with observational data, it should be weighted by luminosity $L$.
For stellar populations, we use V-band luminosity where Mgb line (5143-5206 \AA) is located, and these definitions give slightly different results.
For the ISM, the star formation rate of each gas particle can be used for weighting since it correlates with H$\alpha$ emissions. This weighting gives very different results as there are many non-star forming gas particles in galaxies (see Fig.\,19 of \citealt{kob07}).
It is not obvious which is the best definition.

Within galaxies, central parts are more metal-rich than outskirts, which is called {\it metallicity radial gradients}.
The origin of these gradients is the inside-out formation, where star formation starts from the centre with a higher efficiency and more chemical enrichment cycles.
The star formation is more intense at the centre in the simulated galaxies.
The star formation duration is not necessarily longer in the centre if quenching happens also inside-out by AGN feedback. 
Radial flows steepen the gradients, while galaxy mergers flatten the gradients \citep{kob04}.
Re-distribution of metals by AGN feedback may be important for gas-phase metallicity gradients \citep{tay17}.
Simulations predict a significant scatter of gradients at a given mass \citep{kob04,tay17}, while the average of the gradients become steeper at higher redshifts (\citealt{pik12}; Kobayashi \& Taylor, in prep.).

The metallicity gradient is defined in a log-scale for stellar populations but in a linear scale for gas-phase, and their example values are:
\begin{equation}
\Delta \log Z_*/\Delta \log r \sim 0.3
\end{equation}
for nearby early-type galaxies and
\begin{equation}
\Delta Z_{\rm g}/\Delta r \sim 0.05 \mathrm{~[dex/kpc]}
\end{equation}
for nearby disks, where $r$ is the projected, galactocentric radius.
These definitions give a good fit to slit observations of nearby galaxies \citep[e.g.,][]{zaritsky94,kob99} that cover a large radius ($r \ltsim 2r_{\rm e}$), where $r_{\rm e}$ is the effective radius at which a half of stellar luminosity is enclosed. 
Often, the central part ($r \ltsim 1$ kpc) is excluded from the fitting because it shows a flattening due to a limited seeing in observations, or due to a limited spatial resolution in simulations.
The outer part also shows a sudden decrease in metallicity, or an increase due to satellite galaxies, which is also excluded from the fitting.
However, recent analysis of IFU surveys stellar metallicities is also measured against linear $r$, and it is important to clarify the definition.

\begin{figure}[t]
\begin{center}
  \includegraphics[width=0.35\textwidth]{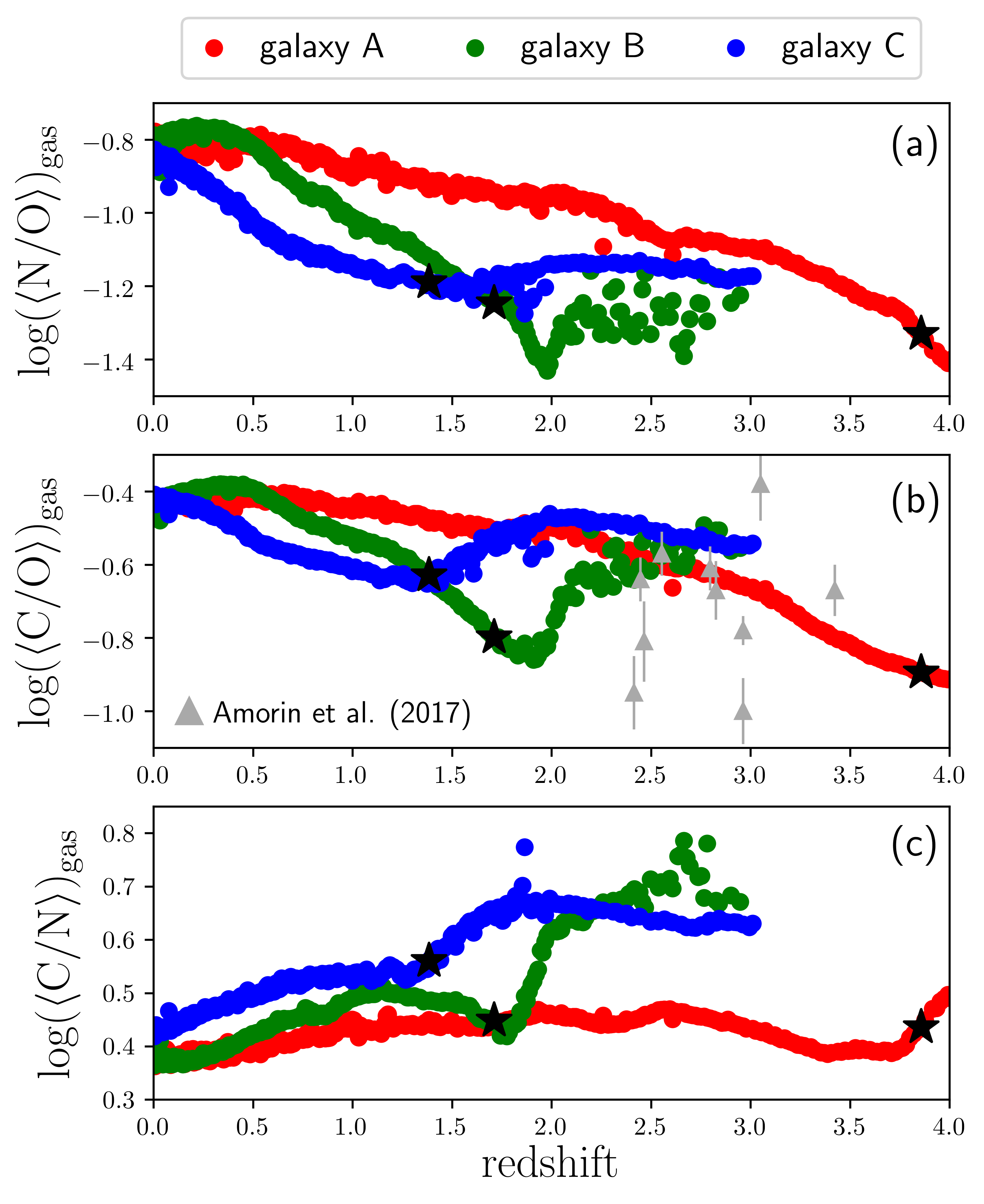}
  \includegraphics[width=0.60\textwidth]{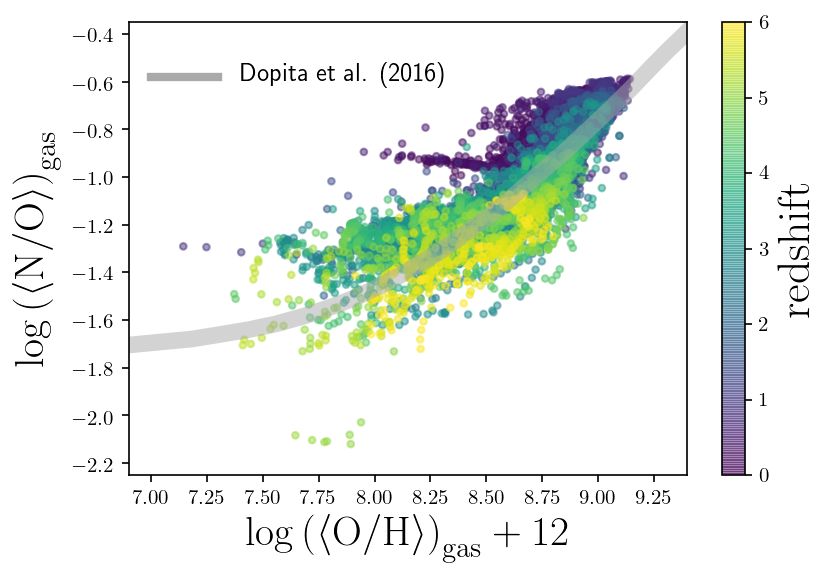}
\caption{(Left) Evolution of gas-phase CNO abundance ratios in our simulated disk galaxies with different star formation timescales, taken from a cosmological simulation.
Figure is taken from \cite{vin18a}.
(Right) Evolution of the gas-phase N/O--O/H relation of our simulated galaxies in a cosmological simulation. The gray bar indicates the compilation of observational data.
Figure is taken from \cite{vin19}.
}
\label{fig:cno}
\end{center}
\end{figure}

Our cosmological simulations also predict elemental abundances and isotopic ratios.
The left panel of Figure \ref{fig:cno} shows theoretical predictions of CNO abundance ratios for simulated disk galaxies that have different star formation timescales. These galaxies are chosen from a cosmological simulation, which is run by a Gadget-3 based code that includes detailed chemical enrichment \citep{kob07}. In the simulation, C is mainly produced from low-mass stars ($\ltsim 4M_\odot$), N is from intermediate-mass stars ($\gtsim 4M_\odot$) as a primary process \citep{kob11agb}, and O is from massive stars ($\gtsim 13M_\odot$). C and N are also produced by massive stars; the N yield depends on the metallicity as a secondary process, and can be greatly enhanced by stellar rotation (as for F).

In the nearby universe, the N/O--O/H relation is known for stellar and ISM abundances, which shows a plateau (N/O $\sim -1.6$) at low metallicities and a rapid increase toward higher metallicities. Damped Ly$\alpha$ systems at higher redshifts also roughly follow the same plateau. This relation was interpreted as the necessity of rotating massive stars by \citet{chi06}. However, this should be studied with hydrodynamical simulations including detailed chemical enrichment, and \citet{kob14iau} first showed the N/O--O/H relation in a chemodynamical simulation. \citet{vin18no} showed that both the global relation, which is obtained for average abundances of the entire galaxies, and the local relation, which is obtained for spatially resolved abundances from IFU data, can be reproduced by the inhomogeneous enrichment from AGB stars. Since N yield increases at higher metallicities, the global relation originates from the mass--metallicity relation of galaxies, while the local relation is caused by radial metallicity gradients within galaxies. 
Moreover, the right panel of Figure \ref{fig:cno} shows a theoretical prediction on the time evolution of the N/O--O/H relation, where galaxies evolve along the relation. Recent observation with KMOS on VLT confirmed a near redshift-invariant N/O-O/H relation \citep[][KLEVER survey]{hay22}. 

\begin{figure}[t]
\begin{center}
\vspace*{-2.5cm}
  \includegraphics[width=0.7\textwidth]{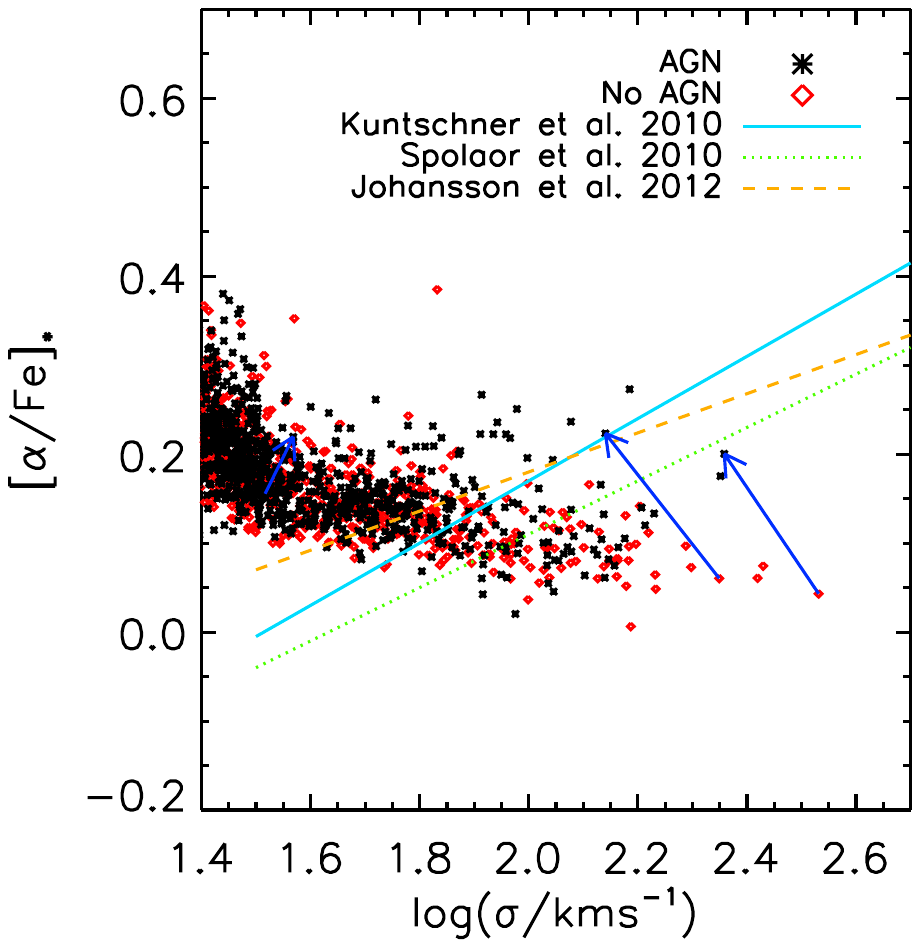}
\vspace*{-2.5cm}
\caption{Stellar [O/Fe] ratios as a function of central velocity dispersion, which is a proxy of galaxy mass, for our simulated galaxies with (black asterisks) and without (red diamonds) AGN feedback at $z=0$. The arrows indicate the differences in two cosmological simulations from the same initial conditions. The solid lines indicate the observed relations. Figure is taken from \citet{tay15}.
}
\label{fig:ofe-cosmic}
\end{center}
\end{figure}

Figure \ref{fig:ofe-cosmic} shows [$\alpha$/Fe] ratios of stellar populations as a function of galaxy mass for our simulated galaxies with and without AGN feedback. Without AGN feedback (red diamonds), since star formation lasts longer in massive galaxies with a deeper potential well, [$\alpha$/Fe] ratios become lower in massive galaxies, which is the opposite compared with the observations (solid lines). With AGN feedback (black asterisks) star formation can be suppressed before the SN Ia enrichment becomes dominant in massive galaxies, so that [$\alpha$/Fe] ratios can stay high. However, the scatter of [$\alpha$/Fe] ratios at a given galaxy mass is still larger than observed.

[$\alpha$/Fe] ratios become higher at higher redshifts in the simulations, but not as much as observed \citep[e.g.,][]{kriek16}.
Some cosmological simulations claimed that they could reproduce the observed [$\alpha$/Fe]--mass relation at $z=0$, but with an ad-hoc SN Ia model. Some even introduced a variation in the IMF or in the binary fraction. 
However, the [$\alpha$/Fe] problem should be discussed with a chemodynamical model that is based on nuclear astrophysics and that can reproduce the observed elemental abundances both in the Milky Way and dSph galaxies. We do not have such a model yet.

\section{Conclusions and Future Prospects}

Thanks to the long-term collaborations between nuclear and astrophysics, we have good understanding on the origin of elements (except for the elements around Ti and a few neutron-capture elements such as Au).
Inhomogeneous enrichment is extremely important for interpreting the elemental abundance trends. It can reproduce the observed N/O--O/H relation only with AGB stars and supernovae (Fig.\,\ref{fig:cno}), but not the observed r-process abundances only with NSMs (Fig.\,\ref{fig:mweu}); an r-process associated with core-collapse supernovae such as magneto-rotational hypernovae is required (Fig.\,\ref{fig:yong}), although the explosion mechanism is unknown.
It is necessary to run chemodynamical simulations from cosmological initial conditions, including detailed chemical enrichment based on nuclear astrophysics.

The impact of stellar rotation, binaries, and magnetic fields during stellar evolution, and the multi-dimensional effects of supernova explosions in nucleosynthesis should be investigated further.
If Wolf-Rayet stars are producing heavy elements on a very short timescale, it might be hard to find very metal-poor or Population III (and dust-free) galaxies at very high redshifts even with JWST.
In that case the elemental abundances can be quite different in very high redshift galaxies (e.g., high (C,N)/O; Fig.\,\ref{fig:wwcno}).
Although there is no observational evidence, pair-instability supernovae could also cause very different abundance pattern (e.g., high (Si,S)/O; Fig.\,\ref{fig:cdla}).
Finally, some metals are locked in a solid state - it is also important to calculate element-by-element dust formation, growth, and destruction, as well as the detailed chemical enrichment.

Galactic archaeology is a powerful approach for reconstructing the formation history of the Milky Way and its satellite galaxies.
APOGEE and HERMES-GALAH surveys have provided homogeneous datasets of many elemental abundances that can be statistically compared with chemodynamical simulations. Future surveys with WEAVE and 4MOST will provide more.
Having said that, the number of EMP stars will not be increased so much in these surveys, and a target survey such as the SkyMapper EMP survey is also needed, in particular for constraining the early chemical enrichment from the first stars.
It is important to increase spectral resolution and wavelength coverage (including UV), to obtain more accurate abundances of more elements (namely neutron-capture elements).

Reflecting the difference in the formation timescale, elemental abundances depend on the location within galaxies.
Although this dependence has been explored toward the Galactic bulge by APOGEE, the dependence at the outer disk is still unknown, which requires 8m-class telescopes such as the PFS on Subaru telescope. Despite the limited spectral resolution of the PFS, $\alpha$/Fe and a small number of elements will be available.
The PFS will also be able to explore the $\alpha$/Fe bimodality in M31; it is not yet known if M31 has a similar $\alpha$/Fe dichotomy or not.

The next step will be to apply the Galactic archaeology approach to external or distant galaxies. Although it became possible to map metallicity, some elemental abundances, and kinematics within galaxies with IFU, the sample and/or spatial resolution are still limited even with JWST.
Integrated physical quantities over galaxies, or stacked quantities at a given mass bin, will also be useful, which can be done with the same MOS developed for Galactic archaeology (although optimal spectral resolutions and wavelength coverages are different).
Spectroscopic surveys with 8m-class telescopes will be useful, and it is a matter of urgency to establish the analysis methods to obtain absolute values of metallicities and elemental abundances from the observational data.
In addition, ALMA has opened a new window for elemental abundances and isotopic ratios in high-redshift galaxies.
Extra-galactic archaeology will become popular in coming years.

\acknowledgement
We thank D. Yong, F. Vincenzo, A. Karakas, M. Lugaro, N. Tominaga, S.-C. Leung, M. Ishigaki, K. Nomoto, L. Kewley, R. Maiolino, A. Bunker for fruitful discussion, and V. Springel for providing Gadget-3.
CK acknowledge funding from the UK Science and Technology Facility Council through grant ST/M000958/1, ST/R000905/1, ST/V000632/1. 
The work was also funded by a Leverhulme Trust Research Project Grant on ``Birth of Elements''.

\bibliography{book}{}

\begin{thebibliography}{}
\expandafter\ifx\csname natexlab\endcsname\relax\def\natexlab#1{#1}\fi
\providecommand{\url}[1]{\href{#1}{#1}}
\providecommand{\dodoi}[1]{doi:~\href{http://doi.org/#1}{\nolinkurl{#1}}}
\providecommand{\doeprint}[1]{\href{http://ascl.net/#1}{\nolinkurl{http://ascl.net/#1}}}
\providecommand{\doarXiv}[1]{\href{https://arxiv.org/abs/#1}{\nolinkurl{https://arxiv.org/abs/#1}}}

\bibitem[{{Amarsi} {et~al.}(2019){Amarsi}, {Nissen}, \&
  {Sk{\'u}lad{\'o}ttir}}]{ama19b}
{Amarsi}, A.~M., {Nissen}, P.~E., \& {Sk{\'u}lad{\'o}ttir}, {\'A}. 2019, \aap,
  630, A104, \dodoi{10.1051/0004-6361/201936265}

\bibitem[{{Arcones} {et~al.}(2007){Arcones}, {Janka}, \& {Scheck}}]{arc07}
{Arcones}, A., {Janka}, H.~T., \& {Scheck}, L. 2007, \aap, 467, 1227,
  \dodoi{10.1051/0004-6361:20066983}

\bibitem[{{Arnaboldi} {et~al.}(2022){Arnaboldi}, {Bhattacharya}, {Gerhard},
  {Kobayashi}, {Freeman}, {Caldwell}, {Hartke}, {McConnachie}, \&
  {Guhathakurta}}]{arnaboldi22}
{Arnaboldi}, M., {Bhattacharya}, S., {Gerhard}, O., {et~al.} 2022, \aap, 666,
  A109, \dodoi{10.1051/0004-6361/202244258}

\bibitem[{{Asplund} {et~al.}(2009){Asplund}, {Grevesse}, {Sauval}, \&
  {Scott}}]{asp09}
{Asplund}, M., {Grevesse}, N., {Sauval}, A.~J., \& {Scott}, P. 2009, \araa, 47,
  481, \dodoi{10.1146/annurev.astro.46.060407.145222}

\bibitem[{{Audouze} \& {Silk}(1995)}]{audouze95}
{Audouze}, J., \& {Silk}, J. 1995, \apjl, 451, L49, \dodoi{10.1086/309687}

\bibitem[{{Barkat} {et~al.}(1967){Barkat}, {Rakavy}, \& {Sack}}]{barkat67}
{Barkat}, Z., {Rakavy}, G., \& {Sack}, N. 1967, \prl, 18, 379,
  \dodoi{10.1103/PhysRevLett.18.379}

\bibitem[{{Bensby} {et~al.}(2003){Bensby}, {Feltzing}, \&
  {Lundstr{\"o}m}}]{ben03}
{Bensby}, T., {Feltzing}, S., \& {Lundstr{\"o}m}, I. 2003, \aap, 410, 527,
  \dodoi{10.1051/0004-6361:20031213}

\bibitem[{{Bensby} {et~al.}(2004){Bensby}, {Feltzing}, \&
  {Lundstr{\"o}m}}]{ben04}
---. 2004, \aap, 415, 155, \dodoi{10.1051/0004-6361:20031655}

\bibitem[{{Blondin} {et~al.}(2022){Blondin}, {Bravo}, {Timmes}, {Dessart}, \&
  {Hillier}}]{blondin22}
{Blondin}, S., {Bravo}, E., {Timmes}, F.~X., {Dessart}, L., \& {Hillier}, D.~J.
  2022, \aap, 660, A96, \dodoi{10.1051/0004-6361/202142323}

\bibitem[{{Brook} {et~al.}(2012){Brook}, {Stinson}, {Gibson}, {Kawata},
  {House}, {Miranda}, {Macci{\`o}}, {Pilkington}, {Ro{\v{s}}kar}, {Wadsley}, \&
  {Quinn}}]{brook12}
{Brook}, C.~B., {Stinson}, G.~S., {Gibson}, B.~K., {et~al.} 2012, \mnras, 426,
  690, \dodoi{10.1111/j.1365-2966.2012.21738.x}

\bibitem[{{Brown} {et~al.}(2001){Brown}, {Heger}, {Langer}, {Lee}, {Wellstein},
  \& {Bethe}}]{brown01}
{Brown}, G.~E., {Heger}, A., {Langer}, N., {et~al.} 2001, \na, 6, 457,
  \dodoi{10.1016/S1384-1076(01)00077-X}

\bibitem[{{Buck} {et~al.}(2020){Buck}, {Obreja}, {Macci{\`o}}, {Minchev},
  {Dutton}, \& {Ostriker}}]{buck20}
{Buck}, T., {Obreja}, A., {Macci{\`o}}, A.~V., {et~al.} 2020, \mnras, 491,
  3461, \dodoi{10.1093/mnras/stz3241}

\bibitem[{{Buder} {et~al.}(2018){Buder}, {Asplund}, {Duong}, {Kos}, {Lind},
  {Ness}, {Sharma}, {Bland-Hawthorn}, {Casey}, {de Silva}, {D'Orazi},
  {Freeman}, {Lewis}, {Lin}, {Martell}, {Schlesinger}, {Simpson}, {Zucker},
  {Zwitter}, {Amarsi}, {Anguiano}, {Carollo}, {Casagrande}, {{\v{C}}otar},
  {Cottrell}, {da Costa}, {Gao}, {Hayden}, {Horner}, {Ireland}, {Kafle},
  {Munari}, {Nataf}, {Nordlander}, {Stello}, {Ting}, {Traven}, {Watson},
  {Wittenmyer}, {Wyse}, {Yong}, {Zinn}, {{\v{Z}}erjal}, \& {Galah
  Collaboration}}]{buder18}
{Buder}, S., {Asplund}, M., {Duong}, L., {et~al.} 2018, \mnras, 478, 4513,
  \dodoi{10.1093/mnras/sty1281}

\bibitem[{{Burkert} \& {Hensler}(1987)}]{bur87}
{Burkert}, A., \& {Hensler}, G. 1987, in Nuclear Astrophysics, ed.
  W.~{Hillebrandt}, R.~{Kuhfuss}, E.~{Mueller}, \& J.~W. {Truran}, Vol. 287,
  159, \dodoi{10.1007/BFb0016576}

\bibitem[{{Burrows} \& {Vartanyan}(2021)}]{bur21}
{Burrows}, A., \& {Vartanyan}, D. 2021, \nat, 589, 29,
  \dodoi{10.1038/s41586-020-03059-w}

\bibitem[{{Busso} {et~al.}(1999){Busso}, {Gallino}, \& {Wasserburg}}]{busso99}
{Busso}, M., {Gallino}, R., \& {Wasserburg}, G.~J. 1999, \araa, 37, 239,
  \dodoi{10.1146/annurev.astro.37.1.239}

\bibitem[{{Cappellari} {et~al.}(2013){Cappellari}, {McDermid}, {Alatalo},
  {Blitz}, {Bois}, {Bournaud}, {Bureau}, {Crocker}, {Davies}, {Davis}, {de
  Zeeuw}, {Duc}, {Emsellem}, {Khochfar}, {Krajnovi{\'c}}, {Kuntschner},
  {Morganti}, {Naab}, {Oosterloo}, {Sarzi}, {Scott}, {Serra}, {Weijmans}, \&
  {Young}}]{cappellari13}
{Cappellari}, M., {McDermid}, R.~M., {Alatalo}, K., {et~al.} 2013, \mnras, 432,
  1862, \dodoi{10.1093/mnras/stt644}

\bibitem[{{Cayrel} {et~al.}(2004){Cayrel}, {Depagne}, {Spite}, {Hill}, {Spite},
  {Fran{\c{c}}ois}, {Plez}, {Beers}, {Primas}, {Andersen}, {Barbuy},
  {Bonifacio}, {Molaro}, \& {Nordstr{\"o}m}}]{cay04}
{Cayrel}, R., {Depagne}, E., {Spite}, M., {et~al.} 2004, \aap, 416, 1117,
  \dodoi{10.1051/0004-6361:20034074}

\bibitem[{{Cen} \& {Ostriker}(1999)}]{cen99}
{Cen}, R., \& {Ostriker}, J.~P. 1999, \apjl, 519, L109, \dodoi{10.1086/312123}

\bibitem[{{Cescutti} \& {Chiappini}(2014)}]{ces14}
{Cescutti}, G., \& {Chiappini}, C. 2014, \aap, 565, A51,
  \dodoi{10.1051/0004-6361/201423432}

\bibitem[{{Cescutti} \& {Kobayashi}(2017)}]{ces17}
{Cescutti}, G., \& {Kobayashi}, C. 2017, \aap, 607, A23,
  \dodoi{10.1051/0004-6361/201731398}

\bibitem[{{Chabrier}(2003)}]{cha03}
{Chabrier}, G. 2003, \pasp, 115, 763, \dodoi{10.1086/376392}

\bibitem[{{Chary} {et~al.}(2007){Chary}, {Berger}, \& {Cowie}}]{chary07}
{Chary}, R., {Berger}, E., \& {Cowie}, L. 2007, \apj, 671, 272,
  \dodoi{10.1086/522692}

\bibitem[{{Chiappini} {et~al.}(2006){Chiappini}, {Hirschi}, {Meynet},
  {Ekstr{\"o}m}, {Maeder}, \& {Matteucci}}]{chi06}
{Chiappini}, C., {Hirschi}, R., {Meynet}, G., {et~al.} 2006, \aap, 449, L27,
  \dodoi{10.1051/0004-6361:20064866}

\bibitem[{{Chiappini} {et~al.}(1997){Chiappini}, {Matteucci}, \&
  {Gratton}}]{chi97}
{Chiappini}, C., {Matteucci}, F., \& {Gratton}, R. 1997, \apj, 477, 765,
  \dodoi{10.1086/303726}

\bibitem[{{Conroy}(2013)}]{conroy13}
{Conroy}, C. 2013, \araa, 51, 393, \dodoi{10.1146/annurev-astro-082812-141017}

\bibitem[{{Cristallo} {et~al.}(2011){Cristallo}, {Piersanti}, {Straniero},
  {Gallino}, {Dom{\'\i}nguez}, {Abia}, {Di Rico}, {Quintini}, \&
  {Bisterzo}}]{cri11}
{Cristallo}, S., {Piersanti}, L., {Straniero}, O., {et~al.} 2011, \apjs, 197,
  17, \dodoi{10.1088/0067-0049/197/2/17}

\bibitem[{{Croton} {et~al.}(2006){Croton}, {Springel}, {White}, {De Lucia},
  {Frenk}, {Gao}, {Jenkins}, {Kauffmann}, {Navarro}, \& {Yoshida}}]{cro06}
{Croton}, D.~J., {Springel}, V., {White}, S. D.~M., {et~al.} 2006, \mnras, 365,
  11, \dodoi{10.1111/j.1365-2966.2005.09675.x}

\bibitem[{{Dalla Vecchia} \& {Schaye}(2008)}]{dal08}
{Dalla Vecchia}, C., \& {Schaye}, J. 2008, \mnras, 387, 1431,
  \dodoi{10.1111/j.1365-2966.2008.13322.x}

\bibitem[{{Dalla Vecchia} \& {Schaye}(2012)}]{dal12}
---. 2012, \mnras, 426, 140, \dodoi{10.1111/j.1365-2966.2012.21704.x}

\bibitem[{{Dav{\'e}} {et~al.}(2019){Dav{\'e}}, {Angl{\'e}s-Alc{\'a}zar},
  {Narayanan}, {Li}, {Rafieferantsoa}, \& {Appleby}}]{dave19}
{Dav{\'e}}, R., {Angl{\'e}s-Alc{\'a}zar}, D., {Narayanan}, D., {et~al.} 2019,
  \mnras, 486, 2827, \dodoi{10.1093/mnras/stz937}

\bibitem[{{De Donder} \& {Vanbeveren}(2004)}]{ded04}
{De Donder}, E., \& {Vanbeveren}, D. 2004, \nar, 48, 861,
  \dodoi{10.1016/j.newar.2004.07.001}

\bibitem[{{Doherty} {et~al.}(2014){Doherty}, {Gil-Pons}, {Lau}, {Lattanzio}, \&
  {Siess}}]{doh14a}
{Doherty}, C.~L., {Gil-Pons}, P., {Lau}, H. H.~B., {Lattanzio}, J.~C., \&
  {Siess}, L. 2014, \mnras, 437, 195, \dodoi{10.1093/mnras/stt1877}

\bibitem[{{Doherty} {et~al.}(2015){Doherty}, {Gil-Pons}, {Siess}, {Lattanzio},
  \& {Lau}}]{doh15}
{Doherty}, C.~L., {Gil-Pons}, P., {Siess}, L., {Lattanzio}, J.~C., \& {Lau}, H.
  H.~B. 2015, \mnras, 446, 2599, \dodoi{10.1093/mnras/stu2180}

\bibitem[{{Dolag} {et~al.}(2017){Dolag}, {Mevius}, \& {Remus}}]{dolag17}
{Dolag}, K., {Mevius}, E., \& {Remus}, R.-S. 2017, Galaxies, 5, 35,
  \dodoi{10.3390/galaxies5030035}

\bibitem[{{Driver} {et~al.}(2018){Driver}, {Andrews}, {da Cunha}, {Davies},
  {Lagos}, {Robotham}, {Vinsen}, {Wright}, {Alpaslan}, {Bland-Hawthorn},
  {Bourne}, {Brough}, {Bremer}, {Cluver}, {Colless}, {Conselice}, {Dunne},
  {Eales}, {Gomez}, {Holwerda}, {Hopkins}, {Kafle}, {Kelvin}, {Loveday},
  {Liske}, {Maddox}, {Phillipps}, {Pimbblet}, {Rowlands}, {Sansom}, {Taylor},
  {Wang}, \& {Wilkins}}]{driver18}
{Driver}, S.~P., {Andrews}, S.~K., {da Cunha}, E., {et~al.} 2018, \mnras, 475,
  2891, \dodoi{10.1093/mnras/stx2728}

\bibitem[{{Dubois} {et~al.}(2012){Dubois}, {Devriendt}, {Slyz}, \&
  {Teyssier}}]{dub12}
{Dubois}, Y., {Devriendt}, J., {Slyz}, A., \& {Teyssier}, R. 2012, \mnras, 420,
  2662, \dodoi{10.1111/j.1365-2966.2011.20236.x}

\bibitem[{{Dubois} {et~al.}(2016){Dubois}, {Peirani}, {Pichon}, {Devriendt},
  {Gavazzi}, {Welker}, \& {Volonteri}}]{dub16}
{Dubois}, Y., {Peirani}, S., {Pichon}, C., {et~al.} 2016, \mnras, 463, 3948,
  \dodoi{10.1093/mnras/stw2265}

\bibitem[{{Federrath} {et~al.}(2011){Federrath}, {Sur}, {Schleicher},
  {Banerjee}, \& {Klessen}}]{fed11}
{Federrath}, C., {Sur}, S., {Schleicher}, D. R.~G., {Banerjee}, R., \&
  {Klessen}, R.~S. 2011, \apj, 731, 62, \dodoi{10.1088/0004-637X/731/1/62}

\bibitem[{{Few} {et~al.}(2014){Few}, {Courty}, {Gibson}, {Michel-Dansac}, \&
  {Calura}}]{few14}
{Few}, C.~G., {Courty}, S., {Gibson}, B.~K., {Michel-Dansac}, L., \& {Calura},
  F. 2014, \mnras, 444, 3845, \dodoi{10.1093/mnras/stu1709}

\bibitem[{{Font} {et~al.}(2020){Font}, {McCarthy}, {Poole-Mckenzie},
  {Stafford}, {Brown}, {Schaye}, {Crain}, {Theuns}, \& {Schaller}}]{font20}
{Font}, A.~S., {McCarthy}, I.~G., {Poole-Mckenzie}, R., {et~al.} 2020, \mnras,
  498, 1765, \dodoi{10.1093/mnras/staa2463}

\bibitem[{{Franco} {et~al.}(2021){Franco}, {Coppin}, {Geach}, {Kobayashi},
  {Chapman}, {Yang}, {Gonz{\'a}lez-Alfonso}, {Spilker}, {Cooray}, \&
  {Micha{\l}owski}}]{franco21}
{Franco}, M., {Coppin}, K.~E.~K., {Geach}, J.~E., {et~al.} 2021, Nature
  Astronomy, 5, 1240, \dodoi{10.1038/s41550-021-01515-9}

\bibitem[{{Frischknecht} {et~al.}(2016){Frischknecht}, {Hirschi}, {Pignatari},
  {Maeder}, {Meynet}, {Chiappini}, {Thielemann}, {Rauscher}, {Georgy}, \&
  {Ekstr{\"o}m}}]{fri16}
{Frischknecht}, U., {Hirschi}, R., {Pignatari}, M., {et~al.} 2016, \mnras, 456,
  1803, \dodoi{10.1093/mnras/stv2723}

\bibitem[{{Fuhrmann}(1998)}]{fuhrmann98}
{Fuhrmann}, K. 1998, \aap, 338, 161

\bibitem[{{Gaia Collaboration} {et~al.}(2022){Gaia Collaboration},
  {Recio-Blanco}, {Kordopatis}, {de Laverny}, {Palicio}, {Spagna}, {Spina},
  {Katz}, {Re Fiorentin}, {Poggio}, \& {et al.}}]{gaiadr3}
{Gaia Collaboration}, {Recio-Blanco}, A., {Kordopatis}, G., {et~al.} 2022,
  arXiv e-prints, arXiv:2206.05534.
\newblock \doarXiv{2206.05534}

\bibitem[{{Grand} {et~al.}(2017){Grand}, {G{\'o}mez}, {Marinacci}, {Pakmor},
  {Springel}, {Campbell}, {Frenk}, {Jenkins}, \& {White}}]{grand17}
{Grand}, R. J.~J., {G{\'o}mez}, F.~A., {Marinacci}, F., {et~al.} 2017, \mnras,
  467, 179, \dodoi{10.1093/mnras/stx071}

\bibitem[{{Grichener} {et~al.}(2022){Grichener}, {Kobayashi}, \&
  {Soker}}]{gri22}
{Grichener}, A., {Kobayashi}, C., \& {Soker}, N. 2022, \apjl, 926, L9,
  \dodoi{10.3847/2041-8213/ac4f68}

\bibitem[{{Grisoni} {et~al.}(2019){Grisoni}, {Matteucci}, {Romano}, \&
  {Fu}}]{gri19}
{Grisoni}, V., {Matteucci}, F., {Romano}, D., \& {Fu}, X. 2019, \mnras, 489,
  3539, \dodoi{10.1093/mnras/stz2428}

\bibitem[{{Grisoni} {et~al.}(2017){Grisoni}, {Spitoni}, {Matteucci},
  {Recio-Blanco}, {de Laverny}, {Hayden}, {Mikolaitis}, \& {Worley}}]{gri17}
{Grisoni}, V., {Spitoni}, E., {Matteucci}, F., {et~al.} 2017, \mnras, 472,
  3637, \dodoi{10.1093/mnras/stx2201}

\bibitem[{{Gunawardhana} {et~al.}(2011){Gunawardhana}, {Hopkins}, {Sharp},
  {Brough}, {Taylor}, {Bland-Hawthorn}, {Maraston}, {Tuffs}, {Popescu},
  {Wijesinghe}, {Jones}, {Croom}, {Sadler}, {Wilkins}, {Driver}, {Liske},
  {Norberg}, {Baldry}, {Bamford}, {Loveday}, {Peacock}, {Robotham}, {Zucker},
  {Parker}, {Conselice}, {Cameron}, {Frenk}, {Hill}, {Kelvin}, {Kuijken},
  {Madore}, {Nichol}, {Parkinson}, {Pimbblet}, {Prescott}, {Sutherland},
  {Thomas}, \& {van Kampen}}]{gunawardhana11}
{Gunawardhana}, M.~L.~P., {Hopkins}, A.~M., {Sharp}, R.~G., {et~al.} 2011,
  \mnras, 415, 1647, \dodoi{10.1111/j.1365-2966.2011.18800.x}

\bibitem[{{Haardt} \& {Madau}(1996)}]{haa96}
{Haardt}, F., \& {Madau}, P. 1996, \apj, 461, 20, \dodoi{10.1086/177035}

\bibitem[{{Hansen} {et~al.}(2016){Hansen}, {Andersen}, {Nordstr{\"o}m},
  {Beers}, {Placco}, {Yoon}, \& {Buchhave}}]{han16}
{Hansen}, T.~T., {Andersen}, J., {Nordstr{\"o}m}, B., {et~al.} 2016, \aap, 586,
  A160, \dodoi{10.1051/0004-6361/201527235}

\bibitem[{{Hayden} {et~al.}(2015){Hayden}, {Bovy}, {Holtzman}, {Nidever},
  {Bird}, {Weinberg}, {Andrews}, {Majewski}, {Allende Prieto}, {Anders},
  {Beers}, {Bizyaev}, {Chiappini}, {Cunha}, {Frinchaboy},
  {Garc{\'\i}a-Her{\'n}andez}, {Garc{\'\i}a P{\'e}rez}, {Girardi}, {Harding},
  {Hearty}, {Johnson}, {M{\'e}sz{\'a}ros}, {Minchev}, {O'Connell}, {Pan},
  {Robin}, {Schiavon}, {Schneider}, {Schultheis}, {Shetrone}, {Skrutskie},
  {Steinmetz}, {Smith}, {Wilson}, {Zamora}, \& {Zasowski}}]{hay15}
{Hayden}, M.~R., {Bovy}, J., {Holtzman}, J.~A., {et~al.} 2015, \apj, 808, 132,
  \dodoi{10.1088/0004-637X/808/2/132}

\bibitem[{{Hayden-Pawson} {et~al.}(2022){Hayden-Pawson}, {Curti}, {Maiolino},
  {Cirasuolo}, {Belfiore}, {Cappellari}, {Concas}, {Cresci}, {Cullen},
  {Kobayashi}, {Mannucci}, {Marconi}, {Meneghetti}, {Mercurio}, {Peng},
  {Swinbank}, \& {Vincenzo}}]{hay22}
{Hayden-Pawson}, C., {Curti}, M., {Maiolino}, R., {et~al.} 2022, \mnras, 512,
  2867, \dodoi{10.1093/mnras/stac584}

\bibitem[{{Haynes} \& {Kobayashi}(2019)}]{hay19}
{Haynes}, C.~J., \& {Kobayashi}, C. 2019, \mnras, 483, 5123,
  \dodoi{10.1093/mnras/sty3389}

\bibitem[{{Herwig}(2005)}]{her05}
{Herwig}, F. 2005, \araa, 43, 435,
  \dodoi{10.1146/annurev.astro.43.072103.150600}

\bibitem[{{Hirschi}(2007)}]{hir07}
{Hirschi}, R. 2007, \aap, 461, 571, \dodoi{10.1051/0004-6361:20065356}

\bibitem[{{Hopkins} {et~al.}(2006){Hopkins}, {Hernquist}, {Cox}, {Di Matteo},
  {Robertson}, \& {Springel}}]{hop06}
{Hopkins}, P.~F., {Hernquist}, L., {Cox}, T.~J., {et~al.} 2006, \apjs, 163, 1,
  \dodoi{10.1086/499298}

\bibitem[{{Hopkins} {et~al.}(2018{\natexlab{a}}){Hopkins}, {Wetzel},
  {Kere{\v{s}}}, {Faucher-Gigu{\`e}re}, {Quataert}, {Boylan-Kolchin}, {Murray},
  {Hayward}, \& {El-Badry}}]{hop18}
{Hopkins}, P.~F., {Wetzel}, A., {Kere{\v{s}}}, D., {et~al.} 2018{\natexlab{a}},
  \mnras, 477, 1578, \dodoi{10.1093/mnras/sty674}

\bibitem[{{Hopkins} {et~al.}(2018{\natexlab{b}}){Hopkins}, {Wetzel},
  {Kere{\v{s}}}, {Faucher-Gigu{\`e}re}, {Quataert}, {Boylan-Kolchin}, {Murray},
  {Hayward}, {Garrison-Kimmel}, {Hummels}, {Feldmann}, {Torrey}, {Ma},
  {Angl{\'e}s-Alc{\'a}zar}, {Su}, {Orr}, {Schmitz}, {Escala}, {Sanderson},
  {Grudi{\'c}}, {Hafen}, {Kim}, {Fitts}, {Bullock}, {Wheeler}, {Chan},
  {Elbert}, \& {Narayanan}}]{hop18fire}
---. 2018{\natexlab{b}}, \mnras, 480, 800, \dodoi{10.1093/mnras/sty1690}

\bibitem[{{Ishigaki} {et~al.}(2018){Ishigaki}, {Tominaga}, {Kobayashi}, \&
  {Nomoto}}]{ish18}
{Ishigaki}, M.~N., {Tominaga}, N., {Kobayashi}, C., \& {Nomoto}, K. 2018, \apj,
  857, 46, \dodoi{10.3847/1538-4357/aab3de}

\bibitem[{{Israelian} {et~al.}(1998){Israelian}, {Garc{\'\i}a L{\'o}pez}, \&
  {Rebolo}}]{isr98}
{Israelian}, G., {Garc{\'\i}a L{\'o}pez}, R.~J., \& {Rebolo}, R. 1998, \apj,
  507, 805, \dodoi{10.1086/306351}

\bibitem[{{Janka}(2012)}]{jan12}
{Janka}, H.-T. 2012, Annual Review of Nuclear and Particle Science, 62, 407,
  \dodoi{10.1146/annurev-nucl-102711-094901}

\bibitem[{{Ji} {et~al.}(2016){Ji}, {Frebel}, {Chiti}, \& {Simon}}]{ji16}
{Ji}, A.~P., {Frebel}, A., {Chiti}, A., \& {Simon}, J.~D. 2016, \nat, 531, 610,
  \dodoi{10.1038/nature17425}

\bibitem[{{Just} {et~al.}(2015){Just}, {Bauswein}, {Ardevol Pulpillo},
  {Goriely}, \& {Janka}}]{jus15}
{Just}, O., {Bauswein}, A., {Ardevol Pulpillo}, R., {Goriely}, S., \& {Janka},
  H.~T. 2015, \mnras, 448, 541, \dodoi{10.1093/mnras/stv009}

\bibitem[{{Karakas}(2010)}]{kar10}
{Karakas}, A.~I. 2010, \mnras, 403, 1413,
  \dodoi{10.1111/j.1365-2966.2009.16198.x}

\bibitem[{{Karakas} \& {Lattanzio}(2014)}]{kar14}
{Karakas}, A.~I., \& {Lattanzio}, J.~C. 2014, \pasa, 31, e030,
  \dodoi{10.1017/pasa.2014.21}

\bibitem[{{Karakas} \& {Lugaro}(2016)}]{kar16}
{Karakas}, A.~I., \& {Lugaro}, M. 2016, \apj, 825, 26,
  \dodoi{10.3847/0004-637X/825/1/26}

\bibitem[{{Katz}(1992)}]{kat92}
{Katz}, N. 1992, \apj, 391, 502, \dodoi{10.1086/171366}

\bibitem[{{Kawata} \& {Gibson}(2003)}]{kaw03}
{Kawata}, D., \& {Gibson}, B.~K. 2003, \mnras, 340, 908,
  \dodoi{10.1046/j.1365-8711.2003.06356.x}

\bibitem[{{Kemp} {et~al.}(2022){Kemp}, {Karakas}, {Casey}, {Kobayashi}, \&
  {Izzard}}]{kem22}
{Kemp}, A.~J., {Karakas}, A.~I., {Casey}, A.~R., {Kobayashi}, C., \& {Izzard},
  R.~G. 2022, \mnras, 509, 1175, \dodoi{10.1093/mnras/stab3103}

\bibitem[{{Kewley} {et~al.}(2019){Kewley}, {Nicholls}, \& {Sutherland}}]{kew19}
{Kewley}, L.~J., {Nicholls}, D.~C., \& {Sutherland}, R.~S. 2019, \araa, 57,
  511, \dodoi{10.1146/annurev-astro-081817-051832}

\bibitem[{{Kobayashi}(2002)}]{kob02}
{Kobayashi}, C. 2002, PhD Thesis, University of Tokyo

\bibitem[{{Kobayashi}(2004)}]{kob04}
---. 2004, \mnras, 347, 740, \dodoi{10.1111/j.1365-2966.2004.07258.x}

\bibitem[{{Kobayashi}(2005)}]{kob05}
---. 2005, \mnras, 361, 1216, \dodoi{10.1111/j.1365-2966.2005.09248.x}

\bibitem[{{Kobayashi}(2010)}]{kob10imf}
{Kobayashi}, C. 2010, in American Institute of Physics Conference Series, Vol.
  1240, Hunting for the Dark: the Hidden Side of Galaxy Formation, ed. V.~P.
  {Debattista} \& C.~C. {Popescu}, 123--126, \dodoi{10.1063/1.3458465}

\bibitem[{{Kobayashi}(2014)}]{kob14iau}
{Kobayashi}, C. 2014, in Setting the scene for Gaia and LAMOST, ed.
  S.~{Feltzing}, G.~{Zhao}, N.~A. {Walton}, \& P.~{Whitelock}, Vol. 298,
  167--178, \dodoi{10.1017/S1743921313006339}

\bibitem[{{Kobayashi}(2016)}]{kob16iau}
{Kobayashi}, C. 2016, in The General Assembly of Galaxy Halos: Structure,
  Origin and Evolution, ed. A.~{Bragaglia}, M.~{Arnaboldi}, M.~{Rejkuba}, \&
  D.~{Romano}, Vol. 317, 57--63, \dodoi{10.1017/S1743921315009783}

\bibitem[{{Kobayashi}(2022)}]{kob22iau}
{Kobayashi}, C. 2022, in The Origin of Outflows in Evolved Stars, ed.
  L.~{Decin}, A.~{Zijlstra}, \& C.~{Gielen}, Vol. 366, 63--82,
  \dodoi{10.1017/S1743921322001132}

\bibitem[{{Kobayashi}(2023)}]{kob22uv}
---. 2023, Experimental Astronomy, 55, 75, \dodoi{10.1007/s10686-022-09862-9}

\bibitem[{{Kobayashi} \& {Arimoto}(1999)}]{kob99}
{Kobayashi}, C., \& {Arimoto}, N. 1999, \apj, 527, 573, \dodoi{10.1086/308092}

\bibitem[{{Kobayashi} {et~al.}(2014){Kobayashi}, {Ishigaki}, {Tominaga}, \&
  {Nomoto}}]{kob14}
{Kobayashi}, C., {Ishigaki}, M.~N., {Tominaga}, N., \& {Nomoto}, K. 2014,
  \apjl, 785, L5, \dodoi{10.1088/2041-8205/785/1/L5}

\bibitem[{{Kobayashi} {et~al.}(2020{\natexlab{a}}){Kobayashi}, {Karakas}, \&
  {Lugaro}}]{kob20sr}
{Kobayashi}, C., {Karakas}, A.~I., \& {Lugaro}, M. 2020{\natexlab{a}}, \apj,
  900, 179, \dodoi{10.3847/1538-4357/abae65}

\bibitem[{{Kobayashi} {et~al.}(2011{\natexlab{a}}){Kobayashi}, {Karakas}, \&
  {Umeda}}]{kob11agb}
{Kobayashi}, C., {Karakas}, A.~I., \& {Umeda}, H. 2011{\natexlab{a}}, \mnras,
  414, 3231, \dodoi{10.1111/j.1365-2966.2011.18621.x}

\bibitem[{{Kobayashi} {et~al.}(2020{\natexlab{b}}){Kobayashi}, {Leung}, \&
  {Nomoto}}]{kob20ia}
{Kobayashi}, C., {Leung}, S.-C., \& {Nomoto}, K. 2020{\natexlab{b}}, \apj, 895,
  138, \dodoi{10.3847/1538-4357/ab8e44}

\bibitem[{{Kobayashi} \& {Nakasato}(2011)}]{kob11mw}
{Kobayashi}, C., \& {Nakasato}, N. 2011, \apj, 729, 16,
  \dodoi{10.1088/0004-637X/729/1/16}

\bibitem[{{Kobayashi} \& {Nomoto}(2009)}]{kob09}
{Kobayashi}, C., \& {Nomoto}, K. 2009, \apj, 707, 1466,
  \dodoi{10.1088/0004-637X/707/2/1466}

\bibitem[{{Kobayashi} {et~al.}(2015){Kobayashi}, {Nomoto}, \&
  {Hachisu}}]{kob15}
{Kobayashi}, C., {Nomoto}, K., \& {Hachisu}, I. 2015, \apjl, 804, L24,
  \dodoi{10.1088/2041-8205/804/1/L24}

\bibitem[{{Kobayashi} {et~al.}(2007){Kobayashi}, {Springel}, \&
  {White}}]{kob07}
{Kobayashi}, C., {Springel}, V., \& {White}, S. D.~M. 2007, \mnras, 376, 1465,
  \dodoi{10.1111/j.1365-2966.2007.11555.x}

\bibitem[{{Kobayashi} {et~al.}(2011{\natexlab{b}}){Kobayashi}, {Tominaga}, \&
  {Nomoto}}]{kob11dla}
{Kobayashi}, C., {Tominaga}, N., \& {Nomoto}, K. 2011{\natexlab{b}}, \apjl,
  730, L14, \dodoi{10.1088/2041-8205/730/2/L14}

\bibitem[{{Kobayashi} {et~al.}(2000){Kobayashi}, {Tsujimoto}, \&
  {Nomoto}}]{kob00}
{Kobayashi}, C., {Tsujimoto}, T., \& {Nomoto}, K. 2000, \apj, 539, 26,
  \dodoi{10.1086/309195}

\bibitem[{{Kobayashi} {et~al.}(1998){Kobayashi}, {Tsujimoto}, {Nomoto},
  {Hachisu}, \& {Kato}}]{kob98}
{Kobayashi}, C., {Tsujimoto}, T., {Nomoto}, K., {Hachisu}, I., \& {Kato}, M.
  1998, \apjl, 503, L155, \dodoi{10.1086/311556}

\bibitem[{{Kobayashi} {et~al.}(2006){Kobayashi}, {Umeda}, {Nomoto}, {Tominaga},
  \& {Ohkubo}}]{kob06}
{Kobayashi}, C., {Umeda}, H., {Nomoto}, K., {Tominaga}, N., \& {Ohkubo}, T.
  2006, \apj, 653, 1145, \dodoi{10.1086/508914}

\bibitem[{{Kobayashi} {et~al.}(2023){Kobayashi}, {Mandel}, {Belczynski},
  {Goriely}, {Janka}, {Just}, {Ruiter}, {Vanbeveren}, {Kruckow}, {Briel},
  {Eldridge}, \& {Stanway}}]{kob22}
{Kobayashi}, C., {Mandel}, I., {Belczynski}, K., {et~al.} 2023, \apjl, 943,
  L12, \dodoi{10.3847/2041-8213/acad82}

\bibitem[{{Kodama} \& {Arimoto}(1997)}]{kod97}
{Kodama}, T., \& {Arimoto}, N. 1997, \aap, 320, 41.
\newblock \doarXiv{astro-ph/9609160}

\bibitem[{{Kraft} {et~al.}(1997){Kraft}, {Sneden}, {Smith}, {Shetrone},
  {Langer}, \& {Pilachowski}}]{kraft97}
{Kraft}, R.~P., {Sneden}, C., {Smith}, G.~H., {et~al.} 1997, \aj, 113, 279,
  \dodoi{10.1086/118251}

\bibitem[{{Kriek} {et~al.}(2016){Kriek}, {Conroy}, {van Dokkum}, {Shapley},
  {Choi}, {Reddy}, {Siana}, {van de Voort}, {Coil}, \& {Mobasher}}]{kriek16}
{Kriek}, M., {Conroy}, C., {van Dokkum}, P.~G., {et~al.} 2016, \nat, 540, 248,
  \dodoi{10.1038/nature20570}

\bibitem[{{Kroupa}(2008)}]{kro08}
{Kroupa}, P. 2008, in Astronomical Society of the Pacific Conference Series,
  Vol. 390, Pathways Through an Eclectic Universe, ed. J.~H. {Knapen}, T.~J.
  {Mahoney}, \& A.~{Vazdekis}, 3.
\newblock \doarXiv{0708.1164}

\bibitem[{{Kroupa} {et~al.}(1993){Kroupa}, {Tout}, \& {Gilmore}}]{kro93}
{Kroupa}, P., {Tout}, C.~A., \& {Gilmore}, G. 1993, \mnras, 262, 545,
  \dodoi{10.1093/mnras/262.3.545}

\bibitem[{{Krumholz} \& {Gnedin}(2011)}]{krumholz11}
{Krumholz}, M.~R., \& {Gnedin}, N.~Y. 2011, \apj, 729, 36,
  \dodoi{10.1088/0004-637X/729/1/36}

\bibitem[{{Leung} \& {Nomoto}(2018)}]{leu18}
{Leung}, S.-C., \& {Nomoto}, K. 2018, \apj, 861, 143,
  \dodoi{10.3847/1538-4357/aac2df}

\bibitem[{{Leung} \& {Nomoto}(2020)}]{leu20}
---. 2020, \apj, 888, 80, \dodoi{10.3847/1538-4357/ab5c1f}

\bibitem[{{Limongi} \& {Chieffi}(2018)}]{lim18}
{Limongi}, M., \& {Chieffi}, A. 2018, \apjs, 237, 13,
  \dodoi{10.3847/1538-4365/aacb24}

\bibitem[{{Lind} {et~al.}(2022){Lind}, {Nordlander}, {Wehrhahn}, {Montelius},
  {Osorio}, {Barklem}, {Af{\c{s}}ar}, {Sneden}, \& {Kobayashi}}]{lind22}
{Lind}, K., {Nordlander}, T., {Wehrhahn}, A., {et~al.} 2022, \aap, 665, A33,
  \dodoi{10.1051/0004-6361/202142195}

\bibitem[{{Madau} \& {Dickinson}(2014)}]{mad14}
{Madau}, P., \& {Dickinson}, M. 2014, \araa, 52, 415,
  \dodoi{10.1146/annurev-astro-081811-125615}

\bibitem[{{Madau} \& {Rees}(2001)}]{madau01}
{Madau}, P., \& {Rees}, M.~J. 2001, \apjl, 551, L27, \dodoi{10.1086/319848}

\bibitem[{{Maeda} \& {Nomoto}(2003)}]{mae03}
{Maeda}, K., \& {Nomoto}, K. 2003, \apj, 598, 1163, \dodoi{10.1086/378948}

\bibitem[{{Magorrian} {et~al.}(1998){Magorrian}, {Tremaine}, {Richstone},
  {Bender}, {Bower}, {Dressler}, {Faber}, {Gebhardt}, {Green}, {Grillmair},
  {Kormendy}, \& {Lauer}}]{magorrian98}
{Magorrian}, J., {Tremaine}, S., {Richstone}, D., {et~al.} 1998, \aj, 115,
  2285, \dodoi{10.1086/300353}

\bibitem[{{Maiolino} \& {Mannucci}(2019)}]{mai19}
{Maiolino}, R., \& {Mannucci}, F. 2019, \aapr, 27, 3,
  \dodoi{10.1007/s00159-018-0112-2}

\bibitem[{{Matteucci}(2001)}]{matteucci01}
{Matteucci}, F. 2001, {The chemical evolution of the Galaxy}, Vol. 253,
  \dodoi{10.1007/978-94-010-0967-6}

\bibitem[{{Matteucci}(2021)}]{matteucci21}
---. 2021, \aapr, 29, 5, \dodoi{10.1007/s00159-021-00133-8}

\bibitem[{{Matteucci} \& {Brocato}(1990)}]{mat90}
{Matteucci}, F., \& {Brocato}, E. 1990, \apj, 365, 539, \dodoi{10.1086/169508}

\bibitem[{{Mennekens} \& {Vanbeveren}(2014)}]{men14}
{Mennekens}, N., \& {Vanbeveren}, D. 2014, \aap, 564, A134,
  \dodoi{10.1051/0004-6361/201322198}

\bibitem[{{Mennekens} \& {Vanbeveren}(2016)}]{men16}
---. 2016, \aap, 589, A64, \dodoi{10.1051/0004-6361/201628193}

\bibitem[{{Meynet} \& {Maeder}(2002)}]{mey02}
{Meynet}, G., \& {Maeder}, A. 2002, \aap, 390, 561,
  \dodoi{10.1051/0004-6361:20020755}

\bibitem[{{Mihos} \& {Hernquist}(1996)}]{mih96}
{Mihos}, J.~C., \& {Hernquist}, L. 1996, \apj, 464, 641, \dodoi{10.1086/177353}

\bibitem[{{M{\"o}sta} {et~al.}(2018){M{\"o}sta}, {Roberts}, {Halevi}, {Ott},
  {Lippuner}, {Haas}, \& {Schnetter}}]{mos18}
{M{\"o}sta}, P., {Roberts}, L.~F., {Halevi}, G., {et~al.} 2018, \apj, 864, 171,
  \dodoi{10.3847/1538-4357/aad6ec}

\bibitem[{{Mukherjee} {et~al.}(2018){Mukherjee}, {Bicknell}, {Wagner},
  {Sutherland}, \& {Silk}}]{dipanjan18}
{Mukherjee}, D., {Bicknell}, G.~V., {Wagner}, A.~Y., {Sutherland}, R.~S., \&
  {Silk}, J. 2018, \mnras, 479, 5544, \dodoi{10.1093/mnras/sty1776}

\bibitem[{{Navarro} \& {White}(1993)}]{nav93}
{Navarro}, J.~F., \& {White}, S.~D.~M. 1993, \mnras, 265, 271,
  \dodoi{10.1093/mnras/265.2.271}

\bibitem[{{Navarro} \& {White}(1994)}]{nav94}
{Navarro}, J.~F., \& {White}, S. D.~M. 1994, \mnras, 267, 401,
  \dodoi{10.1093/mnras/267.2.401}

\bibitem[{{Nishimura} {et~al.}(2015){Nishimura}, {Takiwaki}, \&
  {Thielemann}}]{nis15}
{Nishimura}, N., {Takiwaki}, T., \& {Thielemann}, F.-K. 2015, \apj, 810, 109,
  \dodoi{10.1088/0004-637X/810/2/109}

\bibitem[{{Nomoto}(1987)}]{nom87}
{Nomoto}, K. 1987, \apj, 322, 206, \dodoi{10.1086/165716}

\bibitem[{{Nomoto} {et~al.}(1997){Nomoto}, {Iwamoto}, {Nakasato}, {Thielemann},
  {Brachwitz}, {Tsujimoto}, {Kubo}, \& {Kishimoto}}]{nom97ia}
{Nomoto}, K., {Iwamoto}, K., {Nakasato}, N., {et~al.} 1997, \nphysa, 621, 467,
  \dodoi{10.1016/S0375-9474(97)00291-1}

\bibitem[{{Nomoto} {et~al.}(2013){Nomoto}, {Kobayashi}, \& {Tominaga}}]{nom13}
{Nomoto}, K., {Kobayashi}, C., \& {Tominaga}, N. 2013, \araa, 51, 457,
  \dodoi{10.1146/annurev-astro-082812-140956}

\bibitem[{{Pagel}(1997)}]{pagel97}
{Pagel}, B. E.~J. 1997, {Nucleosynthesis and Chemical Evolution of Galaxies}

\bibitem[{{Pakmor} {et~al.}(2011){Pakmor}, {Bauer}, \& {Springel}}]{pakmor11}
{Pakmor}, R., {Bauer}, A., \& {Springel}, V. 2011, \mnras, 418, 1392,
  \dodoi{10.1111/j.1365-2966.2011.19591.x}

\bibitem[{{Pilkington} {et~al.}(2012){Pilkington}, {Few}, {Gibson}, {Calura},
  {Michel-Dansac}, {Thacker}, {Moll{\'a}}, {Matteucci}, {Rahimi}, {Kawata},
  {Kobayashi}, {Brook}, {Stinson}, {Couchman}, {Bailin}, \& {Wadsley}}]{pik12}
{Pilkington}, K., {Few}, C.~G., {Gibson}, B.~K., {et~al.} 2012, \aap, 540, A56,
  \dodoi{10.1051/0004-6361/201117466}

\bibitem[{{Pillepich} {et~al.}(2018){Pillepich}, {Springel}, {Nelson}, {Genel},
  {Naiman}, {Pakmor}, {Hernquist}, {Torrey}, {Vogelsberger}, {Weinberger}, \&
  {Marinacci}}]{pillepich18a}
{Pillepich}, A., {Springel}, V., {Nelson}, D., {et~al.} 2018, \mnras, 473,
  4077, \dodoi{10.1093/mnras/stx2656}

\bibitem[{{Portinari} {et~al.}(1998){Portinari}, {Chiosi}, \&
  {Bressan}}]{por98}
{Portinari}, L., {Chiosi}, C., \& {Bressan}, A. 1998, \aap, 334, 505.
\newblock \doarXiv{astro-ph/9711337}

\bibitem[{{Prantzos} {et~al.}(2018){Prantzos}, {Abia}, {Limongi}, {Chieffi}, \&
  {Cristallo}}]{pra18}
{Prantzos}, N., {Abia}, C., {Limongi}, M., {Chieffi}, A., \& {Cristallo}, S.
  2018, \mnras, 476, 3432, \dodoi{10.1093/mnras/sty316}

\bibitem[{{Prantzos} {et~al.}(1993){Prantzos}, {Casse}, \&
  {Vangioni-Flam}}]{pra93}
{Prantzos}, N., {Casse}, M., \& {Vangioni-Flam}, E. 1993, \apj, 403, 630,
  \dodoi{10.1086/172233}

\bibitem[{{Reichert} {et~al.}(2021){Reichert}, {Obergaulinger}, {Eichler},
  {Aloy}, \& {Arcones}}]{rei21}
{Reichert}, M., {Obergaulinger}, M., {Eichler}, M., {Aloy}, M.~{\'A}., \&
  {Arcones}, A. 2021, \mnras, 501, 5733, \dodoi{10.1093/mnras/stab029}

\bibitem[{{Roederer} {et~al.}(2014){Roederer}, {Preston}, {Thompson},
  {Shectman}, {Sneden}, {Burley}, \& {Kelson}}]{roe14}
{Roederer}, I.~U., {Preston}, G.~W., {Thompson}, I.~B., {et~al.} 2014, \aj,
  147, 136, \dodoi{10.1088/0004-6256/147/6/136}

\bibitem[{{Romano} {et~al.}(2010){Romano}, {Karakas}, {Tosi}, \&
  {Matteucci}}]{rom10}
{Romano}, D., {Karakas}, A.~I., {Tosi}, M., \& {Matteucci}, F. 2010, \aap, 522,
  A32, \dodoi{10.1051/0004-6361/201014483}

\bibitem[{{Romano} {et~al.}(2019){Romano}, {Matteucci}, {Zhang}, {Ivison}, \&
  {Ventura}}]{rom19}
{Romano}, D., {Matteucci}, F., {Zhang}, Z.-Y., {Ivison}, R.~J., \& {Ventura},
  P. 2019, \mnras, 490, 2838, \dodoi{10.1093/mnras/stz2741}

\bibitem[{{Rowan-Robinson} {et~al.}(2016){Rowan-Robinson}, {Oliver}, {Wang},
  {Farrah}, {Clements}, {Gruppioni}, {Marchetti}, {Rigopoulou}, \&
  {Vaccari}}]{rowanrobinson16}
{Rowan-Robinson}, M., {Oliver}, S., {Wang}, L., {et~al.} 2016, \mnras, 461,
  1100, \dodoi{10.1093/mnras/stw1169}

\bibitem[{{Salpeter}(1955)}]{sal55}
{Salpeter}, E.~E. 1955, \apj, 121, 161, \dodoi{10.1086/145971}

\bibitem[{{Scannapieco} {et~al.}(2012){Scannapieco}, {Wadepuhl}, {Parry},
  {Navarro}, {Jenkins}, {Springel}, {Teyssier}, {Carlson}, {Couchman}, {Crain},
  {Dalla Vecchia}, {Frenk}, {Kobayashi}, {Monaco}, {Murante}, {Okamoto},
  {Quinn}, {Schaye}, {Stinson}, {Theuns}, {Wadsley}, {White}, \&
  {Woods}}]{sca12}
{Scannapieco}, C., {Wadepuhl}, M., {Parry}, O.~H., {et~al.} 2012, \mnras, 423,
  1726, \dodoi{10.1111/j.1365-2966.2012.20993.x}

\bibitem[{{Schaye} {et~al.}(2015){Schaye}, {Crain}, {Bower}, {Furlong},
  {Schaller}, {Theuns}, {Dalla Vecchia}, {Frenk}, {McCarthy}, {Helly},
  {Jenkins}, {Rosas-Guevara}, {White}, {Baes}, {Booth}, {Camps}, {Navarro},
  {Qu}, {Rahmati}, {Sawala}, {Thomas}, \& {Trayford}}]{sch15}
{Schaye}, J., {Crain}, R.~A., {Bower}, R.~G., {et~al.} 2015, \mnras, 446, 521,
  \dodoi{10.1093/mnras/stu2058}

\bibitem[{{Seitenzahl} {et~al.}(2013){Seitenzahl}, {Cescutti}, {R{\"o}pke},
  {Ruiter}, \& {Pakmor}}]{sei13}
{Seitenzahl}, I.~R., {Cescutti}, G., {R{\"o}pke}, F.~K., {Ruiter}, A.~J., \&
  {Pakmor}, R. 2013, \aap, 559, L5, \dodoi{10.1051/0004-6361/201322599}

\bibitem[{{Siegel} {et~al.}(2019){Siegel}, {Barnes}, \& {Metzger}}]{sie19}
{Siegel}, D.~M., {Barnes}, J., \& {Metzger}, B.~D. 2019, \nat, 569, 241,
  \dodoi{10.1038/s41586-019-1136-0}

\bibitem[{{Smartt}(2009)}]{sma09}
{Smartt}, S.~J. 2009, \araa, 47, 63,
  \dodoi{10.1146/annurev-astro-082708-101737}

\bibitem[{{Sneden} {et~al.}(2016){Sneden}, {Cowan}, {Kobayashi}, {Pignatari},
  {Lawler}, {Den Hartog}, \& {Wood}}]{sne16}
{Sneden}, C., {Cowan}, J.~J., {Kobayashi}, C., {et~al.} 2016, \apj, 817, 53,
  \dodoi{10.3847/0004-637X/817/1/53}

\bibitem[{{Spite} {et~al.}(2005){Spite}, {Cayrel}, {Plez}, {Hill}, {Spite},
  {Depagne}, {Fran{\c{c}}ois}, {Bonifacio}, {Barbuy}, {Beers}, {Andersen},
  {Molaro}, {Nordstr{\"o}m}, \& {Primas}}]{spi05}
{Spite}, M., {Cayrel}, R., {Plez}, B., {et~al.} 2005, \aap, 430, 655,
  \dodoi{10.1051/0004-6361:20041274}

\bibitem[{{Spitoni} {et~al.}(2019){Spitoni}, {Silva Aguirre}, {Matteucci},
  {Calura}, \& {Grisoni}}]{spitoni19}
{Spitoni}, E., {Silva Aguirre}, V., {Matteucci}, F., {Calura}, F., \&
  {Grisoni}, V. 2019, \aap, 623, A60, \dodoi{10.1051/0004-6361/201834188}

\bibitem[{{Springel}(2010)}]{spr10}
{Springel}, V. 2010, \mnras, 401, 791, \dodoi{10.1111/j.1365-2966.2009.15715.x}

\bibitem[{{Springel} {et~al.}(2005){Springel}, {Di Matteo}, \&
  {Hernquist}}]{spr05agn}
{Springel}, V., {Di Matteo}, T., \& {Hernquist}, L. 2005, \mnras, 361, 776,
  \dodoi{10.1111/j.1365-2966.2005.09238.x}

\bibitem[{{Springel} \& {Hernquist}(2003)}]{spr03}
{Springel}, V., \& {Hernquist}, L. 2003, \mnras, 339, 289,
  \dodoi{10.1046/j.1365-8711.2003.06206.x}

\bibitem[{{Springel} {et~al.}(2001){Springel}, {Yoshida}, \& {White}}]{spr01}
{Springel}, V., {Yoshida}, N., \& {White}, S. D.~M. 2001, \na, 6, 79,
  \dodoi{10.1016/S1384-1076(01)00042-2}

\bibitem[{{Steinmetz} \& {Mueller}(1994)}]{ste94}
{Steinmetz}, M., \& {Mueller}, E. 1994, \aap, 281, L97.
\newblock \doarXiv{astro-ph/9312010}

\bibitem[{{Sutherland} \& {Dopita}(1993)}]{sut93}
{Sutherland}, R.~S., \& {Dopita}, M.~A. 1993, \apjs, 88, 253,
  \dodoi{10.1086/191823}

\bibitem[{{Taylor} \& {Kobayashi}(2014)}]{tay14}
{Taylor}, P., \& {Kobayashi}, C. 2014, \mnras, 442, 2751,
  \dodoi{10.1093/mnras/stu983}

\bibitem[{{Taylor} \& {Kobayashi}(2015{\natexlab{a}})}]{tay15letter}
---. 2015{\natexlab{a}}, \mnras, 452, L59, \dodoi{10.1093/mnrasl/slv087}

\bibitem[{{Taylor} \& {Kobayashi}(2015{\natexlab{b}})}]{tay15}
---. 2015{\natexlab{b}}, \mnras, 448, 1835, \dodoi{10.1093/mnras/stv139}

\bibitem[{{Taylor} \& {Kobayashi}(2016)}]{tay16}
---. 2016, \mnras, 463, 2465, \dodoi{10.1093/mnras/stw2157}

\bibitem[{{Taylor} \& {Kobayashi}(2017)}]{tay17}
---. 2017, \mnras, 471, 3856, \dodoi{10.1093/mnras/stx1860}

\bibitem[{{Taylor} {et~al.}(2020){Taylor}, {Kobayashi}, \& {Kewley}}]{tay20}
{Taylor}, P., {Kobayashi}, C., \& {Kewley}, L.~J. 2020, \mnras, 496, 4433,
  \dodoi{10.1093/mnras/staa1904}

\bibitem[{{Teyssier}(2002)}]{tey02}
{Teyssier}, R. 2002, \aap, 385, 337, \dodoi{10.1051/0004-6361:20011817}

\bibitem[{{Thomas} {et~al.}(2005){Thomas}, {Maraston}, {Bender}, \& {Mendes de
  Oliveira}}]{tho05}
{Thomas}, D., {Maraston}, C., {Bender}, R., \& {Mendes de Oliveira}, C. 2005,
  \apj, 621, 673, \dodoi{10.1086/426932}

\bibitem[{{Timmes} {et~al.}(1995){Timmes}, {Woosley}, \& {Weaver}}]{tim95}
{Timmes}, F.~X., {Woosley}, S.~E., \& {Weaver}, T.~A. 1995, \apjs, 98, 617,
  \dodoi{10.1086/192172}

\bibitem[{{Tinsley}(1980)}]{tinsley80}
{Tinsley}, B.~M. 1980, \fcp, 5, 287

\bibitem[{{Tominaga}(2009)}]{tom09}
{Tominaga}, N. 2009, \apj, 690, 526, \dodoi{10.1088/0004-637X/690/1/526}

\bibitem[{{Umeda} \& {Nomoto}(2003)}]{ume03}
{Umeda}, H., \& {Nomoto}, K. 2003, \nat, 422, 871, \dodoi{10.1038/nature01571}

\bibitem[{{van de Voort} {et~al.}(2022){van de Voort}, {Pakmor}, {Bieri}, \&
  {Grand}}]{van22}
{van de Voort}, F., {Pakmor}, R., {Bieri}, R., \& {Grand}, R. J.~J. 2022,
  \mnras, 512, 5258, \dodoi{10.1093/mnras/stac710}

\bibitem[{{van de Voort} {et~al.}(2020){van de Voort}, {Pakmor}, {Grand},
  {Springel}, {G{\'o}mez}, \& {Marinacci}}]{van20}
{van de Voort}, F., {Pakmor}, R., {Grand}, R. J.~J., {et~al.} 2020, \mnras,
  494, 4867, \dodoi{10.1093/mnras/staa754}

\bibitem[{{van Dokkum} \& {Conroy}(2010)}]{vandokkum10}
{van Dokkum}, P.~G., \& {Conroy}, C. 2010, \nat, 468, 940,
  \dodoi{10.1038/nature09578}

\bibitem[{{Ventura} \& {D'Antona}(2009)}]{ventura09}
{Ventura}, P., \& {D'Antona}, F. 2009, \aap, 499, 835,
  \dodoi{10.1051/0004-6361/200811139}

\bibitem[{{Vincenzo} \& {Kobayashi}(2018{\natexlab{a}})}]{vin18no}
{Vincenzo}, F., \& {Kobayashi}, C. 2018{\natexlab{a}}, \mnras, 478, 155,
  \dodoi{10.1093/mnras/sty1047}

\bibitem[{{Vincenzo} \& {Kobayashi}(2018{\natexlab{b}})}]{vin18a}
---. 2018{\natexlab{b}}, \aap, 610, L16, \dodoi{10.1051/0004-6361/201732395}

\bibitem[{{Vincenzo} \& {Kobayashi}(2020)}]{vin20}
---. 2020, \mnras, 496, 80, \dodoi{10.1093/mnras/staa1451}

\bibitem[{{Vincenzo} {et~al.}(2019){Vincenzo}, {Kobayashi}, \& {Yuan}}]{vin19}
{Vincenzo}, F., {Kobayashi}, C., \& {Yuan}, T. 2019, \mnras, 488, 4674,
  \dodoi{10.1093/mnras/stz2065}

\bibitem[{{Wallner} {et~al.}(2021){Wallner}, {Froehlich}, {Hotchkis},
  {Kinoshita}, {Paul}, {Martschini}, {Pavetich}, {Tims}, {Kivel}, {Schumann},
  {Honda}, {Matsuzaki}, \& {Yamagata}}]{wal21}
{Wallner}, A., {Froehlich}, M.~B., {Hotchkis}, M.~A.~C., {et~al.} 2021,
  Science, 372, 742, \dodoi{10.1126/science.aax3972}

\bibitem[{{Wallstr{\"o}m} {et~al.}(2019){Wallstr{\"o}m}, {Muller}, {Roueff},
  {Le Gal}, {Black}, \& {G{\'e}rin}}]{wallstrom19}
{Wallstr{\"o}m}, S.~H.~J., {Muller}, S., {Roueff}, E., {et~al.} 2019, \aap,
  629, A128, \dodoi{10.1051/0004-6361/201935860}

\bibitem[{{Wanajo}(2013)}]{wan13nu}
{Wanajo}, S. 2013, \apjl, 770, L22, \dodoi{10.1088/2041-8205/770/2/L22}

\bibitem[{{Wanajo} {et~al.}(2013){Wanajo}, {Janka}, \& {M{\"u}ller}}]{wan13ec}
{Wanajo}, S., {Janka}, H.-T., \& {M{\"u}ller}, B. 2013, \apjl, 767, L26,
  \dodoi{10.1088/2041-8205/767/2/L26}

\bibitem[{{Wanajo} {et~al.}(2009){Wanajo}, {Nomoto}, {Janka}, {Kitaura}, \&
  {M{\"u}ller}}]{wan09}
{Wanajo}, S., {Nomoto}, K., {Janka}, H.~T., {Kitaura}, F.~S., \& {M{\"u}ller},
  B. 2009, \apj, 695, 208, \dodoi{10.1088/0004-637X/695/1/208}

\bibitem[{{Wanajo} {et~al.}(2014){Wanajo}, {Sekiguchi}, {Nishimura}, {Kiuchi},
  {Kyutoku}, \& {Shibata}}]{wan14}
{Wanajo}, S., {Sekiguchi}, Y., {Nishimura}, N., {et~al.} 2014, \apjl, 789, L39,
  \dodoi{10.1088/2041-8205/789/2/L39}

\bibitem[{{Wang} {et~al.}(2019){Wang}, {Taylor}, {Federrath}, \&
  {Kobayashi}}]{wang19}
{Wang}, E.~X., {Taylor}, P., {Federrath}, C., \& {Kobayashi}, C. 2019, \mnras,
  483, 4640, \dodoi{10.1093/mnras/sty3491}

\bibitem[{{Weinberger} {et~al.}(2017){Weinberger}, {Springel}, {Hernquist},
  {Pillepich}, {Marinacci}, {Pakmor}, {Nelson}, {Genel}, {Vogelsberger},
  {Naiman}, \& {Torrey}}]{weinberger17}
{Weinberger}, R., {Springel}, V., {Hernquist}, L., {et~al.} 2017, \mnras, 465,
  3291, \dodoi{10.1093/mnras/stw2944}

\bibitem[{{Wiersma} {et~al.}(2009){Wiersma}, {Schaye}, \& {Smith}}]{wiersma09}
{Wiersma}, R. P.~C., {Schaye}, J., \& {Smith}, B.~D. 2009, \mnras, 393, 99,
  \dodoi{10.1111/j.1365-2966.2008.14191.x}

\bibitem[{{Winteler} {et~al.}(2012){Winteler}, {K{\"a}ppeli}, {Perego},
  {Arcones}, {Vasset}, {Nishimura}, {Liebend{\"o}rfer}, \&
  {Thielemann}}]{win12}
{Winteler}, C., {K{\"a}ppeli}, R., {Perego}, A., {et~al.} 2012, \apjl, 750,
  L22, \dodoi{10.1088/2041-8205/750/1/L22}

\bibitem[{{Woods} {et~al.}(2019){Woods}, {Agarwal}, {Bromm}, {Bunker}, {Chen},
  {Chon}, {Ferrara}, {Glover}, {Haemmerl{\'e}}, {Haiman}, {Hartwig}, {Heger},
  {Hirano}, {Hosokawa}, {Inayoshi}, {Klessen}, {Kobayashi}, {Koliopanos},
  {Latif}, {Li}, {Mayer}, {Mezcua}, {Natarajan}, {Pacucci}, {Rees}, {Regan},
  {Sakurai}, {Salvadori}, {Schneider}, {Surace}, {Tanaka}, {Whalen}, \&
  {Yoshida}}]{woods19}
{Woods}, T.~E., {Agarwal}, B., {Bromm}, V., {et~al.} 2019, \pasa, 36, e027,
  \dodoi{10.1017/pasa.2019.14}

\bibitem[{{Woosley} \& {Weaver}(1995)}]{woo95}
{Woosley}, S.~E., \& {Weaver}, T.~A. 1995, \apjs, 101, 181,
  \dodoi{10.1086/192237}

\bibitem[{{Worthey} {et~al.}(1992){Worthey}, {Faber}, \&
  {Gonzalez}}]{worthey92}
{Worthey}, G., {Faber}, S.~M., \& {Gonzalez}, J.~J. 1992, \apj, 398, 69,
  \dodoi{10.1086/171836}

\bibitem[{{Yong} {et~al.}(2021){Yong}, {Kobayashi}, {Da Costa}, {Bessell},
  {Chiti}, {Frebel}, {Lind}, {Mackey}, {Nordlander}, {Asplund}, {Casey},
  {Marino}, {Murphy}, \& {Schmidt}}]{yon21a}
{Yong}, D., {Kobayashi}, C., {Da Costa}, G.~S., {et~al.} 2021, \nat, 595, 223,
  \dodoi{10.1038/s41586-021-03611-2}

\bibitem[{{Zaritsky} {et~al.}(1994){Zaritsky}, {Kennicutt}, \&
  {Huchra}}]{zaritsky94}
{Zaritsky}, D., {Kennicutt}, Robert~C., J., \& {Huchra}, J.~P. 1994, \apj, 420,
  87, \dodoi{10.1086/173544}

\bibitem[{{Zhang} {et~al.}(2018){Zhang}, {Romano}, {Ivison}, {Papadopoulos}, \&
  {Matteucci}}]{zhang18}
{Zhang}, Z.-Y., {Romano}, D., {Ivison}, R.~J., {Papadopoulos}, P.~P., \&
  {Matteucci}, F. 2018, \nat, 558, 260, \dodoi{10.1038/s41586-018-0196-x}

\bibitem[{{Zhao} {et~al.}(2016){Zhao}, {Mashonkina}, {Yan}, {Alexeeva},
  {Kobayashi}, {Pakhomov}, {Shi}, {Sitnova}, {Tan}, {Zhang}, {Zhang}, {Zhou},
  {Bolte}, {Chen}, {Li}, {Liu}, \& {Zhai}}]{zhao16}
{Zhao}, G., {Mashonkina}, L., {Yan}, H.~L., {et~al.} 2016, \apj, 833, 225,
  \dodoi{10.3847/1538-4357/833/2/225}

\end{thebibliography}
\bibliographystyle{aasjournal}

\end{document}